\documentclass[onecolumn,twoside]{IEEEtran}
\usepackage[T1]{fontenc}
\usepackage[latin9]{inputenc}
\usepackage{amsthm}
\usepackage{amsmath} 
\usepackage{graphicx} 
\usepackage{color} 
\usepackage{cite}
\usepackage{algpseudocode}
\usepackage{algorithm}
\usepackage{amssymb}
\usepackage{subfigure}
\usepackage{esint}
\usepackage{algpseudocode}
\usepackage{epstopdf}
\usepackage{multirow}
\usepackage{makecell}
\usepackage{float}
\usepackage{lipsum}

\newtheorem{definition}{Definition}
\newtheorem{remark}{Remark}

\newtheorem{example}{Example}

\theoremstyle{plain}
\theoremstyle{plain}
\newtheorem{theorem}{Theorem}
\newtheorem{lemma}{Lemma}

\newcommand{\comment}[1]{}

\IEEEoverridecommandlockouts

\allowdisplaybreaks

\begin{document}

\title{A Channel-Aware Combinatorial Approach to Design High Performance Spatially-Coupled Codes~for Magnetic Recording Systems}

\author{Ahmed~Hareedy,~\IEEEmembership{Student~Member,~IEEE,} Ruiyi~Wu,~\IEEEmembership{Student~Member,~IEEE,} and~Lara~Dolecek,~\IEEEmembership{Senior~Member,~IEEE}\vspace{-1.0em}

\thanks{A. Hareedy, R. Wu, and L. Dolecek are with the Department of Electrical and Computer Engineering, University of California, Los Angeles, Los Angeles, CA 90095 USA (e-mail: \{ahareedy, ruiyiwu\}@ucla.edu; dolecek@ee.ucla.edu). This work was supported in part by an NSF CAREER grant and an ASTC-IDEMA grant. Part of the paper was presented at IEEE Global Communications Conference (GLOBECOM) 2018 \cite{ahh_nboo2}.}
}

\maketitle

\begin{abstract}
Because of their capacity-approaching performance and their complexity/latency advantages, spatially-coupled (SC) codes are among the most attractive error-correcting codes for use in modern dense data storage systems. SC codes are constructed by partitioning an underlying block code and coupling the partitioned components. Here, we focus on circulant-based SC codes. Recently, the optimal overlap (OO), circulant power optimizer (CPO) approach was introduced to construct high performance SC codes for additive white Gaussian noise (AWGN) and Flash channels. The OO stage operates on the protograph of the SC code to derive the optimal partitioning that minimizes the number of graphical objects that undermine the performance of SC codes under iterative decoding. Then, the CPO optimizes the circulant powers to further reduce this number. Since the nature of detrimental objects in the graph of a code critically depends on the characteristics of the channel of interest, extending the OO-CPO approach to construct SC codes for channels with intrinsic memory is not a straightforward task. In this paper, we tackle one relevant extension; we construct high performance SC codes for practical 1-D magnetic recording channels, i.e., partial-response (PR) channels. Via combinatorial techniques, we carefully build and solve the optimization problem of the OO partitioning, focusing on the objects of interest in the case of PR channels. Then, we customize the CPO to further reduce the number of these objects in the graph of the code. SC codes designed using the proposed OO-CPO approach for PR channels outperform prior state-of-the-art SC codes by up to around 3 orders of magnitude in frame error rate (FER) and 1.1 dB in signal-to-noise ratio (SNR). More intriguingly, our SC codes outperform structured block codes of the same length and rate by up to around 1.8 orders of magnitude in FER and 0.4 dB in SNR. The performance advantage of SC codes designed using the devised OO-CPO approach over block codes of the same parameters is not only pronounced in the error floor region, but also in the waterfall region.
\end{abstract}

\vspace{-0.2em}
\section{Introduction}\label{sec_intro}

As other data storage systems, magnetic recording (MR) systems operate at very low frame error rate (FER) levels \cite{vas_prc, col_detect, b_ryan, ahh_bas}. Consequently, to ensure high error correction capability in such systems, binary \cite{col_detect, b_ryan, hu_2} and non-binary (NB) \cite{ahh_bas, wat_nb, kui_nb, dec_yuta, chen_2d} graph-based codes are used. Under iterative decoding, the objects that dominate the error floor region of low-density parity-check (LDPC) codes simulated in partial-response (PR) and additive white Gaussian noise (AWGN) systems are different in their combinatorial nature because of the detector-decoder looping and the intrinsic memory in PR systems \cite{ahh_bas}. In particular, the authors in \cite{ahh_bas} introduced balanced absorbing sets (BASs) to characterize the detrimental objects in the case of PR (1-D MR) channels. Moreover, the weight consistency matrix (WCM) framework was introduced to systematically remove any type of absorbing sets (ASs) from the graph of an NB-LDPC code \cite{ahh_jsac, ahh_tit}.

Spatially-coupled (SC) codes \cite{fels_sc, pus_sc, kud_sc} are graph-based codes constructed by partitioning an underlying block code into components of the same size, then rewiring these components multiple times \cite{homa_sc}. Literature works studying the asymptotic performance of SC codes include \cite{kud_sc, lent_asy, andr_asy}. In this work, the underlying block codes, and consequently our constructed finite-length SC codes, are circulant-based (CB) codes. SC codes offer not only complexity/latency gains (if windowed decoding \cite{iye_sc} is used), but also an additional degree of freedom in the code design; this added flexibility is achieved via partitioning of the parity check matrix of the underlying block code. This observation makes SC codes attractive across a range of applications. Contiguous \cite{homa_sc} and non-contiguous \cite{homa_mo, snr_var, mitch_es} partitioning schemes were introduced in the literature for various applications. Recently, the optimal overlap (OO), circulant power optimizer (CPO) approach was introduced to design SC codes with superior performance for AWGN \cite{homa_boo} and practical asymmetric Flash \cite{ahh_nboo} channels. The OO partitioning operates on the protograph to compute the optimal set of overlap parameters that characterizes the partitioning. The CPO operates on the unlabeled graph (edge weights are set to $1$'s) to adjust the circulant powers. The objective is to minimize the number of instances of a common substructure that exists in several detrimental objects. If the SC code is binary, the unlabeled graph is the final graph. If the SC code is non-binary, the WCM framework \cite{ahh_jsac, ahh_tit} is used to optimize the edge weights after applying the OO-CPO approach.

In this paper, we propose an approach based on tools from combinatorics, optimization, and graph theory, to construct high performance time-invariant SC codes for PR channels. Unlike the case of AWGN and Flash channels (see \cite{homa_boo} and \cite{ahh_nboo}), the common substructure, whose number of instances we seek to minimize, in the case of PR channels can appear in different ways in the protograph of the SC code, making the optimization problem considerably more challenging. For that reason, we introduce the concept of the \textit{pattern}, which is a configuration in the protograph that can result in instances of the common substructure in the unlabeled graph of the SC code after lifting. We derive an optimization problem, in which we express the weighted sum of the counts (numbers of instances) of all patterns in terms of the overlap parameters. Then, we compute the optimal set of overlap parameters (OO) that minimizes this sum. Moreover, we propose the necessary modifications to the CPO algorithm presented in \cite{homa_boo} and \cite{ahh_nboo} to make it suitable for the common substructure in the case of PR channels.

We demonstrate the gains achieved by our OO-CPO (-WCM for NB SC codes) approach through tables and performance plots that compare our codes not only with SC codes, but also with CB block codes of the same length and rate. The reduction achieved by the OO-CPO approach in the number of detrimental objects reaches $92\%$ compared with the uncoupled setting and $72\%$ compared with a prior state-of-the-art SC code design technique. Furthermore, the performance gain achieved by the OO-CPO approach reaches $3$ orders of magnitude and $1.1$ dB compared with the prior state-of-the-art. Most interestingly, the proposed SC codes outperform block codes of the same parameters, and the gain reaches $1.8$ orders of magnitude and $0.4$ dB. A code threshold gain of up to $0.25$ dB is also achieved for our SC codes compared with block codes of the same parameters, highlighting that the performance advantage is there even in the early waterfall region.

The rest of the paper is organized as follows. Section~\ref{sec_prelim} introduces the necessary preliminaries. Different patterns of the common substructure are discussed in Section~\ref{sec_pats}. The analysis of the optimization problem is presented in Section~\ref{sec_oo}. The needed modifications over the baseline CPO are detailed in Section~\ref{sec_cpo}. We present our experimental results in Section~\ref{sec_exp}. Finally, the work is concluded in Section~\ref{sec_conc}.

\vspace{-0.2em}
\section{Preliminaries}\label{sec_prelim}

In this section, we review the construction of SC codes and the definitions of the objects of interest. Here, each row (resp., column) in a parity-check matrix corresponds to a check node (CN) (resp., variable node (VN)) in the equivalent graph of the matrix (the graph of the code). Additionally, each non-zero entry in a parity-check matrix corresponds to an edge in the equivalent graph of the matrix.

Since the contribution of this work (the OO-CPO) is to optimize the topology of the underlying graph, we will focus on the unlabeled graphs and binary matrices. Labeled graphs and non-binary matrices will be discussed as needed. Let $\bold{H}$ be the binary parity-check matrix of the underlying regular CB code that has column weight (VN degree) $\gamma$ and row weight (CN degree) $\kappa$. This matrix consists of $\gamma \kappa$ circulants. Each circulant is of the form $\sigma^{f_{i, j}}$, where $0 \leq i \leq \gamma-1$, $0 \leq j \leq \kappa-1$, and $\sigma$ is the $z \times z$ identity matrix after cyclically shifting its columns one unit to the left. Circulant powers are $f_{i, j}$, $\forall i,j$, and they are defined, in addition to $z$, as the lifting parameters. Separable CB (SCB) codes have $f_{i, j} = f(i)f(j)$. The underlying block codes we use to design SC codes in this work are CB codes with no zero circulants and with $z > \kappa$.

\begin{figure}[H]
\vspace{-0.5em}
\centering
\includegraphics[trim={0.25in 2.1in 0.9in 0.9in},clip,width=2.5in]{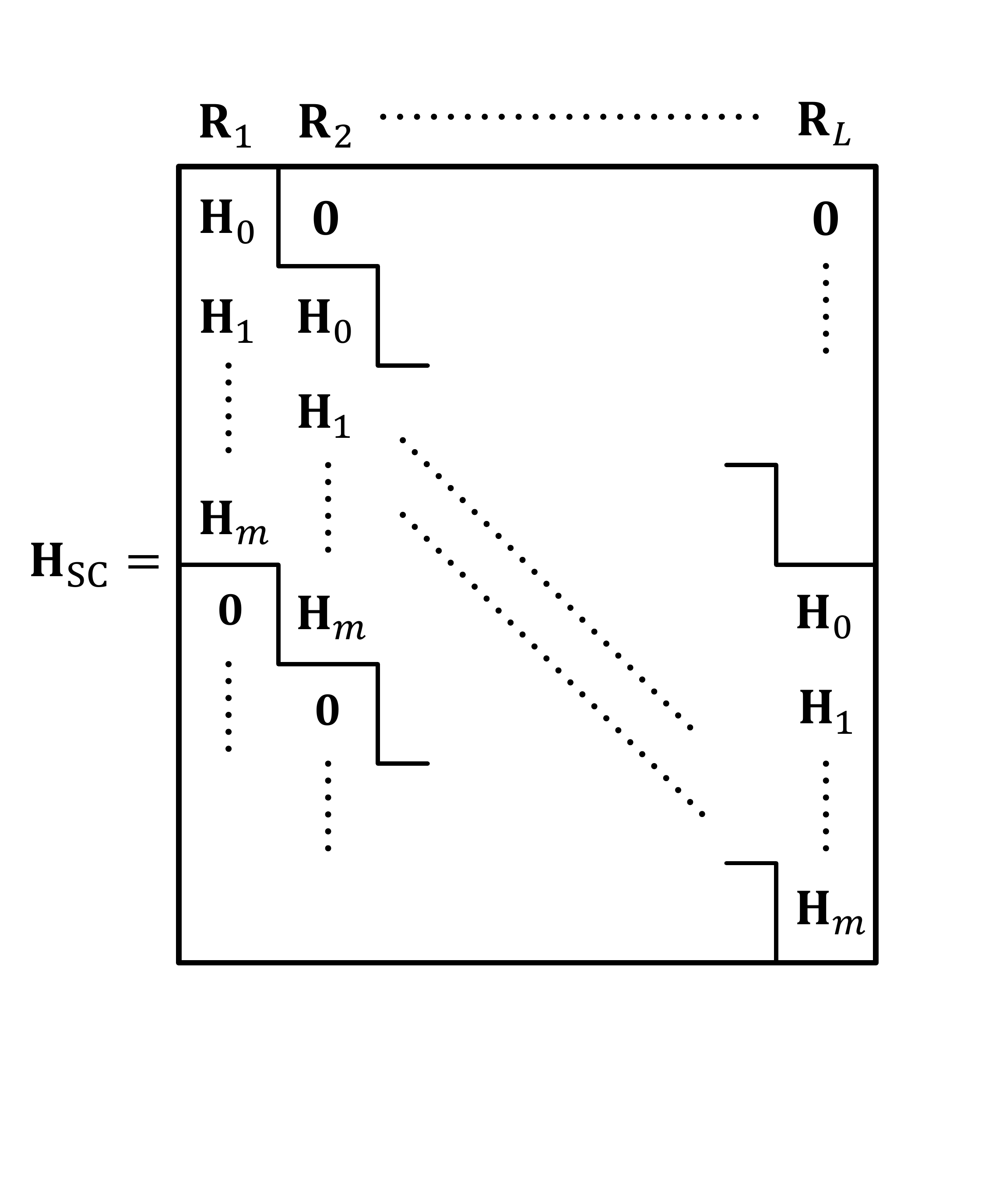}
\caption{The parity-check matrix of an SC code with parameters $m$ and $L$. Replicas are also illustrated.}
\label{Fig_hsc}
\vspace{-0.5em}
\end{figure}

The binary SC code is constructed as follows. First, $\bold{H}$ is partitioned into $(m+1)$ disjoint component matrices (they all have the same size as $\bold{H}$): $\bold{H}_0, \bold{H}_1, \dots, \bold{H}_m$, where $m$ is defined as the memory of the SC code. Each component matrix $\bold{H}_y$, $0 \leq y \leq m$, contains some of the $\gamma \kappa$ circulants of $\bold{H}$ and zero circulants elsewhere such that $\bold{H} = \sum_{y=0}^{m} \bold{H}_y$. Our approach is general; it works for any $m$ and any $\gamma \geq 3$. Then, $\bold{H}_0, \bold{H}_1, \dots, \bold{H}_m$ are coupled $L$ times, as shown in Fig.~\ref{Fig_hsc}, to construct the binary parity-check matrix of the SC code, $\bold{H}_{\textup{SC}}$, which is of size $\gamma z (L+m) \times \kappa z L$. A replica is any $\gamma z (L+m) \times \kappa z$ submatrix of $\bold{H}_{\textup{SC}}$ that contains $\left [\bold{H}^{\textup{T}}_0 \textup{ } \bold{H}^{\textup{T}}_1 \textup{ } \dots \textup{ } \bold{H}^{\textup{T}}_m \right ]^{\textup{T}}$ and zero circulants elsewhere. Replicas are denoted by $\bold{R}_\rho$, $1 \leq \rho \leq L$ (see Fig.~\ref{Fig_hsc}).

The protograph matrix (PM) of a binary CB matrix is the matrix resulting from replacing each $z \times z$ non-zero circulant with $1$, and each $z \times z$ zero circulant with $0$. The PMs of $\bold{H}$ and $\bold{H}_y$, $0 \leq y \leq m$, are $\bold{H}^{\textup{p}}$ and $\bold{H}^{\textup{p}}_y$, respectively, and they are all of size $\gamma \times \kappa$. The PM of $\bold{H}_{\textup{SC}}$ is $\bold{H}^{\textup{p}}_{\textup{SC}}$, and it is of size $\gamma (L+m) \times \kappa L$. This $\bold{H}^{\textup{p}}_{\textup{SC}}$ also has $L$ replicas, $\bold{R}_\rho$, $1 \leq \rho \leq L$, but with $1 \times 1$ circulants. Non-binary SC (NB-SC) codes can be constructed from binary SC codes as described in \cite{ahh_nboo} and guided by \cite{ahh_tit}. NB-SC codes in the finite-length regime are also discussed in \cite{irina_sc}. The NB codes we use in this work have parity-check matrices with their elements in GF($q$), where GF refers to Galois field, $q=2^\lambda$ is the GF size (order), and $\lambda \in \{2, 3, \dots\}$ (in the binary case, $q=2$).

A partitioning is contiguous if the non-zero circulants in any component matrix $\bold{H}_y$, $0 \leq y \leq m$, are contiguous; otherwise, the partitioning is non-contiguous. A technique for contiguously partitioning $\bold{H}$ to construct $\bold{H}_{\textup{SC}}$, namely cutting vector (CV) partitioning, was investigated aiming to generate SC codes for PR channels \cite{homa_sc}. Several non-contiguous partitioning techniques were recently introduced in the literature, e.g., minimum overlap (MO) partitioning \cite{homa_mo, snr_var}, general edge spreading \cite{mitch_es}, in addition to OO partitioning \cite{homa_boo, ahh_nboo}. These non-contiguous partitioning techniques significantly outperform contiguous ones \cite{homa_mo, homa_boo, ahh_nboo}. However, as far as we know, no prior work has proposed non-contiguous techniques in the context of PR channels. The goal of this work is to derive the effective OO-CPO approach for partitioning and lifting to construct high performance SC codes optimized for PR channels.

Consider the graph of an LDPC code. An $(a, b)$ AS in this graph is defined as a set of $a$ VNs with $b$ unsatisfied CNs connected to it such that each VN is connected to strictly more satisfied than unsatisfied CNs, for some set of VN values (these $a$ VNs have non-zero values, while the remaining VNs are set to zero) \cite{lara_as}. For canonical channels, e.g., the AWGN channel, elementary ASs (EASs) are the objects that dominate the error floor region of LDPC codes. EASs have the property that all satisfied CNs are of degree $2$, and all unsatisfied CNs are of degree $1$ \cite{laend_as, ahh_jsac}. Unique characteristics of storage channels (compared with the AWGN channel) result in changing the combinatorial properties of detrimental objects in graph-based codes simulated over such channels \cite{ahh_jsac}.

The intrinsic memory in PR channels \cite{ahh_bas, vas_prc} can result in VN errors having high magnitudes, which is typically not the case for canonical channels. These VN errors with high magnitudes make it very difficult for unsatisfied CNs with degree $> 1$ to participate in correcting an AS error. Consequently, it becomes more likely to have absorbing set errors with unsatisfied CNs having degree $\geq 2$, which are non-elementary absorbing set errors. Moreover, the detector-decoder looping (global iterations) help the decoder correct AS errors with higher numbers of unsatisfied CNs. Thus, the objects that dominate the error floor region of LDPC codes simulated over PR channels can be non-elementary, and they have a fewer number of unsatisfied (particularly degree-$1$) CNs, which is the reason why they are called ``balanced''. Our extensive simulations confirm these combinatorial properties of the detrimental objects in the case of PR channels. BASs and BASs of type two (BASTs) were introduced in \cite{ahh_bas} and \cite{ahh_jsac} to capture such detrimental objects.

We now present the definitions of different objects of interest. Examples of these objects of interest are in Fig.~\ref{Fig_denom}. Let $g=\left \lfloor \frac{\gamma-1}{2} \right \rfloor$, which is the maximum number of unsatisfied CNs a VN can have in an AS.

\begin{definition}\label{def_bas}
Consider a subgraph induced by a subset $\mathcal{V}$ of VNs in the (Tanner) graph of a code. Set all the VNs in $\mathcal{V}$ to values $\in$ GF($q$)$\setminus \{0\}$ and set all other VNs to $0$. The set $\mathcal{V}$ is said to be an $(a, b, d_1, d_2, d_3)$ \textbf{balanced absorbing set of type two (BAST)} over GF($q$) if the size of $\mathcal{V}$ is $a$, the number of unsatisfied CNs connected to $\mathcal{V}$ is $b$, $0 \leq b \leq \lfloor \frac{ag}{2} \rfloor$, the number of degree-$1$ (resp., $2$ and $> 2$) CNs connected to $\mathcal{V}$ is $d_1$ (resp., $d_2$ and  $d_3$), $d_2 > d_3$, all the unsatisfied CNs connected to $\mathcal{V}$ (if any) have either degree $1$ or degree $2$, and each VN in $\mathcal{V}$ is connected to strictly more satisfied than unsatisfied neighboring CNs, for some set of VN values.
\end{definition}

While the above definition was introduced in the context of non-binary codes \cite{ahh_bas, ahh_jsac}, it is valid in the binary case as well (set $q=2$, and $b$ becomes the number of odd-degree CNs). An $(a, d_1, d_2, d_3)$ unlabeled BAST (UBS) is a BAST with the weights of all edges of its graph replaced by $1$'s. All our abbreviations are short-handed for simplicity.

\begin{definition}\label{def_uts_uas}
Let $\mathcal{V}$ be a subset of VNs in the unlabeled graph (all edge weights are $1$'s) of a code. Let $\mathcal{O}$ (resp., $\mathcal{T}$ and $\mathcal{H}$) be the set of degree-$1$ (resp., $2$ and $> 2$) CNs connected to $\mathcal{V}$. This graphical configuration is an $(a, d_1)$ \textbf{unlabeled elementary trapping set (UTS)} if $|\mathcal{V}| = a$, $\vert{\mathcal{O}}\vert=d_1$, and $\vert{\mathcal{H}}\vert=0$. A UTS is an \textbf{unlabeled elementary absorbing set (UAS)} if each VN in $\mathcal{V}$ is connected to strictly more neighbors in $\mathcal{T}$ than in $\mathcal{O}$.
\end{definition}

A binary protograph configuration is also defined by $(a, d_1)$ for simplicity. The WCM framework removes a BAST from the graph of an NB code by careful processing of its edge weights (see \cite{ahh_bas}, \cite{ahh_jsac}, and \cite{ahh_tit} for details).

\section{The Common Substructure and Its Patterns}\label{sec_pats}

\begin{figure}[H]
\vspace{-1.0em}
\center
\includegraphics[width=6.0in]{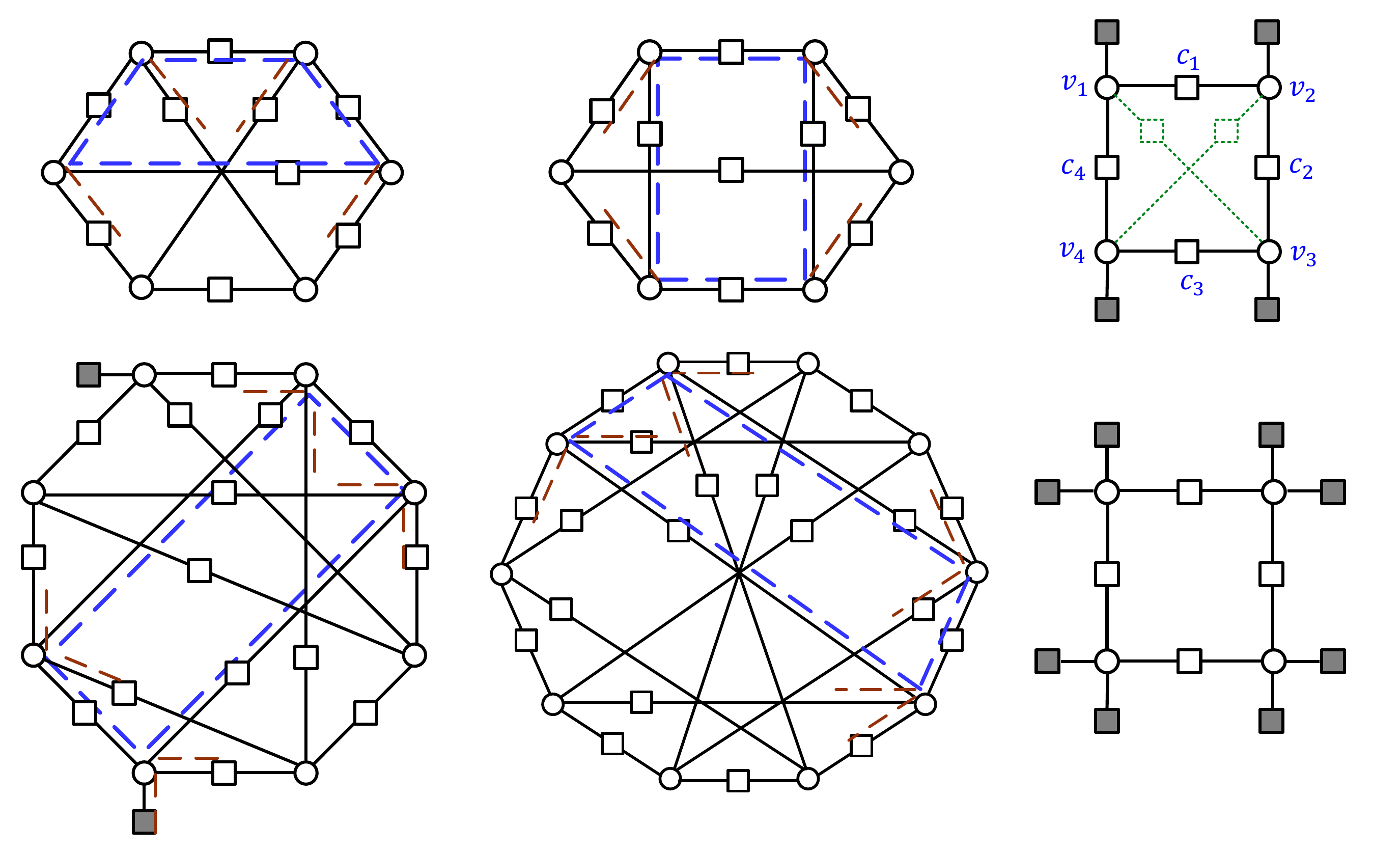}\vspace{-0.5em}
\caption{The UBSs of multiple detrimental BASTs and the associated common substructures. Upper panel ($\gamma=3$): two non-isomorphic $(6, 0, 9, 0)$ UBSs, and the common substructure is the $(4, 4)$ UAS. Lower panel ($\gamma=4$): an $(8, 2, 15, 0)$ UBS and a $(10, 0, 20, 0)$ UBS, and the common substructure is the $(4, 8)$ UTS. Common substructures are marked with dashed blue and dashed brown lines. Internal connections in a cycle of length $8$ are shown in dotted green lines in the $(4, 4)$ UAS.}
\label{Fig_denom}
\vspace{-0.5em}
\end{figure}

The idea of focusing on a common substructure in the design of the unlabeled graph of an SC code simplifies the optimization procedure. Additionally, minimizing the number of instances of the common substructure significantly reduces the multiplicity of several different types of detrimental objects simultaneously \cite{homa_sc, homa_boo}, which is a lot more feasible compared with operating on all these detrimental objects separately (especially for partitioning). It was shown in \cite{homa_sc} that the $(4, 4(\gamma-2))$ UAS/UTS, $\gamma \geq 3$, is the common substructure of interest for PR channels (unlike the case for AWGN \cite{mitch_es, homa_boo} and Flash channels \cite{ahh_nboo}, where the substructure of interest is the $(3, 3(\gamma-2))$). Fig.~\ref{Fig_denom} shows UBSs of multiple detrimental BASTs for codes with $\gamma \in \{3,4\}$ simulated over PR channels, demonstrating that the common substructure of interest is the $(4, 4(\gamma-2))$ UAS/UTS.

\begin{remark}
There are two reasons why we focus on the case of $\gamma \geq 3$ in our analysis:
\begin{enumerate}
\item Codes with $\gamma = 2$ have poor error floor performance since their graphs have high multiplicities of detrimental unlabeled low weight codewords. In fact, each cycle in a code with $\gamma = 2$ is an unlabeled codeword, i.e., an $(a, 0)$ UAS with $a$ being half the cycle length, where $a \geq 3$ if the code has girth $= 6$. In order that these codes can have better error floor performance, high GF sizes should be used in the code design, which significantly increases the complexity of decoding and thus, is not advisable for data storage \cite{dec_yuta}.

\item For codes having $\gamma = 2$, the concept of the common substructure of interest becomes inapplicable. This is because different unlabeled low weight codewords, which are $(a, 0)$ UASs, with different values of $a$ do not share any graphical structure (they are all cycles having different lengths) in these codes.
\end{enumerate}
Having said that the OO-CPO approach can still be useful to some extent for SC codes with $\gamma = 2$. Various versions of the approach, after applying some small modifications, can be used to minimize the number of $(3, 0)$ UASs (see \cite{homa_boo}) or $(4, 0)$ UASs (using the modified OO-CPO approach detailed here).
\end{remark}

We note that the $(4, 4(\gamma-2))$ UAS/UTS is a cycle of length $8$ \textit{\textbf{with no internal connections}} (ignore degree-$1$ CNs). From \cite{fos_cyc} (see also \cite{ahh_nboo}), it is known that each cycle in the unlabeled graph (the graph of $\bold{H}_{\textup{SC}}$) is derived from a configuration in the protograph (the graph of the PM $\bold{H}^{\textup{p}}_{\textup{SC}}$) under specific conditions on the powers of the circulants involved in that cycle. Thus, in the OO stage, we operate on the protograph. Then, in the CPO stage, we operate on the circulant powers.

\begin{remark}
Let $x^{-_a}$ (resp., $x^{-_b}$) be an integer s.t. $2 \leq x^{-_a} \leq x$ (resp., $0 \leq x^{-_b} \leq x$). Note that a $(4^{-_a}, {(4(\gamma-2))}^{-_b})$ configuration in the protograph of the code can result in $(4, 4(\gamma-2))$ UASs/UTSs in the unlabeled graph depending on the circulant power arrangement. Thus, in the OO stage, we operate on all protograph configurations that can result in $(4, 4(\gamma-2))$ UASs/UTSs (cycles of length $8$ with no internal connections) in the unlabeled graph, including the protograph configurations that do have internal connections. Then in the CPO stage, we treat the $(4, 4(\gamma-2))$ UASs/UTSs and the $(4, 4(\gamma-2)-2\delta)$ UASs/UTSs differently, where $\delta \in \{1,2\}$ is the number of existing internal connections in the configuration after lifting.
\end{remark}

The major difference between the $(4, 4(\gamma-2))$ UAS/UTS and the $(3, 3(\gamma-2))$ UAS/UTS is that there are multiple distinct configurations in the protograph, ignoring degree-$1$ CNs and internal connections, that can generate the former object in the unlabeled graph. We call these different configurations \textit{\textbf{patterns}}. A pattern is defined by the dimensions of the matrix of its subgraph. The following lemma investigates the number and nature of these patterns.

\begin{lemma}\label{lem_patcount}
The number of distinct patterns (with different dimensions) in the protograph of a code that can result in $(4, 4(\gamma-2))$ UASs/UTSs in the unlabeled graph of the code after lifting is $9$, in the case of $\gamma \geq 4$. The numbers of CNs and VNs in these $9$ patterns are both in $\{2,3,4\}$. This number of distinct patterns reduces to $7$ in the case of $\gamma=3$.
\end{lemma}

\begin{figure}[H]
\vspace{-1.2em}
\center
\includegraphics[trim={0.6in 0.5in 1.1in 0.5in},clip,width=4.9in]{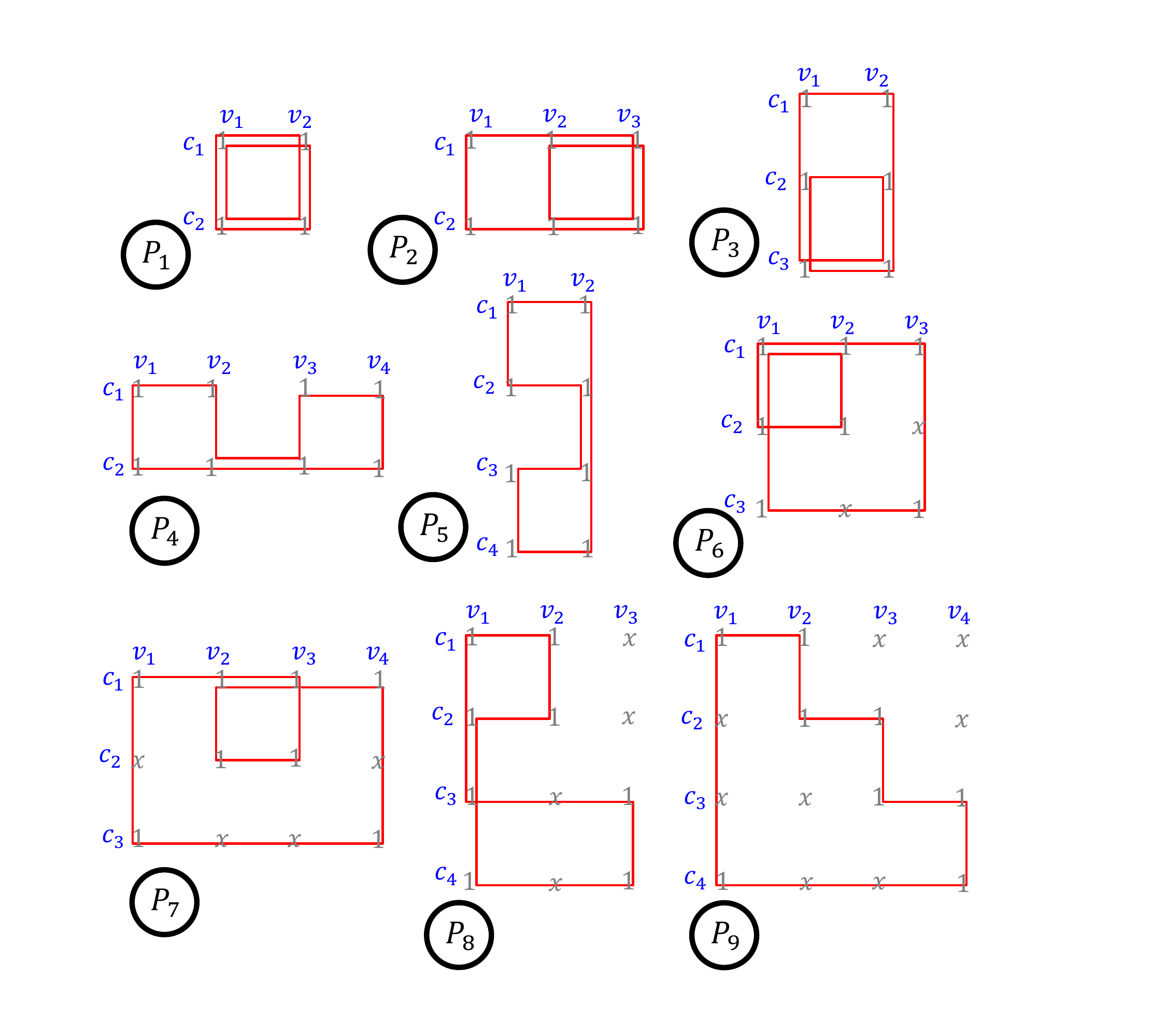}\vspace{-0.5em}
\caption{The $9$ protograph patterns that can result in cycles of length $8$ in the unlabeled graph after lifting. One way of traversing each pattern to generate cycles of length $8$ is depicted in red. Note that only Pattern $P_9$ represents a cycle of length $8$ in the protograph.}
\label{Fig_pat}
\vspace{-0.5em}
\end{figure}

\begin{IEEEproof}
Since the objects of interest in the unlabeled graph are cycles of length $8$ with $4$ CNs and $4$ VNs, a protograph pattern that can generate some of them must have at most $4$ CNs and $4$ VNs. Moreover, to result in cycles of length $8$ after lifting, the pattern must have at least $2$ CNs and $2$ VNs. Combining these two statements yields that the numbers of CNs and VNs of a protograph pattern that can result in $(4, 4(\gamma-2))$ UASs/UTSs in the unlabeled graph must be in $\{2,3,4\}$.

Consequently, in order to have $9$ distinct patterns for the case of $\gamma \geq 4$, we show that selecting any number of CNs in $\{2,3,4\}$ and any number of VNs in $\{2,3,4\}$ can result in a distinct pattern (one or more instances) that is capable of generating cycles of length $8$ in the unlabeled graph. Fig.~\ref{Fig_pat} illustrates this statement, focusing on the matrix representation of patterns and cycles. In the case of $\gamma = 3$, a pattern cannot have $4$ ones in a column, which reduces the number of distinct patterns to $7$.
\end{IEEEproof}

We define the $9$ patterns according to the dimensions of their submatrices in $\bold{H}^{\textup{p}}_{\textup{SC}}$ as follows. Pattern $P_1$ is $2 \times 2$, Pattern $P_2$ is $2 \times 3$, Pattern $P_3$ is $3 \times 2$, Pattern $P_4$ is $2 \times 4$, Pattern $P_5$ is $4 \times 2$, Pattern $P_6$ is $3 \times 3$, Pattern $P_7$ is $3 \times 4$, Pattern $P_8$ is $4 \times 3$, and Pattern $P_9$ is $4 \times 4$ (all illustrated in Fig.~\ref{Fig_pat}).

\begin{remark}
Following the same logic we used in Lemma~\ref{lem_patcount} and its proof for the $(3, 3(\gamma-2))$ UAS/UTS, leads to a possibility to also have patterns for this case, with the number of CNs and VNs in $\{2, 3\}$. However, a careful analysis guides to the fact that only one protograph pattern can result in $(3, 3(\gamma-2))$ UASs/UTSs (cycles of length $6$) after lifting, which is the $3 \times 3$ pattern, and it is itself a cycle of length $6$ \cite{homa_boo, ahh_nboo}.
\end{remark}

The following lemma discusses the relation between different protograph patterns and the resulting cycles after lifting. Define a \textit{\textbf{cycle-$8$ candidate}} of Pattern $P_\ell$ as a way to traverse $P_\ell$ in order to reach cycles of length $8$ in the unlabeled graph of the code after lifting. Some candidates are shown in Fig.~\ref{Fig_pat}.

\begin{lemma}\label{lem_trav}
Let $\zeta_{P_\ell}$ be the number of distinct cycle-$8$ candidates of Pattern $P_\ell$. Then,
\begin{align}\label{eq_cand}
\zeta_{P_\ell}=\left\{\begin{matrix}1, \textup{ } &\ell \in \{1, 6, 9\},
\\ 2, \textup{ } &\ell \in \{7, 8\},
\\ 3, \textup{ } &\ell \in \{2, 3\},
\\ 6, \textup{ } &\ell \in \{4, 5\}.
\end{matrix}\right.
\end{align}
\end{lemma}

\begin{IEEEproof}
We define a cycle-$8$ candidate according to the connectivity as follows: $c_1-v_1-c_2-v_2-c_3-v_3-c_4-v_4$ (each CN connects the next two VNs in a circular fashion, see Fig.~\ref{Fig_denom}). From Fig.~\ref{Fig_pat}, there is only one cycle-$8$ candidate for Pattern $P_1$, which is $c_1-v_1-c_2-v_2-c_1-v_1-c_2-v_2$, and this is the case for all square patterns. Thus, $\zeta_{P_\ell}=1$ for $\ell \in \{1, 6, 9\}$. It can be understood from Fig.~\ref{Fig_pat} that $\zeta_{P_\ell} \neq 1$ for all the remaining patterns. In particular, we have two cycle-$8$ candidates for Pattern $P_7$, that are: $c_1-v_1-c_2-v_2-c_1-v_3-c_3-v_4$ and $c_1-v_1-c_2-v_3-c_1-v_2-c_3-v_4$ (which is the red cycle on $P_7$ in Fig.~\ref{Fig_pat}). The situation is the same for Pattern $P_8$ because it is the transpose of $P_7$. Thus, $\zeta_{P_\ell}=2$ for $\ell \in \{7, 8\}$. The rest of the cases can be derived similarly.
\end{IEEEproof}

Pattern $P_1$ has $\zeta_{P_\ell}=1$ (see (\ref{eq_cand})), and it results in $z/2$ or $0$ cycles of length $8$ after lifting (since $P_1$ is only $2 \times 2$), while all the remaining patterns result in $z$ or $0$ cycles of length $8$ after lifting \cite{fos_cyc, ahh_nboo}. Thus, we define the \textbf{\textit{pattern weight}}, $\beta_{P_\ell}$, which plays an important role in the discrete optimization problem of the OO, as follows:
\begin{align}
\beta_{P_\ell}=\left\{\begin{matrix}1/2, \textup{ } &\ell = 1,
\\ \zeta_{P_\ell}, \textup{ } &\ell \in \{2, 3, 4, 5, 6, 7, 8, 9\}.
\end{matrix}\right.
\end{align}

\section{OO: Building and Solving the Optimization Problem}\label{sec_oo}

Now, we are ready to build the optimization problem. Consider the protograph of an SC code. The \textbf{weighted sum} of the total number of instances of all patterns is given by:
\vspace{-0.2em}\begin{equation}\label{eq_ftot}
F_{\textup{sum}}=\sum_{\ell=1}^{9} \beta_{P_\ell}F_{P_\ell},
\end{equation}
where $F_{P_\ell}$ is the total number of instances of Pattern $P_\ell$. The goal is to express $F_{\textup{sum}}$, through $F_{P_\ell}$, $\forall \ell$, as a function of the overlap parameters, then find the optimal set of overlap parameters that minimizes $F_{\textup{sum}}$ for OO partitioning. We first recall the definition and the properties of overlap parameters. More details can be found in \cite{homa_boo}.

\begin{definition}
For any $m$, let $\bold{\Pi}^1_1=\left[ \bold{H}_0^{\textup{T}} \textup{ } \bold{H}_1^{\textup{T}} \textup{ } \dots \textup{ } \bold{H}_m^{\textup{T}} \right ]^{\textup{T}}$, and let $\bold{\Pi}^{1,\textup{p}}_1$ be its PM (of size $(m+1) \gamma \times \kappa$). A \textbf{degree-$\mu$ overlap} among $\mu$ rows (or CNs) of $\bold{\Pi}^{1,\textup{p}}_1$ indexed by $\{i_1, \dots, i_{\mu}\}$, $1 \leq \mu \leq \gamma$, $0 \leq i_1, \dots, i_{\mu} \leq (m+1) \gamma-1$, is defined as a position (column) in which all these rows have $1$'s simultaneously. A \textbf{degree-$\mu$ overlap parameter}, $t_{\{i_1, \dots, i_{\mu}\}}$, is defined as the number of degree-$\mu$ overlaps among the rows indexed by $\{i_1, \dots, i_{\mu}\}$ in $\bold{\Pi}^{1,\textup{p}}_1$. A degree-$1$ overlap parameter $t_{i_1}$, $0 \leq i_1 \leq (m+1) \gamma-1$, is defined as the number of $1$'s in row $i_1$ of $\bold{\Pi}^{1,\textup{p}}_1$.
\end{definition}

Note that a degree-$\mu$ overlap parameter, if $\mu > 1$, is always zero if in the set $\{i_1, \dots, i_{\mu}\}$ there exists at least one pair of distinct row indices, say $(i_{\tau_1}, i_{\tau_2})$, with the property that $i_{\tau_1} \equiv i_{\tau_2} \textup{ } (\textup{mod } \gamma)$ \cite{homa_boo}. Define the set of all non-zero overlap parameters as $\mathcal{O}$. The parameters in $\mathcal{O}$ are not entirely independent. The set of all independent non-zero overlap parameters, $\mathcal{O}_{\textup{ind}}$, is:
\begin{align}\label{eq_oind}
\mathcal{O}_{\textup{ind}} &= \{t_{\{i_1, \dots, i_{\mu}\}} \textup{ } \vert \textup{ } 1 \leq \mu \leq \gamma, \textup{ } 0 \leq i_1, \dots, i_{\mu} \leq m\gamma-1, 
\nonumber \\ 
&\hspace{2.0em} \forall \{i_{\tau_1}, i_{\tau_2}\} \subseteq \{i_1, \dots, i_{\mu}\} \textup{ } i_{\tau_1} \not\equiv i_{\tau_2} \textup{ } (\textup{mod } \gamma)\}.
\end{align}
The other non-zero overlap parameters in $\mathcal{O} \setminus \mathcal{O}_{\textup{ind}}$ are obtained from the parameters in $\mathcal{O}_{\textup{ind}}$ according to \cite[Lemma~3]{homa_boo}. The cardinality of the set $\mathcal{O}_{\textup{ind}}$, which determines the complexity of the discrete optimization problem of the OO stage, is given by (see also \cite[Lemma~4]{homa_boo} for more details):
\vspace{-0.5em}\begin{equation}
\mathcal{N}_{\textup{ind}}= \vert \mathcal{O}_{\textup{ind}} \vert = \sum_{\mu=1}^{\gamma} m^\mu \binom{\gamma}{\mu} = (m+1)^\gamma-1.
\end{equation}
As demonstrated in Fig.~\ref{Fig_pat}, for all the patterns of interest, the highest overlap degree is $\mu=4$ (a pattern has at most $4$ CNs). Note that while the overlap parameters themselves must be restricted to $\bold{\Pi}^{1,\textup{p}}_1$, the concept of the degree-$\mu$ overlap can be generalized from $\bold{\Pi}^{1,\textup{p}}_1$ to the PM of the SC code, $\bold{H}^{\textup{p}}_{\textup{SC}}$. We will use this generalization in the analysis of patterns.

We aim at expressing $F_{P_\ell}$, $\forall \ell$, in terms of the parameters in $\mathcal{O}_{\textup{ind}}$. Let $\bold{R}_r$ be a replica in which at least one VN of the pattern being studied exists. We call $\bold{R}_r$ the reference replica. Moreover, let the CNs (or rows) of the pattern be of the form $c_x = (r-1)\gamma + i_x$, $1 \leq x \leq 4$. Here, $c_x$ is the index of the row in $\bold{H}^{\textup{p}}_{\textup{SC}}$ corresponding to the CN. In the following, we consider the protograph of an SC code with parameters $\gamma \geq 3$, $\kappa$, $m$, $L$, and $\mathcal{O}$. We define $(x)^+=\max\{x,0\}$, and $F^k_{P_\ell,1}$ as the number of instances of Pattern $P_\ell$ that start at replica $\bold{R}_1$ and span $k$ consecutive replicas. Here, ``start'' and ``span'' are both with respect to the VNs of these instances. Note that each VN in a pattern corresponds to an overlap (see Fig.~\ref{Fig_pat}).

As we shall see later, a Pattern $P_\ell$ spans at most $\chi$ consecutive replicas, where $\chi$ either $= m+1$ or $= 2m+1$, depending on the value of $\ell$. Thus, in the math, we consider the case of $L \geq \chi$.

We say here that $i_x$ is the \textbf{\textit{start of replica $\bold{R}_\rho$}} if $i_x$ is the index of the first non-zero row in $\bold{R}_\rho$ relative to $\bold{R}_r$. We also say that $i_y$ is the \textbf{\textit{end of replica $\bold{R}_\rho$}} if $i_y$ is the index of the last non-zero row in $\bold{R}_\rho$ relative to $\bold{R}_r$. In particular, the start and end of replica $\bold{R}_{r+\nu}$ are $\nu \gamma$ and $(m+\nu+1)\gamma-1$, respectively. For example, the start and end of $\bold{R}_r$ are $0$ and $(m+1)\gamma-1$, respectively, regardless from the value of $r$ since $\bold{R}_r$ is the reference replica. Moreover, the start and end of $\bold{R}_{r+2}$ (resp., $\bold{R}_{r-1}$) are $2\gamma$ and $(m+3)\gamma-1$ (resp., $-\gamma$ and $m\gamma-1$). Furthermore, the indices $1$, $h$, $w$, and $k$ of replicas are always s.t. $1 < k$ for two replicas, $1 < h < k$ for three replicas, and $1 < h < w < k$ for four replicas.

The counts of different existence possibilities of the nine patterns in addition to the final formulas of $F_{P_\ell}$, $\forall \ell$, are presented in the forthcoming subsections. The proofs of all lemmas and theorems in this section are in Appendices \ref{sec_appa}, \ref{sec_appb}, \ref{sec_appc}, \ref{sec_appd}, \ref{sec_appe}, \ref{sec_appf}, \ref{sec_appg}, \ref{sec_apph}, and \ref{sec_appi}.

\subsection{Analysis of Pattern $P_1$ (size $2 \times 2$)}\label{subsec_p1}

This pattern has two VNs that are \textit{adjacent} (connected via at least one path with only one CN). Thus, Pattern $P_1$ has its VNs located in at most two replicas, and the pattern spans (i.e., its VNs span) at most $m+1$ consecutive replicas (see \cite[Lemma~1]{homa_boo}). Suppose $P_1$ has the CNs $c_1$ and $c_2$. The two overlaps forming the pattern are of degree $2$, and they are both $c_1-c_2$ overlaps (among $c_1$ and $c_2$).

\begin{lemma}\label{lem_p1}
Case~1.1: The number of instances of $P_1$ with CNs $c_1$ and $c_2$, and all overlaps in one replica, $\bold{R}_r$, is:
\begin{equation}\label{eq_p1_1}
\mathcal{A}_{P_1}\left (t_{\{i_1,i_2\}} \right ) = \binom{t_{\{i_1,i_2\}}}{2}.
\end{equation}
Case~1.2: The number of instances of $P_1$ with CNs $c_1$ and $c_2$, and overlaps in two replicas, $\bold{R}_r$ and $\bold{R}_e$, $r < e$, is:
\begin{align}\label{eq_p1_2}
\mathcal{B}_{P_1} \left (t_{\{i_1,i_2\}}, t_{\{i_1+(r-e)\gamma, i_2+(r-e)\gamma \}} \right ) = t_{\{i_1,i_2\}}t_{\{i_1+(r-e)\gamma, i_2+(r-e)\gamma \}}.
\end{align}
\end{lemma}

\begin{figure}[H]
\vspace{-1.6em}
\center
\includegraphics[width=3.5in]{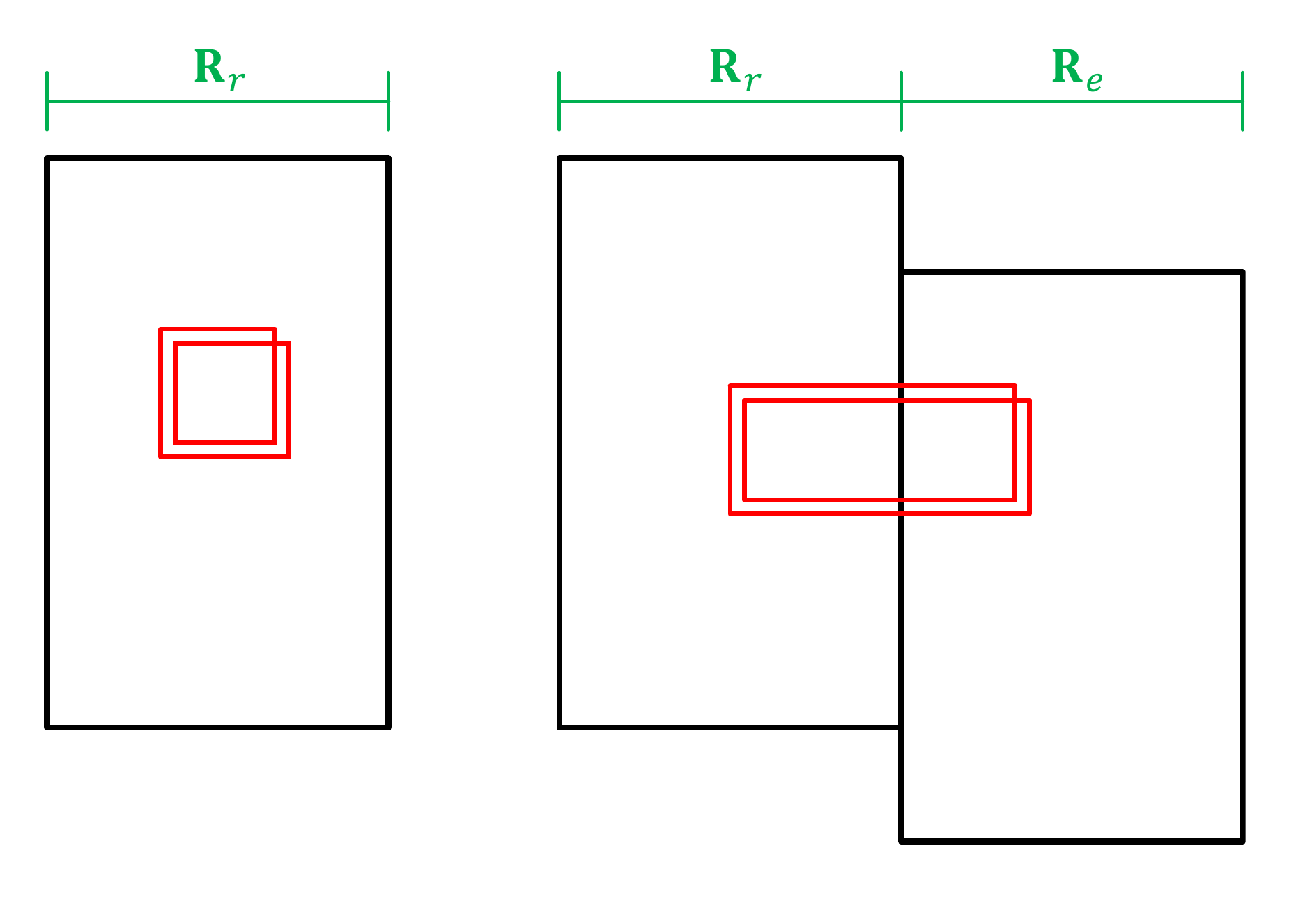}\vspace{-1.5em}
\caption{An instance of Pattern $P_1$ in Case~1.1 and in Case~1.2, from left to right. For simplicity, we have $e=r+1$.}
\label{Fig_pat1}
\vspace{-0.5em}
\end{figure}

The two cases are illustrated in Fig.~\ref{Fig_pat1}.

\begin{theorem}\label{thm_p1}
The total number of instances of Pattern $P_1$ in the binary protograph of an SC code that has parameters $\gamma \geq 3$, $\kappa$, $m$, $L \geq m+1$, and $\mathcal{O}$, is:
\begin{equation}\label{eq_p1_3}
F_{P_1} = \sum_{k=1}^{m+1} (L-k+1) F^k_{P_1,1},
\end{equation}
where $F^k_{P_1,1}$, $k \in \{1, 2, \dots, m+1\}$, are given by:
\vspace{-0.1em}\begin{align} 
F^1_{P_1,1} &= \hspace{-7.0em} \sum_{\hspace{7.0em}\{i_1,i_2\} \subset \{0, \dots, (m+1)\gamma -1\}} \hspace{-6.0em} \mathcal{A}_{P_1}\left (t_{\{i_1,i_2\}} \right ),
\nonumber \\
F^{k \geq 2}_{P_1,1} &= \hspace{-9.0em} \sum_{\hspace{9.0em}\{i_1,i_2\} \subset \{(k-1) \gamma, \dots, (m+1)\gamma -1\}} \hspace{-8.0em} \mathcal{B}_{P_1} \left (t_{\{i_1,i_2\}}, t_{\{i_1+(1-k)\gamma, i_2+(1-k)\gamma \}} \right ),
\end{align}\label{eq_p1_4}
with $\overline{i_1} \neq \overline{i_2}$, and $\overline{i_x}$ is defined by: $\overline{i_x}=(i_x \textup{ mod } \gamma)$.
\end{theorem}

\subsection{Analysis of Pattern $P_2$ (size $2 \times 3$)}\label{subsec_p2}

This pattern has three VNs, with each two of them being adjacent. Thus, $P_2$ spans at most $m+1$ consecutive replicas. Suppose $P_2$ has the CNs $c_1$ and $c_2$. The three overlaps forming $P_2$ are of degree $2$, and they are all $c_1-c_2$ overlaps.

\begin{lemma}\label{lem_p2}
Case~2.1: The number of instances of $P_2$ with CNs $c_1$ and $c_2$, and all overlaps in one replica, $\bold{R}_r$, is:
\begin{equation}\label{eq_p2_1}
\mathcal{A}_{P_2}\left (t_{\{i_1,i_2\}} \right ) = \binom{t_{\{i_1,i_2\}}}{3}.
\end{equation}
Case~2.2: The number of instances of $P_2$ with CNs $c_1$ and $c_2$, and all overlaps in two replicas s.t. two overlaps are in $\bold{R}_r$, and one overlap is in $\bold{R}_e$, is:
\begin{align}\label{eq_p2_2}
\mathcal{B}_{P_2} \left (t_{\{i_1,i_2\}}, t_{\{i_1+(r-e)\gamma, i_2+(r-e)\gamma \}} \right ) = \binom{t_{\{i_1,i_2\}}}{2}t_{\{i_1+(r-e)\gamma, i_2+(r-e)\gamma \}}.
\end{align}
Case~2.3: The number of instances of $P_2$ with CNs $c_1$ and $c_2$, and overlaps in three replicas (one in each), $\bold{R}_r$, $\bold{R}_e$, and $\bold{R}_s$, $r < e < s$, is:
\begin{align}\label{eq_p2_3}
\mathcal{C}_{P_2} \left (t_{\{i_1,i_2\}}, t_{\{i_1+(r-e)\gamma, i_2+(r-e)\gamma \}}, t_{\{i_1+(r-s)\gamma, i_2+(r-s)\gamma \}} \right ) = t_{\{i_1,i_2\}}t_{\{i_1+(r-e)\gamma, i_2+(r-e)\gamma \}}t_{\{i_1+(r-s)\gamma, i_2+(r-s)\gamma \}}.
\end{align}
\end{lemma}

\begin{figure}[H]
\vspace{-1.4em}
\center
\includegraphics[width=5.5in]{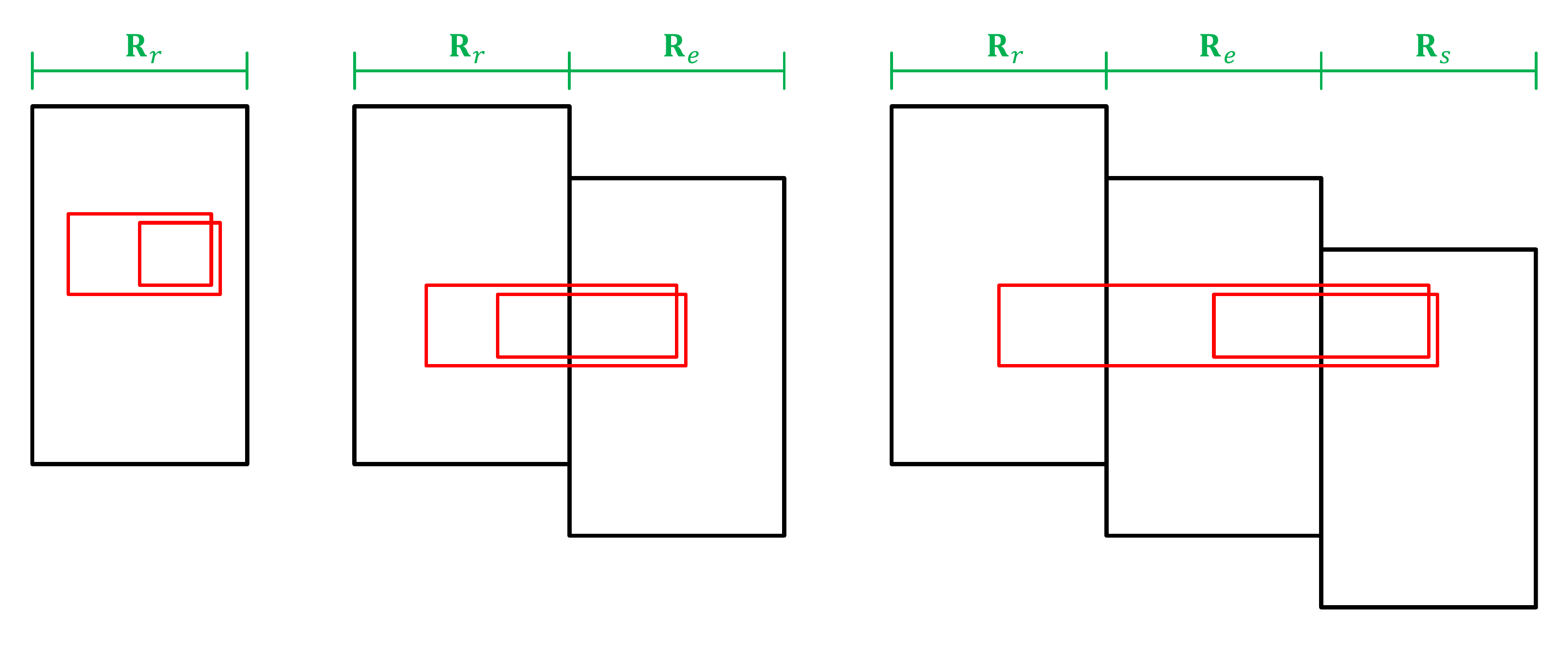}\vspace{-1.5em}
\caption{An instance of Pattern $P_2$ in Case~2.1, in Case~2.2, and in Case~2.3, from left to right. For simplicity, we have $e=r+1$ and $s=e+1$.}
\label{Fig_pat2}
\vspace{-0.5em}
\end{figure}

The three cases are illustrated in Fig \ref{Fig_pat2}.

\begin{theorem}\label{thm_p2}
The total number of instances of Pattern $P_2$ in the binary protograph of an SC code that has parameters $\gamma \geq 3$, $\kappa$, $m$, $L \geq m+1$, and $\mathcal{O}$, is:
\begin{equation}\label{eq_p2_4}
F_{P_2} = \sum_{k=1}^{m+1} (L-k+1) F^k_{P_2,1},
\end{equation}
where $F^k_{P_2,1}$, $k \in \{1, 2, \dots, m+1\}$, are given by:
\begin{align}\label{eq_p2_4}
F^1_{P_2,1} &= \hspace{-7.1em} \sum_{\hspace{7.0em}\{i_1,i_2\} \subset \{0, \dots, (m+1)\gamma -1\}} \hspace{-6.0em} \mathcal{A}_{P_2}\left (t_{\{i_1,i_2\}} \right ), 
\nonumber \\
F^2_{P_2,1} &= \hspace{-7.1em} \sum_{\hspace{7.0em}\{i_1,i_2\} \subset \{\gamma, \dots, (m+1)\gamma -1\}} \hspace{-6.0em} \mathcal{B}_{P_2} \left (t_{\{i_1,i_2\}}, t_{\{i_1-\gamma, i_2-\gamma \}} \right ) 
\nonumber \\ 
&+ \hspace{-5.4em} \sum_{\hspace{5.5em}\{i_1,i_2\} \subset \{0, \dots, m\gamma -1\}} \hspace{-4.3em} \mathcal{B}_{P_2} \left (t_{\{i_1,i_2\}}, t_{\{i_1+\gamma, i_2+\gamma \}} \right ),
\nonumber \\
F^{k \geq 3}_{P_2,1} &= \hspace{-9.1em} \sum_{\hspace{9.0em}\{i_1,i_2\} \subset \{(k-1)\gamma, \dots, (m+1)\gamma -1\}} \hspace{-8.0em} \mathcal{B}_{P_2} \left (t_{\{i_1,i_2\}}, t_{\{i_1+(1-k)\gamma, i_2+(1-k)\gamma \}} \right ) 
\nonumber \\ 
&+ \hspace{-8.0em} \sum_{\hspace{8.0em}\{i_1,i_2\} \subset \{0, \dots, (m-k+2)\gamma -1\}} \hspace{-6.9em} \mathcal{B}_{P_2} \left (t_{\{i_1,i_2\}}, t_{\{i_1+(k-1)\gamma, i_2+(k-1)\gamma \}} \right ) 
\nonumber \\ 
&+ \hspace{0.4em} \sum_{h=2}^{k-1} \hspace{-8.5em} \sum_{\hspace{9.0em}\{i_1,i_2\} \subset \{(k-1)\gamma, \dots, (m+1)\gamma -1\}} \hspace{-8.1em} \mathcal{C}_{P_2} \left (t_{\{i_1,i_2\}}, t_{\{i_1+(1-h)\gamma, i_2+(1-h)\gamma \}}, t_{\{i_1+(1-k)\gamma, i_2+(1-k)\gamma \}} \right ),
\end{align}
with $\overline{i_1} \neq \overline{i_2}$.
\end{theorem}

\subsection{Analysis of Pattern $P_3$ (size $3 \times 2$)}\label{subsec_p3}

This pattern has two VNs that are adjacent. Thus, Pattern $P_3$ spans at most $m+1$ consecutive replicas. Suppose $P_3$ has the CNs $c_1$, $c_2$, and $c_3$. The two overlaps forming $P_3$ are of degree $3$, and they are both $c_1-c_2-c_3$ overlaps.

\begin{lemma}\label{lem_p3}
Case~3.1: The number of instances of $P_3$ with CNs $c_1$, $c_2$, and $c_3$, and all overlaps in one replica, $\bold{R}_r$, is:
\begin{equation}\label{eq_p3_1}
\mathcal{A}_{P_3}\left (t_{\{i_1,i_2,i_3\}} \right ) = \binom{t_{\{i_1,i_2,i_3\}}}{2}.
\end{equation}
Case~3.2: The number of instances of $P_3$ with CNs $c_1$, $c_2$, and $c_3$, and overlaps in two replicas, $\bold{R}_r$ and $\bold{R}_e$, $r < e$, is:
\begin{align}\label{eq_p3_2}
\mathcal{B}_{P_3} \left (t_{\{i_1,i_2,i_3\}}, t_{\{i_1+(r-e)\gamma, i_2+(r-e)\gamma, i_3+(r-e)\gamma \}} \right ) = t_{\{i_1,i_2,i_3\}}t_{\{i_1+(r-e)\gamma, i_2+(r-e)\gamma, i_3+(r-e)\gamma \}}.
\end{align}
\end{lemma}

\begin{figure}[H]
\vspace{-1.5em}
\center
\includegraphics[width=3.5in]{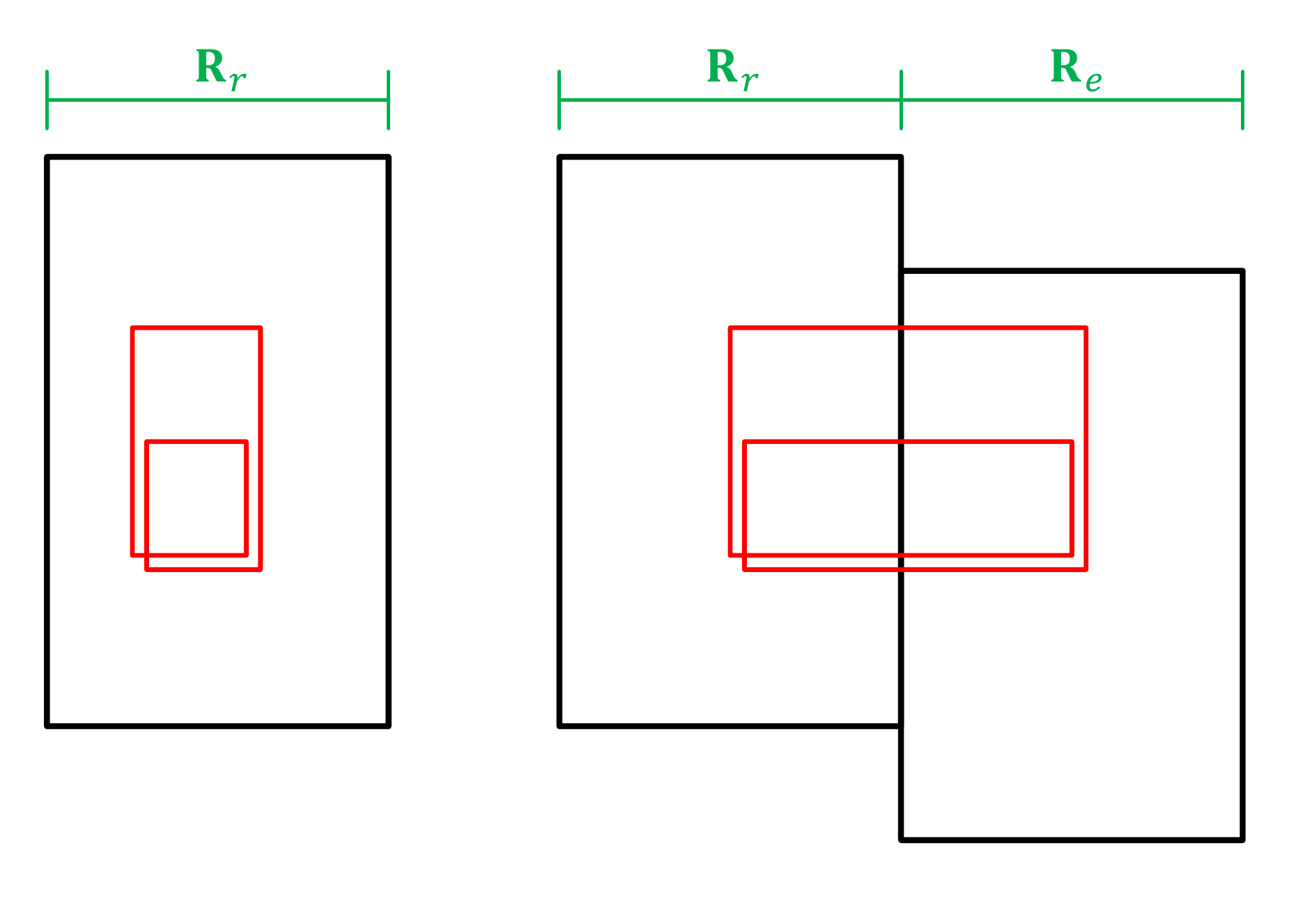}\vspace{-1.5em}
\caption{An instance of Pattern $P_3$ in Case~3.1 and in Case~3.2, from left to right. For simplicity, we have $e=r+1$.}
\label{Fig_pat3}
\vspace{-0.5em}
\end{figure}

The two cases are illustrated in Fig.~\ref{Fig_pat3}.

\begin{theorem}\label{thm_p3}
The total number of instances of Pattern $P_3$ in the binary protograph of an SC code that has parameters $\gamma \geq 3$, $\kappa$, $m$, $L \geq m+1$, and $\mathcal{O}$, is:
\begin{equation}\label{eq_p3_3}
F_{P_3} = \sum_{k=1}^{m+1} (L-k+1) F^k_{P_3,1},
\end{equation}
where $F^k_{P_3,1}$, $k \in \{1, 2, \dots, m+1\}$, are given by:
\begin{align} 
F^1_{P_3,1} &= \hspace{-8.0em} \sum_{\hspace{8.0em}\{i_1,i_2,i_3\} \subset \{0, \dots, (m+1)\gamma -1\}} \hspace{-7.0em} \mathcal{A}_{P_3}\left (t_{\{i_1,i_2,i_3\}} \right ), 
\nonumber \\
F^{k \geq 2}_{P_3,1} &= \hspace{-10.0em} \sum_{\hspace{10.0em}\{i_1,i_2,i_3\} \subseteq \{(k-1)\gamma, \dots, (m+1)\gamma -1\}} \hspace{-9.0em} \mathcal{B}_{P_3} \left (t_{\{i_1,i_2,i_3\}}, t_{\{i_1+(1-k)\gamma, i_2+(1-k)\gamma, i_3+(1-k)\gamma \}} \right ),
\end{align}\label{eq_p3_4}
with $\overline{i_1} \neq \overline{i_2}$, $\overline{i_1} \neq \overline{i_3}$, and $\overline{i_2} \neq \overline{i_3}$.
\end{theorem}

\subsection{Analysis of Pattern $P_4$ (size $2 \times 4$)}\label{subsec_p4}

This pattern has four VNs, with each two of them being adjacent. Consequently, $P_4$ spans at most $m+1$ consecutive replicas. Suppose $P_4$ has the CNs $c_1$ and $c_2$. The four overlaps forming $P_4$ are of degree $2$, and they are all $c_1-c_2$ overlaps.

\begin{lemma}\label{lem_p4}
Case~4.1: The number of instances of $P_4$ with CNs $c_1$ and $c_2$, and all overlaps in one replica, $\bold{R}_r$, is:
\begin{equation}\label{eq_p4_1}
\mathcal{A}_{P_4}\left (t_{\{i_1,i_2\}} \right ) = \binom{t_{\{i_1,i_2\}}}{4}.
\end{equation}
Case~4.2: The number of instances of $P_4$ with CNs $c_1$ and $c_2$, and all overlaps in two replicas s.t. three overlaps are in $\bold{R}_r$, and one overlap is in $\bold{R}_e$, is:
\begin{align}\label{eq_p4_2}
\mathcal{B}_{P_4} \left (t_{\{i_1,i_2\}}, t_{\{i_1+(r-e)\gamma, i_2+(r-e)\gamma \}} \right ) = \binom{t_{\{i_1,i_2\}}}{3}t_{\{i_1+(r-e)\gamma, i_2+(r-e)\gamma \}}.
\end{align}
Case~4.3: The number of instances of $P_4$ with CNs $c_1$ and $c_2$, and all overlaps in two replicas s.t. two overlaps are in $\bold{R}_r$, and two overlaps are in $\bold{R}_e$, $r < e$, is:
\begin{align}\label{eq_p4_3}
\mathcal{C}_{P_4} \left (t_{\{i_1,i_2\}}, t_{\{i_1+(r-e)\gamma, i_2+(r-e)\gamma \}} \right ) = \binom{t_{\{i_1,i_2\}}}{2} \binom{t_{\{i_1+(r-e)\gamma, i_2+(r-e)\gamma \}}}{2}.
\end{align}
Case~4.4: The number of instances of $P_4$ with CNs $c_1$ and $c_2$, and all overlaps in three replicas s.t. two overlaps are in $\bold{R}_r$, one overlap is in $\bold{R}_e$, and one overlap is in $\bold{R}_s$, $e < s$, is:
\begin{align}\label{eq_p4_4}
\mathcal{D}_{P_4} \left (t_{\{i_1,i_2\}}, t_{\{i_1+(r-e)\gamma, i_2+(r-e)\gamma \}}, t_{\{i_1+(r-s)\gamma, i_2+(r-s)\gamma \}} \right ) = \binom{t_{\{i_1,i_2\}}}{2}t_{\{i_1+(r-e)\gamma, i_2+(r-e)\gamma \}}t_{\{i_1+(r-s)\gamma, i_2+(r-s)\gamma \}}.
\end{align}
Case~4.5: The number of instances of $P_4$ with CNs $c_1$ and $c_2$, and overlaps in four replicas, $\bold{R}_r$, $\bold{R}_e$, $\bold{R}_s$, and $\bold{R}_u$, $r < e < s < u$,~is:
\begin{align}\label{eq_p4_5}
\mathcal{E}_{P_4} &\left (t_{\{i_1,i_2\}}, t_{\{i_1+(r-e)\gamma, i_2+(r-e)\gamma \}}, t_{\{i_1+(r-s)\gamma, i_2+(r-s)\gamma \}}, t_{\{i_1+(r-u)\gamma, i_2+(r-u)\gamma \}} \right ) 
\nonumber \\ 
&= t_{\{i_1,i_2\}}t_{\{i_1+(r-e)\gamma, i_2+(r-e)\gamma \}}t_{\{i_1+(r-s)\gamma, i_2+(r-s)\gamma \}}t_{\{i_1+(r-u)\gamma, i_2+(r-u)\gamma \}}.
\end{align}
\end{lemma}

\begin{figure}[H]
\vspace{-1.5em}
\center
\includegraphics[width=7.0in]{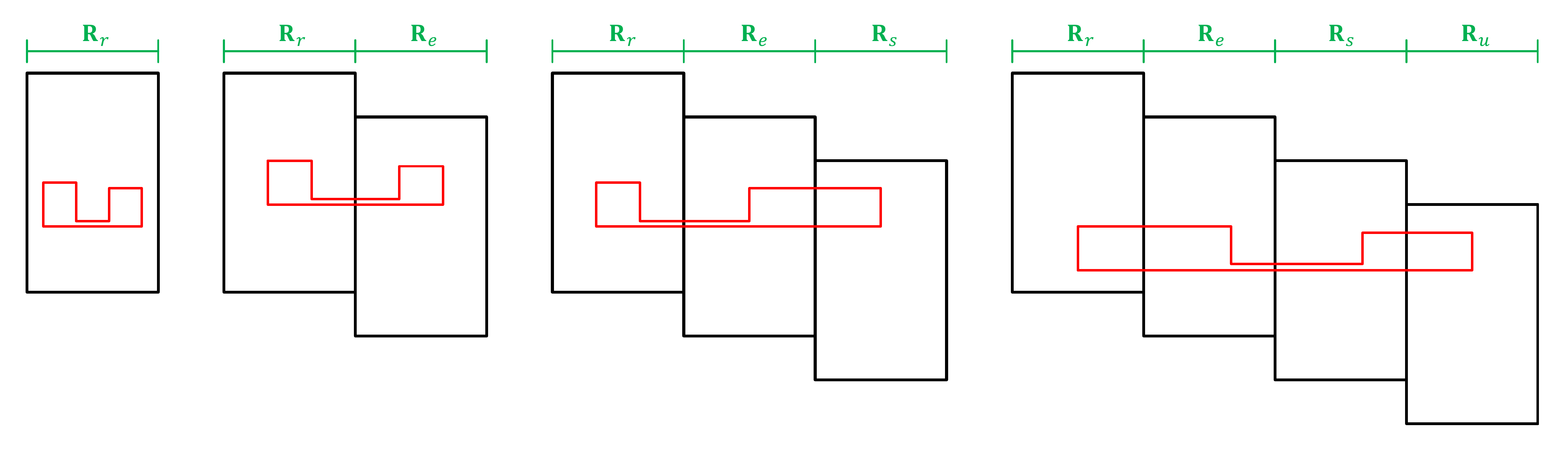}\vspace{-1.5em}
\caption{An instance of Pattern $P_4$ in Case~4.1, in Case~4.3, in Case~4.4, and in Case~4.5, from left to right. For simplicity, we have $e=r+1$, $s=e+1$, and $u=s+1$.}
\label{Fig_pat4}
\vspace{-0.5em}
\end{figure}

Four of the five cases are illustrated in Fig.~\ref{Fig_pat4}.

\begin{theorem}\label{thm_p4}
The total number of instances of Pattern $P_4$ in the binary protograph of an SC code that has parameters $\gamma \geq 3$, $\kappa$, $m$, $L \geq m+1$, and $\mathcal{O}$, is:
\begin{equation}\label{eq_p4_6}
F_{P_4} = \sum_{k=1}^{m+1} (L-k+1) F^k_{P_4,1},
\end{equation}
where $F^k_{P_4,1}$, $k \in \{1, 2, \dots, m+1\}$, are given by:
\begin{align} \label{eq_p4_7}
F^1_{P_4,1} &= \hspace{-7.2em} \sum_{\hspace{7.0em} \{i_1,i_2\} \subset \{0, \dots, (m+1)\gamma -1\}} \hspace{-6.5em} \mathcal{A}_{P_4}\left (t_{\{i_1,i_2\}} \right ), 
\nonumber \\
F^2_{P_4,1} &= \hspace{-7.2em} \sum_{\hspace{7.0em}\{i_1,i_2\} \subset \{\gamma, \dots, (m+1)\gamma -1\}} \hspace{-6.5em} \mathcal{B}_{P_4} \left (t_{\{i_1,i_2\}}, t_{\{i_1-\gamma, i_2-\gamma \}} \right ) 
\nonumber \\ 
&+ \hspace{-5.4em} \sum_{\hspace{5.5em}\{i_1,i_2\} \subset \{0, \dots, m\gamma -1\}} \hspace{-5.0em} \mathcal{B}_{P_4} \left (t_{\{i_1,i_2\}}, t_{\{i_1+\gamma, i_2+\gamma \}} \right ) 
\nonumber \\ 
&+ \hspace{-7.0em} \sum_{\hspace{7.0em}\{i_1,i_2\} \subset \{\gamma, \dots, (m+1)\gamma -1\}} \hspace{-6.5em} \mathcal{C}_{P_4} \left (t_{\{i_1,i_2\}}, t_{\{i_1-\gamma, i_2-\gamma \}} \right ),
\nonumber \\
F^3_{P_4,1} &= \hspace{-7.4em} \sum_{\hspace{7.2em}\{i_1,i_2\} \subset \{2\gamma, \dots, (m+1)\gamma -1\}} \hspace{-6.7em} \mathcal{B}_{P_4} \left (t_{\{i_1,i_2\}}, t_{\{i_1-2\gamma, i_2-2\gamma \}} \right ) 
\nonumber \\ 
&+ \hspace{-6.9em} \sum_{\hspace{6.8em}\{i_1,i_2\} \subset \{0, \dots, (m-1)\gamma -1\}} \hspace{-6.2em} \mathcal{B}_{P_4} \left (t_{\{i_1,i_2\}}, t_{\{i_1+2\gamma, i_2+2\gamma \}} \right )
\nonumber \\ 
&+ \hspace{-7.3em} \sum_{\hspace{7.2em}\{i_1,i_2\} \subset \{2\gamma, \dots, (m+1)\gamma -1\}} \hspace{-6.7em} \mathcal{C}_{P_4} \left (t_{\{i_1,i_2\}}, t_{\{i_1-2\gamma, i_2-2\gamma \}} \right )
\nonumber \\ 
&+ \hspace{-7.3em} \sum_{\hspace{7.2em}\{i_1,i_2\} \subset \{2\gamma, \dots, (m+1)\gamma -1\}} \hspace{-6.8em} \mathcal{D}_{P_4} \left (t_{\{i_1,i_2\}}, t_{\{i_1-\gamma, i_2-\gamma \}}, t_{\{i_1-2\gamma, i_2-2\gamma \}} \right )
\nonumber \\ 
&+ \hspace{-5.4em} \sum_{\hspace{5.5em}\{i_1,i_2\} \subset \{\gamma, \dots, m\gamma -1\}} \hspace{-5.0em} \mathcal{D}_{P_4} \left (t_{\{i_1,i_2\}}, t_{\{i_1+\gamma, i_2+\gamma \}}, t_{\{i_1-\gamma, i_2-\gamma \}} \right )
\nonumber \\ 
&+ \hspace{-7.0em} \sum_{\hspace{7.2em}\{i_1,i_2\} \subset \{0, \dots, (m-1)\gamma -1\}} \hspace{-6.6em} \mathcal{D}_{P_4} \left (t_{\{i_1,i_2\}}, t_{\{i_1+2\gamma, i_2+2\gamma \}}, t_{\{i_1+\gamma, i_2+\gamma \}} \right ),
\nonumber \\
F^{k \geq 4}_{P_4,1} &= \hspace{-9.1em} \sum_{\hspace{9.0em}\{i_1,i_2\} \subset \{(k-1)\gamma, \dots, (m+1)\gamma -1\}} \hspace{-8.5em} \mathcal{B}_{P_4} \left (t_{\{i_1,i_2\}}, t_{\{i_1+(1-k)\gamma, i_2+(1-k)\gamma \}} \right ) 
\nonumber \\ 
&+ \hspace{-7.9em} \sum_{\hspace{8.0em}\{i_1,i_2\} \subset \{0, \dots, (m-k+2)\gamma -1\}} \hspace{-7.3em} \mathcal{B}_{P_4} \left (t_{\{i_1,i_2\}}, t_{\{i_1+(k-1)\gamma, i_2+(k-1)\gamma \}} \right ) 
\nonumber \\ 
&+ \hspace{-9.0em} \sum_{\hspace{9.0em}\{i_1,i_2\} \subset \{(k-1)\gamma, \dots, (m+1)\gamma -1\}} \hspace{-8.3em} \mathcal{C}_{P_4} \left (t_{\{i_1,i_2\}}, t_{\{i_1+(1-k)\gamma, i_2+(1-k)\gamma \}} \right )
\nonumber \\ 
&+ \hspace{0.4em} \sum_{h=2}^{k-1} \hspace{-8.6em} \sum_{\hspace{9.0em}\{i_1,i_2\} \subset \{(k-1)\gamma, \dots, (m+1)\gamma -1\}} \hspace{-8.5em} \mathcal{D}_{P_4} \left (t_{\{i_1,i_2\}}, t_{\{i_1+(1-h)\gamma, i_2+(1-h)\gamma \}}, t_{\{i_1+(1-k)\gamma, i_2+(1-k)\gamma \}} \right )
\nonumber \\ 
&+ \hspace{0.4em} \sum_{h=2}^{k-1} \hspace{-9.7em} \sum_{\hspace{10.0em}\{i_1,i_2\} \subset \{(k-h)\gamma, \dots, (m-h+2)\gamma -1\}} \hspace{-9.5em} \mathcal{D}_{P_4} \left (t_{\{i_1,i_2\}}, t_{\{i_1+(h-1)\gamma, i_2+(h-1)\gamma \}}, t_{\{i_1+(h-k)\gamma, i_2+(h-k)\gamma \}} \right )
\nonumber \\ 
&+ \hspace{0.4em} \sum_{h=2}^{k-1} \hspace{-7.3em} \sum_{\hspace{7.5em}\{i_1,i_2\} \subset \{0, \dots, (m-k+2)\gamma -1\}} \hspace{-7.1em} \mathcal{D}_{P_4} \left (t_{\{i_1,i_2\}}, t_{\{i_1+(k-1)\gamma, i_2+(k-1)\gamma \}}, t_{\{i_1+(k-h)\gamma, i_2+(k-h)\gamma \}} \right )
\nonumber \\ 
&+ \hspace{0.4em} \sum_{h=2}^{k-2} \hspace{0.3em} \sum_{w=h+1}^{k-1} \hspace{-8.5em} \sum_{\hspace{8.7em}\{i_1,i_2\} \subset \{(k-1)\gamma, \dots, (m+1)\gamma -1\}} \hspace{-8.3em} \mathcal{E}_{P_4} \big (t_{\{i_1,i_2\}}, t_{\{i_1+(1-h)\gamma, i_2+(1-h)\gamma \}}, t_{\{i_1+(1-w)\gamma, i_2+(1-w)\gamma \}} 
\nonumber \\ 
&\hspace{11.5em}, t_{\{i_1+(1-k)\gamma, i_2+(1-k)\gamma \}} \big ),
\end{align}
with $\overline{i_1} \neq \overline{i_2}$.
\end{theorem}

\subsection{Analysis of Pattern $P_5$ (size $4 \times 2$)}\label{subsec_p5}

This pattern has two adjacent VNs. Thus, Pattern $P_5$ spans at most $m+1$ consecutive replicas. Pattern $P_5$ does not exist in the case of $\gamma=3$. Suppose $P_5$ has the CNs $c_1$, $c_2$, $c_3$, and $c_4$. The two overlaps forming $P_5$ are of degree $4$, and they are both $c_1-c_2-c_3-c_4$ overlaps.

\begin{lemma}\label{lem_p5}
Case~5.1: The number of instances of $P_5$ with CNs $c_1$, $c_2$, $c_3$, and $c_4$, and all overlaps in one replica, $\bold{R}_r$, is:
\begin{equation}\label{eq_p5_1}
\mathcal{A}_{P_5}\left (t_{\{i_1,i_2,i_3,i_4\}} \right ) = \binom{t_{\{i_1,i_2,i_3,i_4\}}}{2}.
\end{equation}
Case~5.2: The number of instances of $P_5$ with $c_1$, $c_2$, $c_3$, and $c_4$, and overlaps in two replicas, $\bold{R}_r$ and $\bold{R}_e$, $r < e$, is:
\begin{align}\label{eq_p5_2}
\mathcal{B}_{P_5} \left (t_{\{i_1,i_2,i_3,i_4\}}, t_{\{i_1+(r-e)\gamma, i_2+(r-e)\gamma, i_3+(r-e)\gamma, i_4+(r-e)\gamma \}} \right ) = t_{\{i_1,i_2,i_3,i_4\}}t_{\{i_1+(r-e)\gamma, i_2+(r-e)\gamma, i_3+(r-e)\gamma, i_4+(r-e)\gamma \}}.
\end{align}
\end{lemma}

\begin{figure}[H]
\vspace{-1.0em}
\center
\includegraphics[width=3.5in]{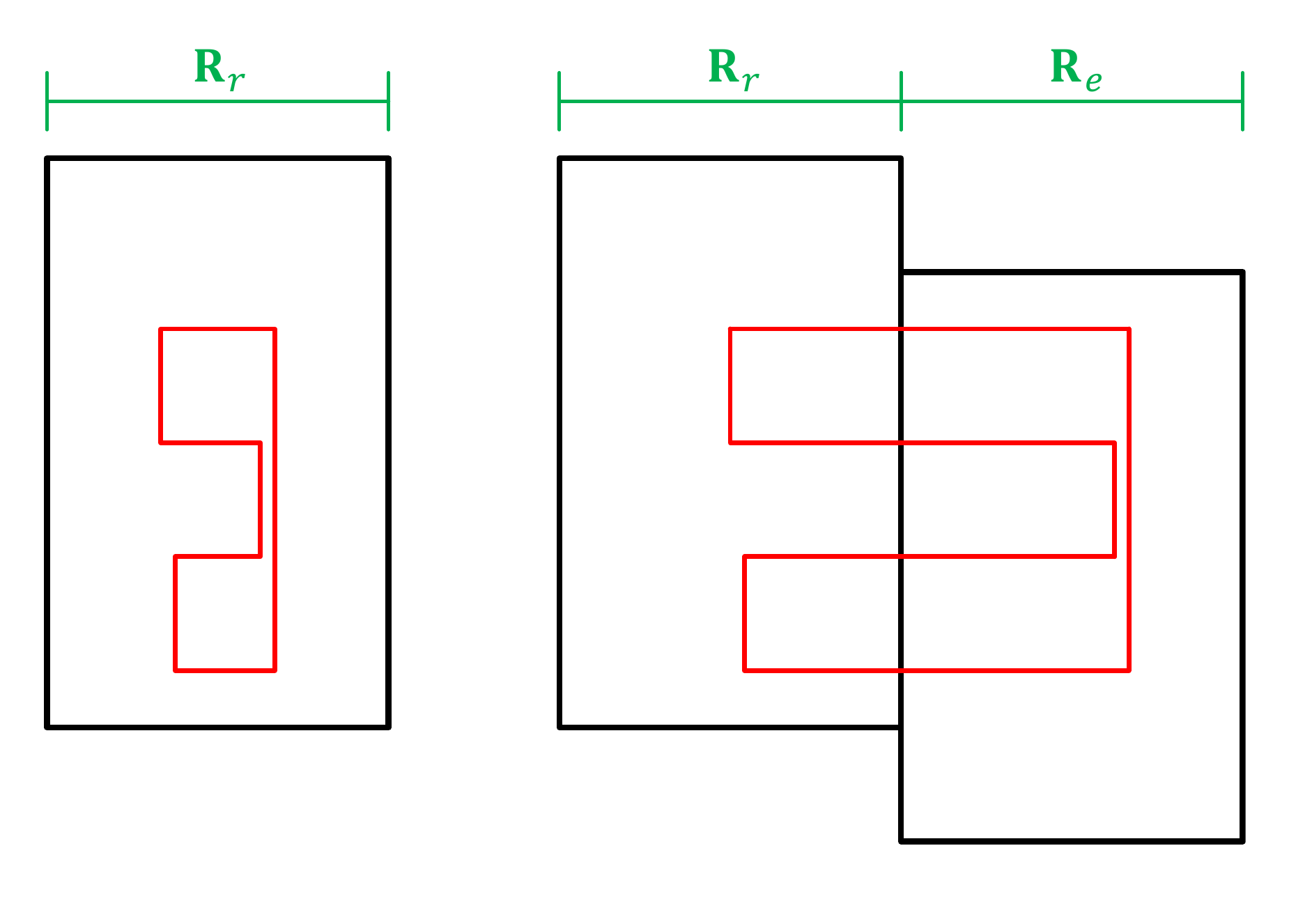}\vspace{-1.5em}
\caption{An instance of Pattern $P_5$ in Case~5.1 and in Case~5.2, from left to right. For simplicity, we have $e=r+1$.}
\label{Fig_pat5}
\vspace{-0.5em}
\end{figure}

The two cases are illustrated in Fig.~\ref{Fig_pat5}.

\begin{theorem}\label{thm_p5}
The total number of instances of Pattern $P_5$ in the binary protograph of an SC code that has parameters $\gamma \geq 4$, $\kappa$, $m$, $L \geq m+1$, and $\mathcal{O}$, is:
\begin{equation}\label{eq_p5_3}
F_{P_5} = \sum_{k=1}^{m+1} (L-k+1) F^k_{P_5,1},
\end{equation}
where $F^k_{P_5,1}$, $k \in \{1, 2, \dots, m+1\}$, are given by:
\vspace{-0.1em}\begin{align} \label{eq_p5_4}
F^1_{P_5,1} &= \hspace{-8.8em} \sum_{\hspace{8.7em}\{i_1,i_2,i_3,i_4\} \subset \{0, \dots, (m+1)\gamma -1\}} \hspace{-8.0em} \mathcal{A}_{P_5}\left (t_{\{i_1,i_2,i_3,i_4\}} \right ), 
\nonumber \\
F^{k \geq 2}_{P_5,1} &= \hspace{-11.0em} \sum_{\hspace{11.0em}\{i_1,i_2,i_3,i_4\} \subseteq \{(k-1)\gamma, \dots, (m+1)\gamma -1\}} \hspace{-10.1em} \mathcal{B}_{P_5} \left (t_{\{i_1,i_2,i_3,i_4\}}, t_{\{i_1+(1-k)\gamma, i_2+(1-k)\gamma, i_3+(1-k)\gamma, i_4+(1-k)\gamma \}} \right ),
\end{align}
with $\overline{i_1} \neq \overline{i_2}$, $\overline{i_1} \neq \overline{i_3}$, $\overline{i_1} \neq \overline{i_4}$, $\overline{i_2} \neq \overline{i_3}$, $\overline{i_2} \neq \overline{i_4}$, and $\overline{i_3} \neq \overline{i_4}$.
\end{theorem}

\subsection{Analysis of Pattern $P_6$ (size $3 \times 3$)}\label{subsec_p6}

This pattern has three VNs, with each two of them being adjacent. Thus, $P_6$ spans at most $m+1$ consecutive replicas. Suppose $P_6$ has the CNs $c_1$, $c_2$, and $c_3$. Define \textit{\textbf{distinct overlaps}} to be overlaps from different families, i.e., overlaps among different sets of CNs. Pattern $P_6$ is formed of three overlaps; two (distinct) of degree-$2$ and one of degree-$3$. Define $c_1$ as the CN connecting the three VNs. Thus, the overlaps are $c_1-c_2$, $c_1-c_3$, and $c_1-c_2-c_3$ (see $P_6$ in Fig.~\ref{Fig_pat}). Again, each VN corresponds to an overlap.

\begin{lemma}\label{lem_p6}
Case~6.1: The number of instances of $P_6$ with CNs $c_1$, $c_2$, and $c_3$ as defined in the previous paragraph, and all overlaps in one replica, $\bold{R}_r$, is:
\begin{align}\label{eq_p6_1}
\mathcal{A}_{P_6} & \left (t_{\{i_1,i_2\}}, t_{\{i_1,i_3\}}, t_{\{i_1,i_2,i_3\}} \right ) = t_{\{i_1,i_2,i_3\}} \left( t_{\{i_1,i_2,i_3\}}-1 \right)^+ \left( t_{\{i_1,i_3\}}-2 \right)^+ 
\nonumber \\ 
&+ t_{\{i_1,i_2,i_3\}} \left( t_{\{i_1,i_2\}}-t_{\{i_1,i_2,i_3\}} \right) \left( t_{\{i_1,i_3\}}-1 \right)^+.
\end{align}
Case~6.2: The number of instances of $P_6$ with CNs $c_1$, $c_2$, and $c_3$ as defined in the previous paragraph, and all overlaps in two replicas s.t. the two degree-$2$ overlaps are in $\bold{R}_r$, and the degree-$3$ overlap is in $\bold{R}_e$, is:
\begin{align}\label{eq_p6_2}
\mathcal{B}_{P_6} & \left (t_{\{i_1,i_2\}}, t_{\{i_1,i_3\}}, t_{\{i_1,i_2,i_3\}}, t_{\{i_1+(r-e)\gamma,i_2+(r-e)\gamma,i_3+(r-e)\gamma\}} \right ) 
\nonumber \\ 
& = \Big[ t_{\{i_1,i_2,i_3\}} \left( t_{\{i_1,i_3\}}-1 \right)^+ + \left ( t_{\{i_1,i_2\}} - t_{\{i_1,i_2,i_3\}} \right ) t_{\{i_1,i_3\}} \Big] t_{\{i_1+(r-e)\gamma,i_2+(r-e)\gamma,i_3+(r-e)\gamma\}}.
\end{align}
Case~6.3: The number of instances of $P_6$ with CNs $c_1$, $c_2$, and $c_3$ as defined in the previous paragraph, and all overlaps in two replicas s.t. the degree-$3$ overlap and the $c_1-c_2$ overlap are in $\bold{R}_r$, and the $c_1-c_3$ overlap is in $\bold{R}_e$, is:
\begin{align}\label{eq_p6_3}
\mathcal{C}_{P_6} \left (t_{\{i_1,i_2\}}, t_{\{i_1,i_2,i_3\}}, t_{\{i_1+(r-e)\gamma,i_3+(r-e)\gamma\}} \right ) = t_{\{i_1,i_2,i_3\}} \left( t_{\{i_1,i_2\}}-1 \right)^+ t_{\{i_1+(r-e)\gamma,i_3+(r-e)\gamma\}}.
\end{align}
Case~6.4: The number of instances of $P_6$ with CNs $c_1$, $c_2$, and $c_3$ as defined in the previous paragraph, and overlaps in three replicas s.t. the $c_1-c_2$ overlap is in $\bold{R}_r$, the $c_1-c_3$ overlap is in $\bold{R}_e$, and the degree-$3$ overlap is in $\bold{R}_s$, $r < e$, is:
\begin{align}\label{eq_p6_4}
\mathcal{D}_{P_6} &\left (t_{\{i_1,i_2\}}, t_{\{i_1+(r-e)\gamma,i_3+(r-e)\gamma\}}, t_{\{i_1+(r-s)\gamma,i_2+(r-s)\gamma,i_3+(r-s)\gamma\}} \right ) 
\nonumber \\ 
&= t_{\{i_1,i_2\}}t_{\{i_1+(r-e)\gamma,i_3+(r-e)\gamma\}}t_{\{i_1+(r-s)\gamma,i_2+(r-s)\gamma,i_3+(r-s)\gamma\}}.
\end{align}
\end{lemma}

\begin{figure}[H]
\vspace{-1.5em}
\center
\includegraphics[width=5.5in]{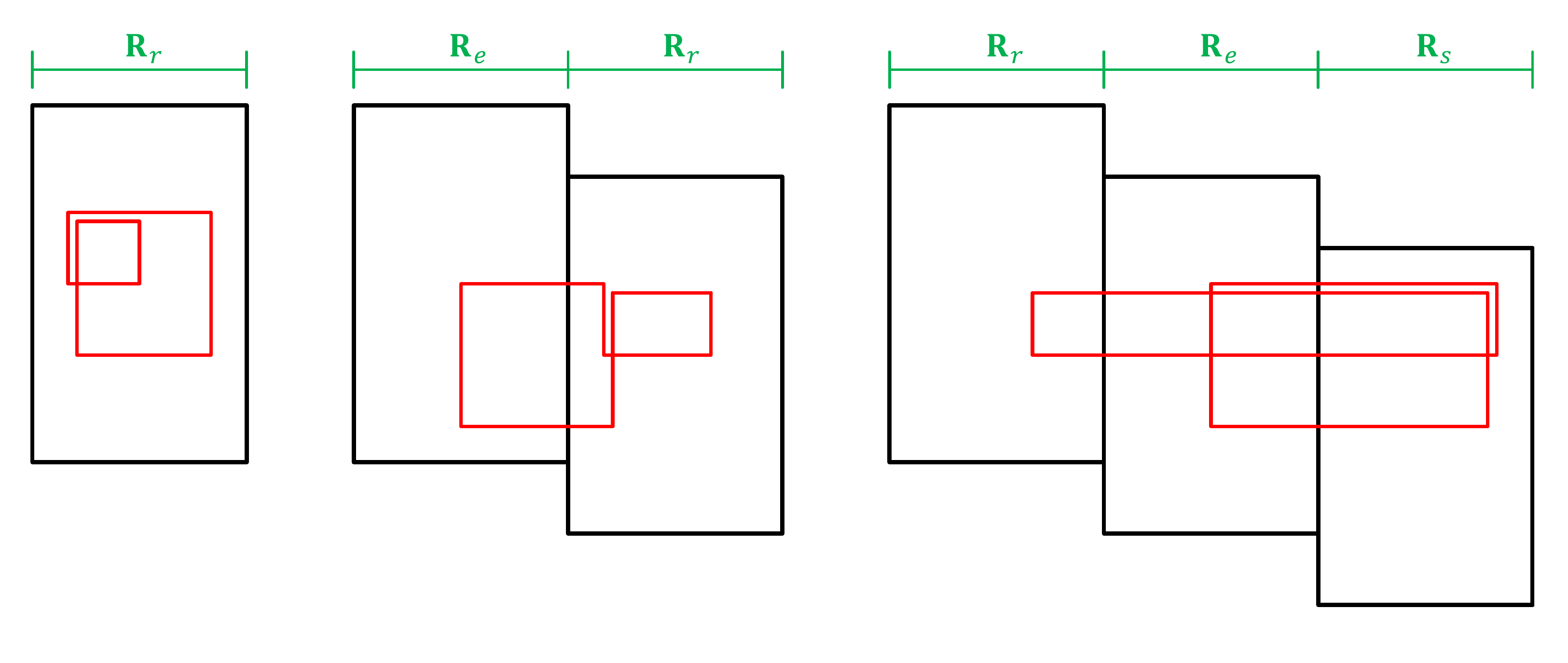}\vspace{-1.5em}
\caption{An instance of Pattern $P_6$ in Case~6.1, in Case~6.3, and in Case~6.4, from left to right. For simplicity, we have $e=r+y$, where $y \in \{-1,1\}$, and $s=e+1$.}
\label{Fig_pat6}
\vspace{-0.5em}
\end{figure}

Three of the four cases are illustrated in Fig.~\ref{Fig_pat6}.

\begin{theorem}\label{thm_p6}
The total number of instances of Pattern $P_6$ in the binary protograph of an SC code that has parameters $\gamma \geq 3$, $\kappa$, $m$, $L \geq m+1$, and $\mathcal{O}$, is:
\begin{equation}\label{eq_p6_5}
F_{P_6} = \sum_{k=1}^{m+1} (L-k+1) F^k_{P_6,1},
\end{equation}
where $F^k_{P_6,1}$, $k \in \{1, 2, \dots, m+1\}$, are given by:
\vspace{-0.1em}\begin{align}\label{eq_p6_5}
F^1_{P_6,1} &= \hspace{-14.6em} \sum_{\hspace{14.8em} i_1 \in \{0, \dots, (m+1)\gamma -1\}, \{i_2,i_3\} \subset \{0, \dots, (m+1)\gamma -1\}} \hspace{-14.1em} \mathcal{A}_{P_6}\left (t_{\{i_1,i_2\}}, t_{\{i_1,i_3\}}, t_{\{i_1,i_2,i_3\}} \right ), 
\nonumber \\
F^2_{P_6,1} &= \hspace{-14.7em} \sum_{\hspace{14.8em} i_1 \in \{\gamma, \dots, (m+1)\gamma -1\}, \{i_2,i_3\} \subset \{\gamma, \dots, (m+1)\gamma -1\}} \hspace{-14.0em} \mathcal{B}_{P_6} \left (t_{\{i_1,i_2\}}, t_{\{i_1,i_3\}}, t_{\{i_1,i_2,i_3\}}, t_{\{i_1-\gamma,i_2-\gamma,i_3-\gamma\}} \right ) 
\nonumber \\ 
&+ \hspace{-11.3em} \sum_{\hspace{11.7em} i_1 \in \{0, \dots, m\gamma -1\}, \{i_2,i_3\} \subset \{0, \dots, m\gamma -1\}} \hspace{-10.7em} \mathcal{B}_{P_6} \left (t_{\{i_1,i_2\}}, t_{\{i_1,i_3\}}, t_{\{i_1,i_2,i_3\}}, t_{\{i_1+\gamma,i_2+\gamma,i_3+\gamma\}} \right ) 
\nonumber \\ 
&+ \hspace{-20.5em} \sum_{\hspace{20.8em} i_1 \in \{\gamma, \dots, (m+1)\gamma -1\}, i_2 \in \{0, \dots, (m+1)\gamma -1\}, i_3 \in \{\gamma, \dots, (m+1)\gamma -1\}} \hspace{-19.9em} \mathcal{C}_{P_6} \left (t_{\{i_1,i_2\}}, t_{\{i_1,i_2,i_3\}}, t_{\{i_1-\gamma,i_3-\gamma\}} \right ) 
\nonumber \\ 
&+ \hspace{-17.2em} \sum_{\hspace{17.5em}  i_1 \in \{0, \dots, m\gamma -1\}, i_2 \in \{0, \dots, (m+1)\gamma -1\}, i_3 \in \{0, \dots, m\gamma -1\}} \hspace{-16.4em} \mathcal{C}_{P_6} \left (t_{\{i_1,i_2\}}, t_{\{i_1,i_2,i_3\}}, t_{\{i_1+\gamma,i_3+\gamma\}} \right ),
\nonumber \\
F^{k \geq 3}_{P_6,1} &= \hspace{-18.9em} \sum_{\hspace{19.0em} i_1 \in \{(k-1)\gamma, \dots, (m+1)\gamma -1\}, \{i_2,i_3\} \subset \{(k-1)\gamma, \dots, (m+1)\gamma -1\}} \hspace{-18.0em} \mathcal{B}_{P_6} \left (t_{\{i_1,i_2\}}, t_{\{i_1,i_3\}}, t_{\{i_1,i_2,i_3\}}, t_{\{i_1+(1-k)\gamma,i_2+(1-k)\gamma,i_3+(1-k)\gamma\}} \right ) 
\nonumber \\ 
&+ \hspace{-16.7em} \sum_{\hspace{17.0em} i_1 \in \{0, \dots, (m-k+2)\gamma -1\}, \{i_2,i_3\} \subset \{0, \dots, (m-k+2)\gamma -1\}} \hspace{-16.0em} \mathcal{B}_{P_6} \left (t_{\{i_1,i_2\}}, t_{\{i_1,i_3\}}, t_{\{i_1,i_2,i_3\}}, t_{\{i_1+(k-1)\gamma,i_2+(k-1)\gamma,i_3+(k-1)\gamma\}} \right ) 
\nonumber \\ 
&+ \hspace{-24.8em} \sum_{\hspace{25.0em} i_1 \in \{(k-1)\gamma, \dots, (m+1)\gamma -1\}, i_2 \in \{0, \dots, (m+1)\gamma -1\}, i_3 \in \{(k-1)\gamma, \dots, (m+1)\gamma -1\}} \hspace{-24.0em} \mathcal{C}_{P_6} \left (t_{\{i_1,i_2\}}, t_{\{i_1,i_2,i_3\}}, t_{\{i_1+(1-k)\gamma,i_3+(1-k)\gamma\}} \right ) 
\nonumber \\ 
&+ \hspace{-22.7em} \sum_{\hspace{23.0em}  i_1 \in \{0, \dots, (m-k+2)\gamma -1\}, i_2 \in \{0, \dots, (m+1)\gamma -1\}, i_3 \in \{0, \dots, (m-k+2)\gamma -1\}} \hspace{-21.9em} \mathcal{C}_{P_6} \left (t_{\{i_1,i_2\}}, t_{\{i_1,i_2,i_3\}}, t_{\{i_1+(k-1)\gamma,i_3+(k-1)\gamma\}} \right )
\nonumber \\ 
&+ \hspace{0.5em} \sum_{h=2}^{k-1} \hspace{-26.5em} \sum_{\hspace{27.0em} i_1 \in \{(k-1)\gamma, \dots, (m+1)\gamma -1\}, i_2 \in \{(k-1)\gamma, \dots, (m+1)\gamma -1\}, i_3 \in \{(k-1)\gamma, \dots, (m+h)\gamma -1\}} \hspace{-26.4em} \mathcal{D}_{P_6} \left (t_{\{i_1,i_2\}}, t_{\{i_1+(1-h)\gamma,i_3+(1-h)\gamma\}}, t_{\{i_1+(1-k)\gamma,i_2+(1-k)\gamma,i_3+(1-k)\gamma\}} \right ) 
\nonumber \\ 
&+ \hspace{0.5em} \sum_{h=2}^{k-1} \hspace{-26.1em} \sum_{\hspace{26.6em} i_1 \in \{(k-1)\gamma, \dots, (m+1)\gamma -1\}, i_2 \in \{(h-1), \dots, (m+1)\gamma -1\}, i_3 \in \{(k-1)\gamma, \dots, (m+h)\gamma -1\}} \hspace{-25.9em} \mathcal{D}_{P_6} \left (t_{\{i_1,i_2\}}, t_{\{i_1+(1-k)\gamma,i_3+(1-k)\gamma\}}, t_{\{i_1+(1-h)\gamma,i_2+(1-h)\gamma,i_3+(1-h)\gamma\}} \right ) 
\nonumber \\ 
&+ \hspace{0.5em} \sum_{h=2}^{k-1} \hspace{-27.6em} \sum_{\hspace{28.2em} i_1 \in \{(k-h)\gamma, \dots, (m-h+2)\gamma -1\}, i_2 \in \{0, \dots, (m-h+2)\gamma -1\}, i_3 \in \{(k-h)\gamma, \dots, (m-h+2)\gamma -1\}} \hspace{-27.6em} \mathcal{D}_{P_6} \left (t_{\{i_1,i_2\}}, t_{\{i_1+(h-k)\gamma,i_3+(h-k)\gamma\}}, t_{\{i_1+(h-1)\gamma,i_2+(h-1)\gamma,i_3+(h-1)\gamma\}} \right ),
\end{align}
with $\overline{i_1} \neq \overline{i_2}$, $\overline{i_1} \neq \overline{i_3}$, and $\overline{i_2} \neq \overline{i_3}$.
\end{theorem}

\subsection{Analysis of Pattern $P_7$ (size $3 \times 4$)}\label{subsec_p7}

This pattern has four VNs, with each two of them being adjacent. Consequently, $P_7$ spans at most $m+1$ consecutive replicas. Suppose $P_7$ has the CNs $c_1$, $c_2$, and $c_3$. The pattern is formed of four degree-$2$ overlaps that are evenly distributed over two different families. Define $c_1$ as the CN connecting the four VNs. Thus, the overlaps are two $c_1-c_2$ and two $c_1-c_3$ overlaps (see $P_7$ in Fig.~\ref{Fig_pat} for clarification).

\begin{lemma}\label{lem_p7}
Case~7.1: The number of instances of $P_7$ with CNs $c_1$, $c_2$, and $c_3$ as defined in the previous paragraph, and all overlaps in one replica, $\bold{R}_r$, is:
\begin{align}\label{eq_p7_1}
\mathcal{A}_{P_7} & \left (t_{\{i_1,i_2\}}, t_{\{i_1,i_3\}}, t_{\{i_1,i_2,i_3\}} \right ) = \binom{t_{\{i_1,i_2,i_3\}}}{2} \binom{\left (t_{\{i_1,i_3\}}-2 \right )^+}{2} 
\nonumber \\ 
&+ t_{\{i_1,i_2,i_3\}} \left( t_{\{i_1,i_2\}}-t_{\{i_1,i_2,i_3\}} \right) \binom{\left (t_{\{i_1,i_3\}}-1 \right )^+}{2} + \binom{t_{\{i_1,i_2\}}-t_{\{i_1,i_2,i_3\}}}{2} \binom{t_{\{i_1,i_3\}}}{2}.
\end{align}
Case~7.2: The number of instances of $P_7$ with CNs $c_1$, $c_2$, and $c_3$ as defined in the previous paragraph, and all overlaps in two replicas s.t. three overlaps are in $\bold{R}_r$, and one $c_1-c_3$ overlap is in $\bold{R}_e$, is:
\vspace{-0.1em}\begin{align}\label{eq_p7_2}
\mathcal{B}_{P_7} & \left (t_{\{i_1,i_2\}}, t_{\{i_1,i_3\}}, t_{\{i_1,i_2,i_3\}}, t_{\{i_1+(r-e)\gamma,i_3+(r-e)\gamma\}} \right ) = \bigg[ \binom{t_{\{i_1,i_2,i_3\}}}{2} \left (t_{\{i_1,i_3\}}-2 \right )^+ 
\nonumber \\ 
&+ t_{\{i_1,i_2,i_3\}} \left( t_{\{i_1,i_2\}}-t_{\{i_1,i_2,i_3\}} \right) \left (t_{\{i_1,i_3\}}-1 \right )^+ + \binom{t_{\{i_1,i_2\}}-t_{\{i_1,i_2,i_3\}}}{2} t_{\{i_1,i_3\}} \bigg] t_{\{i_1+(r-e)\gamma,i_3+(r-e)\gamma\}}.
\end{align}
Case~7.3: The number of instances of $P_7$ with CNs $c_1$, $c_2$, and $c_3$ as defined in the previous paragraph, and all overlaps in two replicas s.t. the two $c_1-c_2$ overlaps are in $\bold{R}_r$, and the two $c_1-c_3$ overlaps are in $\bold{R}_e$, $r<e$, is:
\begin{align}\label{eq_p7_3}
\mathcal{C}_{P_7} \left (t_{\{i_1,i_2\}}, t_{\{i_1+(r-e)\gamma,i_3+(r-e)\gamma\}} \right ) = \binom{t_{\{i_1,i_2\}}}{2} \binom{t_{\{i_1+(r-e)\gamma,i_3+(r-e)\gamma\}}}{2}.
\end{align}
Case~7.4: The number of instances of $P_7$ with CNs $c_1$, $c_2$, and $c_3$ as defined in the previous paragraph, and all overlaps in two replicas s.t. two distinct overlaps (from different families) are in $\bold{R}_r$, and two distinct overlaps are in $\bold{R}_e$, $r<e$, is:
\begin{align}\label{eq_p7_4}
\mathcal{D}_{P_7} & \big( t_{\{i_1,i_2\}}, t_{\{i_1,i_3\}}, t_{\{i_1,i_2,i_3\}}, t_{\{i_1+(r-e)\gamma,i_2+(r-e)\gamma\}}, t_{\{i_1+(r-e)\gamma,i_3+(r-e)\gamma\}}, t_{\{i_1+(r-e)\gamma,i_2+(r-e)\gamma,i_3+(r-e)\gamma\}} \big) 
\nonumber \\ 
&= \Big[ t_{\{i_1,i_2,i_3\}} \left (t_{\{i_1,i_3\}}-1 \right )^+ + \left (t_{\{i_1,i_2\}}-t_{\{i_1,i_2,i_3\}} \right ) t_{\{i_1,i_3\}} \Big] \Big[ t_{\{i_1+(r-e)\gamma,i_2+(r-e)\gamma,i_3+(r-e)\gamma\}} \left (t_{\{i_1+(r-e)\gamma,i_3+(r-e)\gamma\}}-1 \right )^+ 
\nonumber \\ 
& \hspace{1.3em}+ \left (t_{\{i_1+(r-e)\gamma,i_2+(r-e)\gamma\}}-t_{\{i_1+(r-e)\gamma,i_2+(r-e)\gamma,i_3+(r-e)\gamma\}} \right ) t_{\{i_1+(r-e)\gamma,i_3+(r-e)\gamma\}} \Big].
\end{align}
Case~7.5: The number of instances of $P_7$ with CNs $c_1$, $c_2$, and $c_3$ as defined in the previous paragraph, and all overlaps in three replicas s.t. the two $c_1-c_2$ overlaps are in $\bold{R}_r$, and the $c_1-c_3$ overlaps are in $\bold{R}_e$ and $\bold{R}_s$, $e<s$, is:
\begin{align}\label{eq_p7_5}
\mathcal{E}_{P_7} \left (t_{\{i_1,i_2\}}, t_{\{i_1+(r-e)\gamma,i_3+(r-e)\gamma\}}, t_{\{i_1+(r-s)\gamma,i_3+(r-s)\gamma\}} \right ) = \binom{t_{\{i_1,i_2\}}}{2} t_{\{i_1+(r-e)\gamma,i_3+(r-e)\gamma\}}t_{\{i_1+(r-s)\gamma,i_3+(r-s)\gamma\}}.
\end{align}
Case~7.6: The number of instances of $P_7$ with CNs $c_1$, $c_2$, and $c_3$ as defined in the previous paragraph, and all overlaps in three replicas s.t. two distinct overlaps (from different families) are in $\bold{R}_r$, one $c_1-c_2$ overlap is in $\bold{R}_e$, and one $c_1-c_3$ overlap is in $\bold{R}_s$, $e<s$, is:
\begin{align}\label{eq_p7_6}
\mathcal{G}_{P_7} & \big( t_{\{i_1,i_2\}}, t_{\{i_1,i_3\}}, t_{\{i_1,i_2,i_3\}}, t_{\{i_1+(r-e)\gamma,i_2+(r-e)\gamma\}}, t_{\{i_1+(r-s)\gamma,i_3+(r-s)\gamma\}} \big) 
\nonumber \\ 
&= \Big[ t_{\{i_1,i_2,i_3\}} \left (t_{\{i_1,i_3\}}-1 \right )^+ + \left (t_{\{i_1,i_2\}}-t_{\{i_1,i_2,i_3\}} \right ) t_{\{i_1,i_3\}} \Big] t_{\{i_1+(r-e)\gamma,i_2+(r-e)\gamma\}}t_{\{i_1+(r-s)\gamma,i_3+(r-s)\gamma\}}.
\end{align}
Case~7.7: The number of instances of $P_7$ with CNs $c_1$, $c_2$, and $c_3$ as defined in the previous paragraph, and overlaps in four replicas s.t. the two $c_1-c_2$ overlaps are in $\bold{R}_r$ and $\bold{R}_e$, and the two $c_1-c_3$ overlaps are in $\bold{R}_s$ and $\bold{R}_u$, $r<e$, $r<s$, and $s<u$, is:
\begin{align}\label{eq_p7_7}
\mathcal{I}_{P_7} &\left (t_{\{i_1,i_2\}}, t_{\{i_1+(r-e)\gamma,i_2+(r-e)\gamma\}}, t_{\{i_1+(r-s)\gamma,i_3+(r-s)\gamma\}}, t_{\{i_1+(r-u)\gamma,i_3+(r-u)\gamma\}} \right ) 
\nonumber \\ 
&= t_{\{i_1,i_2\}}t_{\{i_1+(r-e)\gamma,i_2+(r-e)\gamma\}}t_{\{i_1+(r-s)\gamma,i_3+(r-s)\gamma\}}t_{\{i_1+(r-u)\gamma,i_3+(r-u)\gamma\}}.
\end{align}
\end{lemma}

\begin{figure}[H]
\vspace{-1.5em}
\center
\includegraphics[width=7.0in]{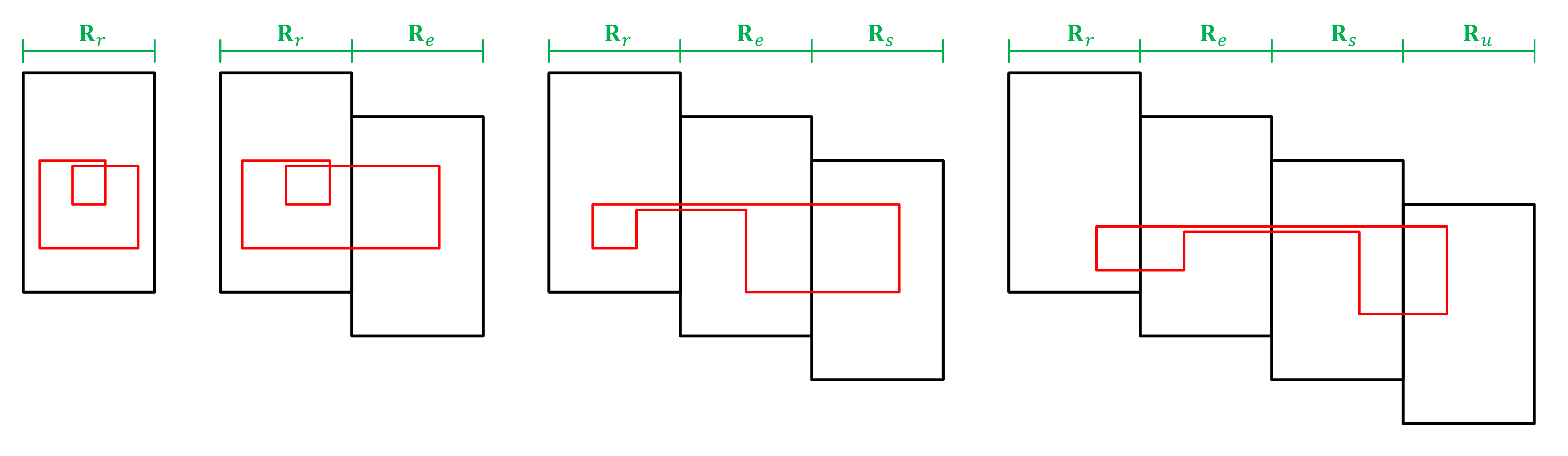}\vspace{-1.5em}
\caption{An instance of Pattern $P_7$ in Case~7.1, in Case~7.2, in Case~7.5, and in Case~7.7, from left to right. For simplicity, we have $e=r+1$, $s=e+1$, and $u=s+1$.}
\label{Fig_pat7}
\vspace{-0.5em}
\end{figure}

Four of the seven cases are illustrated in Fig.~\ref{Fig_pat7}.

\begin{theorem}\label{thm_p7}
The total number of instances of Pattern $P_7$ in the binary protograph of an SC code that has parameters $\gamma \geq 3$, $\kappa$, $m$, $L \geq m+1$, and $\mathcal{O}$, is:
\begin{equation}\label{eq_p7_8}
F_{P_7} = \sum_{k=1}^{m+1} (L-k+1) F^k_{P_7,1},
\end{equation}
where $F^k_{P_7,1}$, $k \in \{1, 2, \dots, m+1\}$, are given by:
\begin{align}\label{eq_p7_9}
F^1_{P_7,1} &= \hspace{-14.8em} \sum_{\hspace{15.0em} i_1 \in \{0, \dots, (m+1)\gamma -1\}, \{i_2,i_3\} \subset \{0, \dots, (m+1)\gamma -1\}} \hspace{-14.8em} \mathcal{A}_{P_7}\left (t_{\{i_1,i_2\}}, t_{\{i_1,i_3\}}, t_{\{i_1,i_2,i_3\}} \right ),  
\nonumber \\ 
F^2_{P_7,1} &= \hspace{-20.7em} \sum_{\hspace{20.8em} i_1 \in \{\gamma, \dots, (m+1)\gamma -1\}, i_2 \in \{0, \dots, (m+1)\gamma -1\}, i_3 \in \{\gamma, \dots, (m+1)\gamma -1\}} \hspace{-20.6em} \mathcal{B}_{P_7} \left (t_{\{i_1,i_2\}}, t_{\{i_1,i_3\}}, t_{\{i_1,i_2,i_3\}}, t_{\{i_1-\gamma,i_3-\gamma\}} \right ) 
\nonumber \\ 
&+ \hspace{-17.3em} \sum_{\hspace{17.7em} i_1 \in \{0, \dots, m\gamma -1\}, i_2 \in \{0, \dots, (m+1)\gamma -1\}, i_3 \in \{0, \dots, m\gamma -1\}} \hspace{-17.3em} \mathcal{B}_{P_7} \left (t_{\{i_1,i_2\}}, t_{\{i_1,i_3\}}, t_{\{i_1,i_2,i_3\}}, t_{\{i_1+\gamma,i_3+\gamma\}} \right ) 
\nonumber \\ 
&+ \hspace{-20.6em} \sum_{\hspace{21.0em} i_1 \in \{\gamma, \dots, (m+1)\gamma -1\}, i_2 \in \{0, \dots, (m+1)\gamma -1\}, i_3 \in \{\gamma, \dots, (m+2)\gamma -1\}} \hspace{-20.6em} \mathcal{C}_{P_7} \left (t_{\{i_1,i_2\}}, t_{\{i_1-\gamma,i_3-\gamma\}} \right ) 
\nonumber \\ 
&+ \hspace{-14.7em} \sum_{\hspace{15.0em}  i_1 \in \{\gamma, \dots, (m+1)\gamma -1\}, \{i_2,i_3\} \subset \{\gamma, \dots, (m+1)\gamma -1\}} \hspace{-14.8em} \mathcal{D}_{P_7} \big(t_{\{i_1,i_2\}}, t_{\{i_1,i_3\}}, t_{\{i_1,i_2,i_3\}}, t_{\{i_1-\gamma,i_2-\gamma\}}, t_{\{i_1-\gamma,i_3-\gamma\}}, t_{\{i_1-\gamma,i_2-\gamma,i_3-\gamma\}} \big),
\nonumber \\ 
F^3_{P_7,1} &= \hspace{-21.4em} \sum_{\hspace{21.5em} i_1 \in \{2\gamma, \dots, (m+1)\gamma -1\}, i_2 \in \{0, \dots, (m+1)\gamma -1\}, i_3 \in \{2\gamma, \dots, (m+1)\gamma -1\}} \hspace{-21.2em} \mathcal{B}_{P_7} \left (t_{\{i_1,i_2\}}, t_{\{i_1,i_3\}}, t_{\{i_1,i_2,i_3\}}, t_{\{i_1-2\gamma,i_3-2\gamma\}} \right ) 
\nonumber \\ 
&+ \hspace{-20.4em} \sum_{\hspace{20.6em} i_1 \in \{0, \dots, (m-1)\gamma -1\}, i_2 \in \{0, \dots, (m+1)\gamma -1\}, i_3 \in \{0, \dots, (m-1)\gamma -1\}} \hspace{-20.3em} \mathcal{B}_{P_7} \left (t_{\{i_1,i_2\}}, t_{\{i_1,i_3\}}, t_{\{i_1,i_2,i_3\}}, t_{\{i_1+2\gamma,i_3+2\gamma\}} \right ) 
\nonumber \\ 
&+ \hspace{-21.3em} \sum_{\hspace{21.5em} i_1 \in \{2\gamma, \dots, (m+1)\gamma -1\}, i_2 \in \{0, \dots, (m+1)\gamma -1\}, i_3 \in \{2\gamma, \dots, (m+3)\gamma -1\}} \hspace{-21.2em} \mathcal{C}_{P_7} \left (t_{\{i_1,i_2\}}, t_{\{i_1-2\gamma,i_3-2\gamma\}} \right ) 
\nonumber \\ 
&+ \hspace{-15.4em} \sum_{\hspace{15.7em}  i_1 \in \{2\gamma, \dots, (m+1)\gamma -1\}, \{i_2,i_3\} \subset \{2\gamma, \dots, (m+1)\gamma -1\}} \hspace{-15.4em} \mathcal{D}_{P_7} \big(t_{\{i_1,i_2\}}, t_{\{i_1,i_3\}}, t_{\{i_1,i_2,i_3\}}, t_{\{i_1-2\gamma,i_2-2\gamma\}}, t_{\{i_1-2\gamma,i_3-2\gamma\}}, t_{\{i_1-2\gamma,i_2-2\gamma,i_3-2\gamma\}} \big) 
\nonumber \\ 
&+ \hspace{-21.2em} \sum_{\hspace{21.5em}  i_1 \in \{2\gamma, \dots, (m+1)\gamma -1\}, i_2 \in \{0, \dots, (m+1)\gamma -1\}, i_3 \in \{2\gamma, \dots, (m+2)\gamma -1\}} \hspace{-21.2em} \mathcal{E}_{P_7} \big(t_{\{i_1,i_2\}}, t_{\{i_1-\gamma,i_3-\gamma\}}, t_{\{i_1-2\gamma,i_3-2\gamma\}} \big) 
\nonumber \\ 
&+ \hspace{-17.2em} \sum_{\hspace{17.5em}  i_1 \in \{\gamma, \dots, m\gamma -1\}, i_2 \in \{0, \dots, (m+1)\gamma -1\}, i_3 \in \{\gamma, \dots, m\gamma -1\}} \hspace{-17.1em} \mathcal{E}_{P_7} \big(t_{\{i_1,i_2\}}, t_{\{i_1+\gamma,i_3+\gamma\}}, t_{\{i_1-\gamma,i_3-\gamma\}} \big) 
\nonumber \\ 
&+ \hspace{-21.0em} \sum_{\hspace{21.3em}  i_1 \in \{0, \dots, (m-1)\gamma -1\}, i_2 \in \{0, \dots, (m+1)\gamma -1\}, i_3 \in \{-\gamma, \dots, (m-1)\gamma -1\}} \hspace{-21.0em} \mathcal{E}_{P_7} \big(t_{\{i_1,i_2\}}, t_{\{i_1+2\gamma,i_3+2\gamma\}}, t_{\{i_1+\gamma,i_3+\gamma\}} \big) 
\nonumber \\ 
&+ \hspace{-21.4em} \sum_{\hspace{21.8em}  i_1 \in \{2\gamma, \dots, (m+1)\gamma -1\}, i_2 \in \{\gamma, \dots, (m+1)\gamma -1\}, i_3 \in \{2\gamma, \dots, (m+1)\gamma -1\}} \hspace{-21.3em} \mathcal{G}_{P_7} \big( t_{\{i_1,i_2\}}, t_{\{i_1,i_3\}}, t_{\{i_1,i_2,i_3\}}, t_{\{i_1-\gamma,i_2-\gamma\}}, t_{\{i_1-2\gamma,i_3-2\gamma\}} \big) 
\nonumber \\ 
&+ \hspace{-17.3em} \sum_{\hspace{17.6em}  i_1 \in \{\gamma, \dots, m\gamma -1\}, i_2 \in \{0, \dots, m\gamma -1\}, i_3 \in \{\gamma, \dots, (m+1)\gamma -1\}} \hspace{-17.1em} \mathcal{G}_{P_7} \big( t_{\{i_1,i_2\}}, t_{\{i_1,i_3\}}, t_{\{i_1,i_2,i_3\}}, t_{\{i_1+\gamma,i_2+\gamma\}}, t_{\{i_1-\gamma,i_3-\gamma\}} \big) 
\nonumber \\ 
&+ \hspace{-18.9em} \sum_{\hspace{19.4em}  i_1 \in \{0, \dots, (m-1)\gamma -1\}, i_2 \in \{0, \dots, (m-1)\gamma -1\}, i_3 \in \{0, \dots, m\gamma -1\}} \hspace{-18.8em} \mathcal{G}_{P_7} \big( t_{\{i_1,i_2\}}, t_{\{i_1,i_3\}}, t_{\{i_1,i_2,i_3\}}, t_{\{i_1+2\gamma,i_2+2\gamma\}}, t_{\{i_1+\gamma,i_3+\gamma\}} \big), 
\nonumber \\ 
F^{k \geq 4}_{P_7,1} &= \hspace{-24.8em} \sum_{\hspace{25.0em} i_1 \in \{(k-1)\gamma, \dots, (m+1)\gamma -1\}, i_2 \in \{0, \dots, (m+1)\gamma -1\}, i_3 \in \{(k-1)\gamma, \dots, (m+1)\gamma -1\}} \hspace{-24.6em} \mathcal{B}_{P_7} \left (t_{\{i_1,i_2\}}, t_{\{i_1,i_3\}}, t_{\{i_1,i_2,i_3\}}, t_{\{i_1+(1-k)\gamma,i_3+(1-k)\gamma\}} \right ) 
\nonumber \\ 
&+ \hspace{-22.6em} \sum_{\hspace{23.0em} i_1 \in \{0, \dots, (m-k+2)\gamma -1\}, i_2 \in \{0, \dots, (m+1)\gamma -1\}, i_3 \in \{0, \dots, (m-k+2)\gamma -1\}} \hspace{-22.5em} \mathcal{B}_{P_7} \left (t_{\{i_1,i_2\}}, t_{\{i_1,i_3\}}, t_{\{i_1,i_2,i_3\}}, t_{\{i_1+(k-1)\gamma,i_3+(k-1)\gamma\}} \right ) 
\nonumber \\  
&+ \hspace{-24.7em} \sum_{\hspace{25.0em} i_1 \in \{(k-1)\gamma, \dots, (m+1)\gamma -1\}, i_2 \in \{0, \dots, (m+1)\gamma -1\}, i_3 \in \{(k-1)\gamma, \dots, (m+k)\gamma -1\}} \hspace{-24.5em} \mathcal{C}_{P_7} \left (t_{\{i_1,i_2\}}, t_{\{i_1+(1-k)\gamma,i_3+(1-k)\gamma\}} \right ) 
\nonumber \\ 
&+ \hspace{-18.8em} \sum_{\hspace{19.2em}  i_1 \in \{(k-1)\gamma, \dots, (m+1)\gamma -1\}, \{i_2,i_3\} \subset \{(k-1)\gamma, \dots, (m+1)\gamma -1\}} \hspace{-18.7em} \mathcal{D}_{P_7} \big(t_{\{i_1,i_2\}}, t_{\{i_1,i_3\}}, t_{\{i_1,i_2,i_3\}}, t_{\{i_1+(1-k)\gamma,i_2+(1-k)\gamma\}}, t_{\{i_1+(1-k)\gamma,i_3+(1-k)\gamma\}} 
\nonumber \\
& \hspace{6.1em} , t_{\{i_1+(1-k)\gamma,i_2+(1-k)\gamma,i_3+(1-k)\gamma\}} \big) 
\nonumber \\ 
&+ \hspace{0.5em} \sum_{h=2}^{k-1} \hspace{-24.5em} \sum_{\hspace{25.0em}  i_1 \in \{(k-1)\gamma, \dots, (m+1)\gamma -1\}, i_2 \in \{0, \dots, (m+1)\gamma -1\}, i_3 \in \{(k-1)\gamma, \dots, (m+h)\gamma -1\}} \hspace{-24.5em} \mathcal{E}_{P_7} \big(t_{\{i_1,i_2\}}, t_{\{i_1+(1-h)\gamma,i_3+(1-h)\gamma\}}, t_{\{i_1+(1-k)\gamma,i_3+(1-k)\gamma\}} \big) 
\nonumber \\ 
&+ \hspace{0.5em} \sum_{h=2}^{k-1} \hspace{-26.7em} \sum_{\hspace{27.2em}  i_1 \in \{(k-h)\gamma, \dots, (m-h+2)\gamma -1\}, i_2 \in \{0, \dots, (m+1)\gamma -1\}, i_3 \in \{(k-h)\gamma, \dots, (m-h+2)\gamma -1\}} \hspace{-26.8em} \mathcal{E}_{P_7} \big(t_{\{i_1,i_2\}}, t_{\{i_1+(h-1)\gamma,i_3+(h-1)\gamma\}}, t_{\{i_1+(h-k)\gamma,i_3+(h-k)\gamma\}} \big) 
\nonumber \\ 
&+ \hspace{0.5em} \sum_{h=2}^{k-1} \hspace{-24.4em} \sum_{\hspace{25.0em}  i_1 \in \{0, \dots, (m-k+2)\gamma -1\}, i_2 \in \{0, \dots, (m+1)\gamma -1\}, i_3 \in \{(h-k)\gamma, \dots, (m-k+2)\gamma -1\}} \hspace{-24.6em} \mathcal{E}_{P_7} \big(t_{\{i_1,i_2\}}, t_{\{i_1+(k-1)\gamma,i_3+(k-1)\gamma\}}, t_{\{i_1+(k-h)\gamma,i_3+(k-h)\gamma\}} \big) 
\nonumber \\ 
&+ \hspace{0.5em} \sum_{h=2}^{k-1} \hspace{-26.7em} \sum_{\hspace{27.5em} i_1 \in \{(k-1)\gamma, \dots, (m+1)\gamma -1\}, i_2 \in \{(h-1)\gamma, \dots, (m+1)\gamma -1\}, i_3 \in \{(k-1)\gamma, \dots, (m+1)\gamma -1\}} \hspace{-26.9em} \mathcal{G}_{P_7} \big( t_{\{i_1,i_2\}}, t_{\{i_1,i_3\}}, t_{\{i_1,i_2,i_3\}}, t_{\{i_1+(1-h)\gamma,i_2+(1-h)\gamma\}}, t_{\{i_1+(1-k)\gamma,i_3+(1-k)\gamma\}} \big) 
\nonumber \\ 
&+ \hspace{0.5em} \sum_{h=2}^{k-1} \hspace{-26.7em} \sum_{\hspace{27.3em}  i_1 \in \{(k-h)\gamma, \dots, (m-h+2)\gamma -1\}, i_2 \in \{0, \dots, (m-h+2)\gamma -1\}, i_3 \in \{(k-h)\gamma, \dots, (m+1)\gamma -1\}} \hspace{-26.8em} \mathcal{G}_{P_7} \big( t_{\{i_1,i_2\}}, t_{\{i_1,i_3\}}, t_{\{i_1,i_2,i_3\}}, t_{\{i_1+(h-1)\gamma,i_2+(h-1)\gamma\}}, t_{\{i_1+(h-k)\gamma,i_3+(h-k)\gamma\}} \big)
\nonumber \\ 
&+ \hspace{0.5em} \sum_{h=2}^{k-1} \hspace{-24.4em} \sum_{\hspace{25.0em}  i_1 \in \{0, \dots, (m-k+2)\gamma -1\}, i_2 \in \{0, \dots, (m-k+2)\gamma -1\}, i_3 \in \{0, \dots, (m-k+h+1)\gamma -1\}} \hspace{-24.5em} \mathcal{G}_{P_7} \big( t_{\{i_1,i_2\}}, t_{\{i_1,i_3\}}, t_{\{i_1,i_2,i_3\}}, t_{\{i_1+(k-1)\gamma,i_2+(k-1)\gamma\}}, t_{\{i_1+(k-h)\gamma,i_3+(k-h)\gamma\}} \big)  
\nonumber \\ 
&+ \hspace{0.5em} \sum_{h=2}^{k-2} \hspace{0.3em} \sum_{w=h+1}^{k-1} \hspace{-27.1em} \sum_{\hspace{27.5em}  i_1 \in \{(k-1)\gamma, \dots, (m+1)\gamma -1\}, i_2 \in \{(h-1)\gamma, \dots, (m+1)\gamma -1\}, i_3 \in \{(k-1)\gamma, \dots, (m+w)\gamma -1\}} \hspace{-27.1em} \mathcal{I}_{P_7} \big( t_{\{i_1,i_2\}}, t_{\{i_1+(1-h)\gamma,i_2+(1-h)\gamma\}}, t_{\{i_1+(1-w)\gamma,i_3+(1-w)\gamma\}}, t_{\{i_1+(1-k)\gamma,i_3+(1-k)\gamma\}} \big)
\nonumber \\ 
&+ \hspace{0.5em} \sum_{h=2}^{k-2} \hspace{0.3em} \sum_{w=h+1}^{k-1} \hspace{-27.1em} \sum_{\hspace{27.5em}  i_1 \in \{(k-1)\gamma, \dots, (m+1)\gamma -1\}, i_2 \in \{(w-1)\gamma, \dots, (m+1)\gamma -1\}, i_3 \in \{(k-1)\gamma, \dots, (m+h)\gamma -1\}} \hspace{-27.1em} \mathcal{I}_{P_7} \big( t_{\{i_1,i_2\}}, t_{\{i_1+(1-w)\gamma,i_2+(1-w)\gamma\}}, t_{\{i_1+(1-h)\gamma,i_3+(1-h)\gamma\}}, t_{\{i_1+(1-k)\gamma,i_3+(1-k)\gamma\}} \big)
\nonumber \\ 
&+ \hspace{0.5em} \sum_{h=2}^{k-2} \hspace{0.3em} \sum_{w=h+1}^{k-1} \hspace{-27.1em} \sum_{\hspace{27.5em}  i_1 \in \{(k-1)\gamma, \dots, (m+1)\gamma -1\}, i_2 \in \{(k-1)\gamma, \dots, (m+1)\gamma -1\}, i_3 \in \{(w-1)\gamma, \dots, (m+h)\gamma -1\}} \hspace{-27.1em} \mathcal{I}_{P_7} \big( t_{\{i_1,i_2\}}, t_{\{i_1+(1-k)\gamma,i_2+(1-k)\gamma\}}, t_{\{i_1+(1-h)\gamma,i_3+(1-h)\gamma\}}, t_{\{i_1+(1-w)\gamma,i_3+(1-w)\gamma\}} \big),
\end{align}
with $\overline{i_1} \neq \overline{i_2}$, $\overline{i_1} \neq \overline{i_3}$, and $i_2 \neq i_3$.
\end{theorem}

\subsection{Analysis of Pattern $P_8$ (size $4 \times 3$)}\label{subsec_p8}

This pattern has three VNs, and the adjacent pairs are $v_1-v_2$ and $v_1-v_3$ (not all pairs) according to $P_8$ in Fig.~\ref{Fig_pat}. Thus, $P_8$ spans at most $2m+1$ consecutive replicas (see \cite[Lemma~1]{homa_boo}). Pattern $P_8$ does not exist in the case of $\gamma=3$. Suppose $P_8$ has the CNs $c_1$, $c_2$, $c_3$, and $c_4$. The pattern is formed of three overlaps, two of degree-$2$ and one of degree-$4$. The degree-$2$ overlaps are not only distinct, but also mutually exclusive (i.e., they do not share any CNs). Define the CNs such that $c_1$ and $c_2$ are directly connected twice, which is the same for $c_3$ and $c_4$. Thus, the overlaps are $c_1-c_2$, $c_3-c_4$, and $c_1-c_2-c_3-c_4$ (see also $P_8$ in Fig.~\ref{Fig_pat}).

\begin{lemma}\label{lem_p8}
Case~8.1: The number of instances of $P_8$ with CNs $c_1$, $c_2$, $c_3$, and $c_4$ as defined in the previous paragraph, and all overlaps in one replica, $\bold{R}_r$, is:
\begin{align}\label{eq_p8_1}
\mathcal{A}_{P_8} & \left (t_{\{i_1,i_2\}}, t_{\{i_3,i_4\}}, t_{\{i_1,i_2,i_3,i_4\}} \right ) = t_{\{i_1,i_2,i_3,i_4\}} \left( t_{\{i_1,i_2,i_3,i_4\}}-1 \right)^+ \left( t_{\{i_3,i_4\}}-2 \right)^+ 
\nonumber \\ 
&+ t_{\{i_1,i_2,i_3,i_4\}} \left( t_{\{i_1,i_2\}}-t_{\{i_1,i_2,i_3,i_4\}} \right) \left( t_{\{i_3,i_4\}}-1 \right)^+.
\end{align}
Case~8.2: The number of instances of $P_8$ with CNs $c_1$, $c_2$, $c_3$, and $c_4$ as defined in the previous paragraph, and all overlaps in two replicas s.t. the two degree-$2$ overlaps are in $\bold{R}_r$, and the degree-$4$ overlap is in $\bold{R}_e$, is:
\begin{align}\label{eq_p8_2}
\mathcal{B}_{P_8} & \big(t_{\{i_1,i_2\}}, t_{\{i_3,i_4\}}, t_{\{i_1,i_2,i_3,i_4\}}, t_{\{i_1+(r-e)\gamma,i_2+(r-e)\gamma,i_3+(r-e)\gamma,i_4+(r-e)\gamma\}} \big) 
\nonumber \\ 
& = \Big[ t_{\{i_1,i_2,i_3,i_4\}} \left( t_{\{i_3,i_4\}}-1 \right)^+ + \left (t_{\{i_1,i_2\}}-t_{\{i_1,i_2,i_3,i_4\}} \right ) t_{\{i_3,i_4\}} \Big] t_{\{i_1+(r-e)\gamma,i_2+(r-e)\gamma,i_3+(r-e)\gamma,i_4+(r-e)\gamma\}}.
\end{align}
Case~8.3: The number of instances of $P_8$ with CNs $c_1$, $c_2$, $c_3$, and $c_4$ as defined in the previous paragraph, and all overlaps in two replicas s.t. the degree-$4$ overlap and the $c_1-c_2$ overlap are in $\bold{R}_r$, and the $c_3-c_4$ overlap is in $\bold{R}_e$, is:
\begin{align}\label{eq_p8_3}
\mathcal{C}_{P_8} \left(t_{\{i_1,i_2\}}, t_{\{i_1,i_2,i_3,i_4\}}, t_{\{i_3+(r-e)\gamma,i_4+(r-e)\gamma\}} \right ) = t_{\{i_1,i_2,i_3,i_4\}} \left( t_{\{i_1,i_2\}}-1 \right)^+ t_{\{i_3+(r-e)\gamma,i_4+(r-e)\gamma\}}.
\end{align}
Case~8.4: The number of instances of $P_8$ with $c_1$, $c_2$, $c_3$, and $c_4$ as defined previously, and overlaps in three replicas s.t. the $c_1-c_2$ overlap is in $\bold{R}_r$, the $c_3-c_4$ overlap is in $\bold{R}_e$, and the degree-$4$ overlap is in $\bold{R}_s$, $r < e$, is:
\begin{align}\label{eq_p8_4}
\mathcal{D}_{P_8} & \big(t_{\{i_1,i_2\}}, t_{\{i_3+(r-e)\gamma,i_4+(r-e)\gamma\}}, t_{\{i_1+(r-s)\gamma,i_2+(r-s)\gamma,i_3+(r-s)\gamma,i_4+(r-s)\gamma\}} \big) 
\nonumber \\ 
&= t_{\{i_1,i_2\}} t_{\{i_3+(r-e)\gamma,i_4+(r-e)\gamma\}} t_{\{i_1+(r-s)\gamma,i_2+(r-s)\gamma,i_3+(r-s)\gamma,i_4+(r-s)\gamma\}}.
\end{align}
\end{lemma}

\begin{figure}[H]
\vspace{-1.5em}
\center
\includegraphics[width=5.5in]{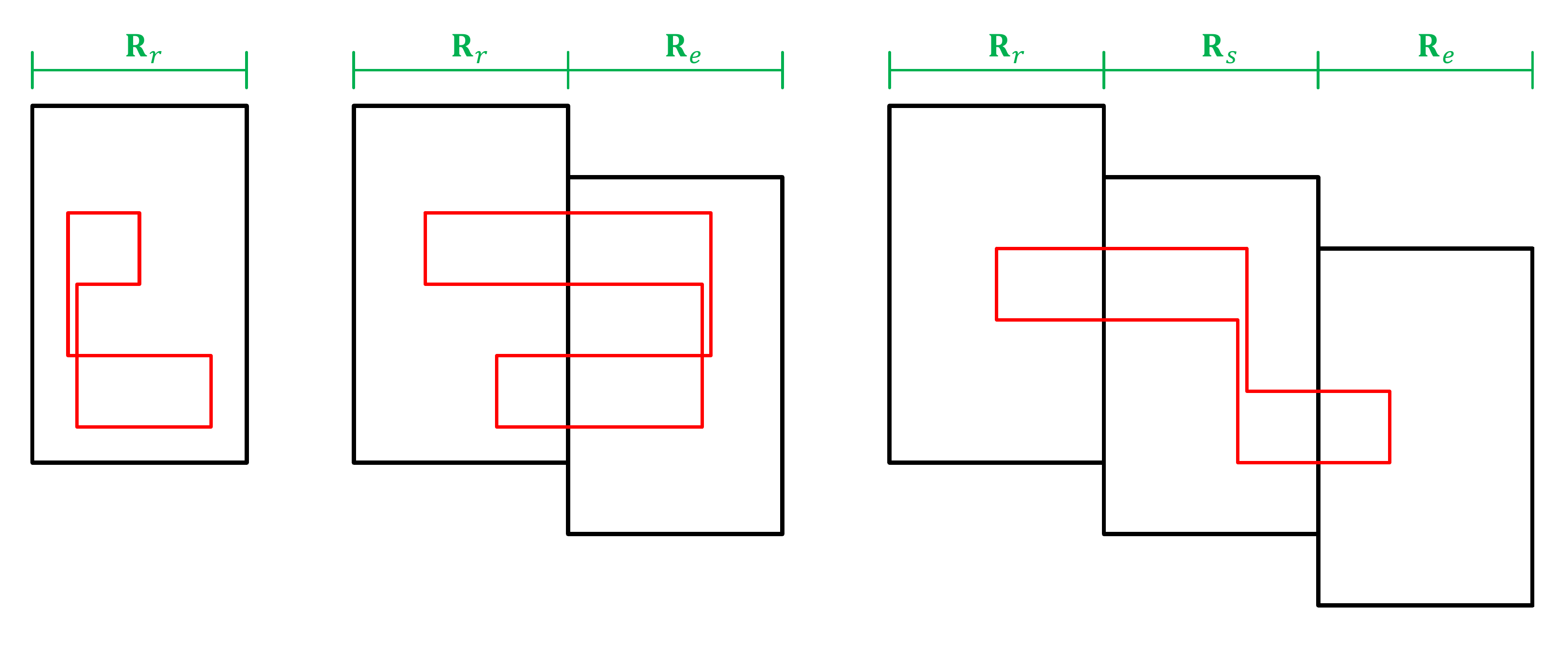}\vspace{-1.5em}
\caption{An instance of Pattern $P_8$ in Case~8.1, in Case~8.2, and in Case~8.4, from left to right. For simplicity, we have $e=r+y$, where $y \in \{1,2\}$, and $s=e-1$.}
\label{Fig_pat8}
\vspace{-0.5em}
\end{figure}

Three of the four cases are illustrated in Fig.~\ref{Fig_pat8}.

\begin{theorem}\label{thm_p8}
The total number of instances of Pattern $P_8$ in the binary protograph of an SC code that has parameters $\gamma \geq 4$, $\kappa$, $m$, $L \geq 2m+1$, and $\mathcal{O}$, is:
\begin{equation}\label{eq_p8_5}
F_{P_8} = \sum_{k=1}^{2m+1} (L-k+1) F^k_{P_8,1},
\end{equation}
where $F^k_{P_8,1}$, $k \in \{1, 2, \dots, 2m+1\}$, are given by:
\vspace{-0.1em}\begin{align}\label{eq_p8_6}
F^1_{P_8,1} &= \hspace{0.3em} \frac{1}{2} \hspace{-16.1em} \sum_{\hspace{16.5em} \{i_1,i_2\} \subset \{0, \dots, (m+1)\gamma -1\}, \{i_3,i_4\} \subset \{0, \dots, (m+1)\gamma -1\}} \hspace{-16.0em} \mathcal{A}_{P_8}\left (t_{\{i_1,i_2\}}, t_{\{i_3,i_4\}}, t_{\{i_1,i_2,i_3,i_4\}} \right ), 
\nonumber \\
F^2_{P_8,1} &= \hspace{0.3em} \frac{1}{2} \hspace{-16.1em} \sum_{\hspace{16.5em} \{i_1,i_2\} \subset \{\gamma, \dots, (m+1)\gamma -1\}, \{i_3,i_4\} \subset \{\gamma, \dots, (m+1)\gamma -1\}} \hspace{-16.0em} \mathcal{B}_{P_8} \big (t_{\{i_1,i_2\}}, t_{\{i_3,i_4\}}, t_{\{i_1,i_2,i_3,i_4\}}, t_{\{i_1-\gamma,i_2-\gamma,i_3-\gamma,i_4-\gamma\}} \big ) 
\nonumber \\ 
&+ \hspace{0.5em} \frac{1}{2} \hspace{-12.9em} \sum_{\hspace{13.5em} \{i_1,i_2\} \subset \{0, \dots, m\gamma -1\}, \{i_3,i_4\} \subset \{0, \dots, m\gamma -1\}} \hspace{-12.7em} \mathcal{B}_{P_8} \big (t_{\{i_1,i_2\}}, t_{\{i_3,i_4\}}, t_{\{i_1,i_2,i_3,i_4\}}, t_{\{i_1+\gamma,i_2+\gamma,i_3+\gamma,i_4+\gamma\}} \big ) 
\nonumber \\ 
&+ \hspace{-16.3em} \sum_{\hspace{16.5em} \{i_1,i_2\} \subset \{0, \dots, (m+1)\gamma -1\}, \{i_3,i_4\} \subset \{\gamma, \dots, (m+1)\gamma -1\}} \hspace{-15.9em} \mathcal{C}_{P_8} \left (t_{\{i_1,i_2\}}, t_{\{i_1,i_2,i_3,i_4\}}, t_{\{i_3-\gamma,i_4-\gamma\}} \right ) 
\nonumber \\ 
&+ \hspace{-14.7em} \sum_{\hspace{15.0em}  \{i_1,i_2\} \subset \{0, \dots, (m+1)\gamma -1\}, \{i_3,i_4\} \subset \{0, \dots, m\gamma -1\}} \hspace{-14.2em} \mathcal{C}_{P_8} \left (t_{\{i_1,i_2\}}, t_{\{i_1,i_2,i_3,i_4\}}, t_{\{i_3+\gamma,i_4+\gamma\}} \right ), 
\nonumber \\ 
F^{k \geq 3}_{P_8,1} &= \hspace{0.3em} \frac{1}{2} \hspace{-20.2em} \sum_{\hspace{20.5em} \{i_1,i_2\} \subset \{(k-1)\gamma, \dots, (m+1)\gamma -1\}, \{i_3,i_4\} \subset \{(k-1)\gamma, \dots, (m+1)\gamma -1\}} \hspace{-20.0em} \mathcal{B}_{P_8} \big (t_{\{i_1,i_2\}}, t_{\{i_3,i_4\}}, t_{\{i_1,i_2,i_3,i_4\}}, t_{\{i_1+(1-k)\gamma,i_2+(1-k)\gamma,i_3+(1-k)\gamma,i_4+(1-k)\gamma\}} \big ) 
\nonumber \\ 
&+ \hspace{0.5em} \frac{1}{2} \hspace{-18.1em} \sum_{\hspace{18.5em} \{i_1,i_2\} \subset \{0, \dots, (m-k+2)\gamma -1\}, \{i_3,i_4\} \subset \{0, \dots, (m-k+2)\gamma -1\}} \hspace{-17.8em} \mathcal{B}_{P_8} \big (t_{\{i_1,i_2\}}, t_{\{i_3,i_4\}}, t_{\{i_1,i_2,i_3,i_4\}}, t_{\{i_1+(k-1)\gamma,i_2+(k-1)\gamma,i_3+(k-1)\gamma,i_4+(k-1)\gamma\}} \big ) 
\nonumber \\ 
&+ \hspace{-18.4em} \sum_{\hspace{18.5em} \{i_1,i_2\} \subset \{0, \dots, (m+1)\gamma -1\}, \{i_3,i_4\} \subset \{(k-1)\gamma, \dots, (m+1)\gamma -1\}} \hspace{-17.9em} \mathcal{C}_{P_8} \left (t_{\{i_1,i_2\}}, t_{\{i_1,i_2,i_3,i_4\}}, t_{\{i_3+(1-k)\gamma,i_4+(1-k)\gamma\}} \right ) 
\nonumber \\ 
&+ \hspace{-17.3em} \sum_{\hspace{17.5em}  \{i_1,i_2\} \subset \{0, \dots, (m+1)\gamma -1\}, \{i_3,i_4\} \subset \{0, \dots, (m-k+2)\gamma -1\}} \hspace{-16.9em} \mathcal{C}_{P_8} \left (t_{\{i_1,i_2\}}, t_{\{i_1,i_2,i_3,i_4\}}, t_{\{i_3+(k-1)\gamma,i_4+(k-1)\gamma\}} \right )
\nonumber \\
&+ \hspace{0.5em} \sum_{h=2}^{k-1} \hspace{-20.3em} \sum_{\hspace{21.0em}  \{i_1,i_2\} \subset \{(k-1)\gamma, \dots, (m+1)\gamma -1\}, \{i_3,i_4\} \subset \{(k-1)\gamma, \dots, (m+h)\gamma -1\}} \hspace{-20.3em} \mathcal{D}_{P_8} \big (t_{\{i_1,i_2\}}, t_{\{i_3+(1-h)\gamma,i_4+(1-h)\gamma\}}, t_{\{i_1+(1-k)\gamma,i_2+(1-k)\gamma,i_3+(1-k)\gamma,i_4+(1-k)\gamma\}} \big )
\nonumber \\
&+ \hspace{0.5em} \sum_{h=2}^{k-1} \hspace{-20.3em} \sum_{\hspace{21.0em}  \{i_1,i_2\} \subset \{(h-1)\gamma, \dots, (m+1)\gamma -1\}, \{i_3,i_4\} \subset \{(k-1)\gamma, \dots, (m+h)\gamma -1\}} \hspace{-20.3em} \mathcal{D}_{P_8} \big (t_{\{i_1,i_2\}}, t_{\{i_3+(1-k)\gamma,i_4+(1-k)\gamma\}}, t_{\{i_1+(1-h)\gamma,i_2+(1-h)\gamma,i_3+(1-h)\gamma,i_4+(1-h)\gamma\}} \big )
\nonumber \\
&+ \hspace{0.5em} \sum_{h=2}^{k-1} \hspace{-20.3em} \sum_{\hspace{21.0em}  \{i_1,i_2\} \subset \{0, \dots, (m-h+2)\gamma -1\}, \{i_3,i_4\} \subset \{(k-h)\gamma, \dots, (m-h+2)\gamma -1\}} \hspace{-20.3em} \mathcal{D}_{P_8} \big (t_{\{i_1,i_2\}}, t_{\{i_3+(h-k)\gamma,i_4+(h-k)\gamma\}}, t_{\{i_1+(h-1)\gamma,i_2+(h-1)\gamma,i_3+(h-1)\gamma,i_4+(h-1)\gamma\}} \big ),
\end{align}
with $\overline{i_1} \neq \overline{i_2}$, $\overline{i_1} \neq \overline{i_3}$, $\overline{i_1} \neq \overline{i_4}$, $\overline{i_2} \neq \overline{i_3}$, $\overline{i_2} \neq \overline{i_4}$, and $\overline{i_3} \neq \overline{i_4}$.
\end{theorem}

\subsection{Analysis of Pattern $P_9$ (size $4 \times 4$)}\label{subsec_p9}

This pattern has four VNs, and the adjacent pairs are $v_1-v_2$, $v_2-v_3$, $v_3-v_4$, and $v_1-v_4$ (not all pairs) according to $P_9$ in Fig.~\ref{Fig_pat}. Thus, $P_9$ also spans at most $2m+1$ consecutive replicas. Suppose $P_9$ has the CNs $c_1$, $c_2$, $c_3$, and $c_4$. The pattern is formed of four distinct degree-$2$ overlaps. Define the CNs such that the adjacent pairs (connected via at least one path with only one VN) are $c_1-c_2$, $c_2-c_3$, $c_3-c_4$, and $c_1-c_4$. This definition already implies what the overlaps are.

\begin{lemma}\label{lem_p9}
Case~9.1: The number of instances of $P_9$ with CNs $c_1$, $c_2$, $c_3$, and $c_4$ as defined in the previous paragraph, and all overlaps in one replica, $\bold{R}_r$, is:
\vspace{-0.1em}\begin{align}\label{eq_p9_1}
\mathcal{A}_{P_9} & \big (t_{\{i_1,i_2\}}, t_{\{i_2,i_3\}}, t_{\{i_3,i_4\}}, t_{\{i_1,i_4\}}, t_{\{i_1,i_2,i_3\}}, t_{\{i_1,i_2,i_4\}}, t_{\{i_1,i_3,i_4\}}, t_{\{i_2,i_3,i_4\}}, t_{\{i_1,i_2,i_3,i_4\}} \big ) 
\nonumber \\ 
&= \mathcal{A}_{P_9,1}+\mathcal{A}_{P_9,2}+\mathcal{A}_{P_9,3}+\mathcal{A}_{P_9,4},
\end{align}
\begin{align}
\mathcal{A}_{P_9,1} &= t_{\{i_1,i_2,i_3,i_4\}} \left( t_{\{i_1,i_2,i_3,i_4\}}-1 \right)^+ \left( t_{\{i_1,i_3,i_4\}}-2 \right)^+ \left( t_{\{i_1,i_4\}}-3 \right)^+ 
\nonumber \\ 
&+ t_{\{i_1,i_2,i_3,i_4\}} \left( t_{\{i_1,i_2,i_3,i_4\}}-1 \right)^+ \left( t_{\{i_3,i_4\}}-t_{\{i_1,i_3,i_4\}} \right) \left( t_{\{i_1,i_4\}}-2 \right)^+ 
\nonumber \\ 
&+ t_{\{i_1,i_2,i_3,i_4\}} \left(t_{\{i_2,i_3,i_4\}}-t_{\{i_1,i_2,i_3,i_4\}} \right) \left( t_{\{i_1,i_3,i_4\}}-1 \right)^+ \left( t_{\{i_1,i_4\}}-2 \right)^+ 
\nonumber \\ 
&+ t_{\{i_1,i_2,i_3,i_4\}} \left(t_{\{i_2,i_3,i_4\}}-t_{\{i_1,i_2,i_3,i_4\}} \right) \left( t_{\{i_3,i_4\}}-t_{\{i_1,i_3,i_4\}}-1\right)^+ \left( t_{\{i_1,i_4\}}-1 \right)^+ 
\nonumber \\ 
&+ t_{\{i_1,i_2,i_3,i_4\}} \left(t_{\{i_2,i_3\}}-t_{\{i_2,i_3,i_4\}} \right) \left( t_{\{i_1,i_3,i_4\}}-1 \right)^+ \left( t_{\{i_1,i_4\}}-2 \right)^+ 
\nonumber \\ 
&+ t_{\{i_1,i_2,i_3,i_4\}} \left(t_{\{i_2,i_3\}}-t_{\{i_2,i_3,i_4\}} \right) \left( t_{\{i_3,i_4\}}-t_{\{i_1,i_3,i_4\}} \right) \left( t_{\{i_1,i_4\}}-1 \right)^+, 
\nonumber
\end{align}
\begin{align}
\mathcal{A}_{P_9,2} &= \left (t_{\{i_1,i_2,i_3\}}-t_{\{i_1,i_2,i_3,i_4\}} \right) t_{\{i_1,i_2,i_3,i_4\}} \left( t_{\{i_1,i_3,i_4\}}-1 \right)^+ \left( t_{\{i_1,i_4\}}-2 \right)^+ 
\nonumber \\ 
&+ \left (t_{\{i_1,i_2,i_3\}}-t_{\{i_1,i_2,i_3,i_4\}} \right) t_{\{i_1,i_2,i_3,i_4\}} \left( t_{\{i_3,i_4\}}-t_{\{i_1,i_3,i_4\}} \right) \left( t_{\{i_1,i_4\}}-1 \right)^+ 
\nonumber \\ 
&+ \left (t_{\{i_1,i_2,i_3\}}-t_{\{i_1,i_2,i_3,i_4\}} \right) \left(t_{\{i_2,i_3,i_4\}}-t_{\{i_1,i_2,i_3,i_4\}}\right) t_{\{i_1,i_3,i_4\}} \left( t_{\{i_1,i_4\}}-1 \right)^+ 
\nonumber \\ 
&+ \left (t_{\{i_1,i_2,i_3\}}-t_{\{i_1,i_2,i_3,i_4\}} \right) \left(t_{\{i_2,i_3,i_4\}}-t_{\{i_1,i_2,i_3,i_4\}} \right) \left( t_{\{i_3,i_4\}}-t_{\{i_1,i_3,i_4\}}-1\right)^+ t_{\{i_1,i_4\}} 
\nonumber \\ 
&+ \left (t_{\{i_1,i_2,i_3\}}-t_{\{i_1,i_2,i_3,i_4\}} \right) \left(t_{\{i_2,i_3\}}-t_{\{i_2,i_3,i_4\}}-1 \right)^+ t_{\{i_1,i_3,i_4\}} \left( t_{\{i_1,i_4\}}-1 \right )^+ 
\nonumber \\ 
&+ \left (t_{\{i_1,i_2,i_3\}}-t_{\{i_1,i_2,i_3,i_4\}} \right) \left(t_{\{i_2,i_3\}}-t_{\{i_2,i_3,i_4\}}-1 \right)^+ \left( t_{\{i_3,i_4\}}-t_{\{i_1,i_3,i_4\}} \right) t_{\{i_1,i_4\}} 
\nonumber,
\end{align}
\begin{align}
\mathcal{A}_{P_9,3} &= \left (t_{\{i_1,i_2,i_4\}}-t_{\{i_1,i_2,i_3,i_4\}} \right) t_{\{i_1,i_2,i_3,i_4\}} \left( t_{\{i_1,i_3,i_4\}}-1 \right)^+ \left( t_{\{i_1,i_4\}}-3 \right)^+ 
\nonumber \\ 
&+ \left (t_{\{i_1,i_2,i_4\}}-t_{\{i_1,i_2,i_3,i_4\}} \right) t_{\{i_1,i_2,i_3,i_4\}} \left( t_{\{i_3,i_4\}}-t_{\{i_1,i_3,i_4\}} \right) \left( t_{\{i_1,i_4\}}-2 \right)^+
\nonumber \\ 
&+ \left (t_{\{i_1,i_2,i_4\}}-t_{\{i_1,i_2,i_3,i_4\}} \right) \left(t_{\{i_2,i_3,i_4\}}-t_{\{i_1,i_2,i_3,i_4\}} \right) t_{\{i_1,i_3,i_4\}} \left (t_{\{i_1,i_4\}}-2 \right)^+ 
\nonumber \\ 
&+ \left (t_{\{i_1,i_2,i_4\}}-t_{\{i_1,i_2,i_3,i_4\}} \right) \left(t_{\{i_2,i_3,i_4\}}-t_{\{i_1,i_2,i_3,i_4\}} \right) \left( t_{\{i_3,i_4\}}-t_{\{i_1,i_3,i_4\}}-1\right)^+ \left (t_{\{i_1,i_4\}}-1 \right)^+ 
\nonumber \\ 
&+ \left (t_{\{i_1,i_2,i_4\}}-t_{\{i_1,i_2,i_3,i_4\}} \right) \left(t_{\{i_2,i_3\}}-t_{\{i_2,i_3,i_4\}} \right) t_{\{i_1,i_3,i_4\}} \left( t_{\{i_1,i_4\}}-2 \right)^+ 
\nonumber \\ 
&+ \left (t_{\{i_1,i_2,i_4\}}-t_{\{i_1,i_2,i_3,i_4\}} \right) \left(t_{\{i_2,i_3\}}-t_{\{i_2,i_3,i_4\}} \right) \left( t_{\{i_3,i_4\}}-t_{\{i_1,i_3,i_4\}} \right) \left (t_{\{i_1,i_4\}}-1 \right)^+, 
\nonumber
\end{align}
\begin{align}\label{eq_p9_2}
\mathcal{A}_{P_9,4} &= \left (t_{\{i_1,i_2\}}-t_{\{i_1,i_2,i_3\}}-t_{\{i_1,i_2,i_4\}}+t_{\{i_1,i_2,i_3,i_4\}} \right) t_{\{i_1,i_2,i_3,i_4\}} \left( t_{\{i_1,i_3,i_4\}}-1 \right)^+ \left( t_{\{i_1,i_4\}}-2 \right)^+ 
\nonumber \\ 
&+ \left (t_{\{i_1,i_2\}}-t_{\{i_1,i_2,i_3\}}-t_{\{i_1,i_2,i_4\}}+t_{\{i_1,i_2,i_3,i_4\}} \right) t_{\{i_1,i_2,i_3,i_4\}} \left( t_{\{i_3,i_4\}}-t_{\{i_1,i_3,i_4\}} \right) \left( t_{\{i_1,i_4\}}-1 \right)^+ 
\nonumber \\ 
&+ \left (t_{\{i_1,i_2\}}-t_{\{i_1,i_2,i_3\}}-t_{\{i_1,i_2,i_4\}}+t_{\{i_1,i_2,i_3,i_4\}} \right) \left(t_{\{i_2,i_3,i_4\}}-t_{\{i_1,i_2,i_3,i_4\}} \right) t_{\{i_1,i_3,i_4\}} \left (t_{\{i_1,i_4\}}-1 \right)^+ 
\nonumber \\ 
&+ \left (t_{\{i_1,i_2\}}-t_{\{i_1,i_2,i_3\}}-t_{\{i_1,i_2,i_4\}}+t_{\{i_1,i_2,i_3,i_4\}} \right) \left(t_{\{i_2,i_3,i_4\}}-t_{\{i_1,i_2,i_3,i_4\}} \right) \left( t_{\{i_3,i_4\}}-t_{\{i_1,i_3,i_4\}}-1\right)^+ t_{\{i_1,i_4\}} 
\nonumber \\ 
&+ \left (t_{\{i_1,i_2\}}-t_{\{i_1,i_2,i_3\}}-t_{\{i_1,i_2,i_4\}}+t_{\{i_1,i_2,i_3,i_4\}} \right) \left(t_{\{i_2,i_3\}}-t_{\{i_2,i_3,i_4\}} \right) t_{\{i_1,i_3,i_4\}} \left( t_{\{i_1,i_4\}}-1 \right)^+ 
\nonumber \\ 
&+ \left (t_{\{i_1,i_2\}}-t_{\{i_1,i_2,i_3\}}-t_{\{i_1,i_2,i_4\}}+t_{\{i_1,i_2,i_3,i_4\}} \right) \left(t_{\{i_2,i_3\}}-t_{\{i_2,i_3,i_4\}} \right) \left( t_{\{i_3,i_4\}}-t_{\{i_1,i_3,i_4\}} \right) t_{\{i_1,i_4\}}.
\end{align}
Case~9.2: The number of instances of $P_9$ with CNs $c_1$, $c_2$, $c_3$, and $c_4$ as defined in the previous paragraph, and all overlaps in two replicas s.t. three overlaps are in $\bold{R}_r$, and the $c_1-c_4$ overlap is in $\bold{R}_e$, is:
\begin{align}\label{eq_p9_3}
\mathcal{B}_{P_9} &\big (t_{\{i_1,i_2\}}, t_{\{i_2,i_3\}}, t_{\{i_3,i_4\}}, t_{\{i_1,i_2,i_3\}}, t_{\{i_2,i_3,i_4\}}, t_{\{i_1,i_2,i_3,i_4\}}, t_{\{i_1+(r-e)\gamma,i_4+(r-e)\gamma\}} \big ) 
\nonumber \\ 
&= \Big[ t_{\{i_1,i_2,i_3,i_4\}} \left( t_{\{i_2,i_3,i_4\}}-1 \right)^+ \left( t_{\{i_3,i_4\}}-2 \right)^+ 
\nonumber \\ 
& \hspace{1.1em}+ t_{\{i_1,i_2,i_3,i_4\}} \left( t_{\{i_2,i_3\}}-t_{\{i_2,i_3,i_4\}} \right) \left( t_{\{i_3,i_4\}}-1 \right)^+ 
\nonumber \\ 
& \hspace{1.1em}+ \left (t_{\{i_1,i_2,i_3\}}-t_{\{i_1,i_2,i_3,i_4\}} \right) t_{\{i_2,i_3,i_4\}} \left( t_{\{i_3,i_4\}}-1 \right)^+ 
\nonumber \\ 
& \hspace{1.1em}+ \left (t_{\{i_1,i_2,i_3\}}-t_{\{i_1,i_2,i_3,i_4\}} \right) \left( t_{\{i_2,i_3\}}-t_{\{i_2,i_3,i_4\}}-1\right)^+ t_{\{i_3,i_4\}} 
\nonumber \\ 
& \hspace{1.1em}+ \left (t_{\{i_1,i_2\}}-t_{\{i_1,i_2,i_3\}} \right) t_{\{i_2,i_3,i_4\}} \left ( t_{\{i_3,i_4\}} -1 \right)^+ 
\nonumber \\ 
& \hspace{1.1em}+ \left (t_{\{i_1,i_2\}}-t_{\{i_1,i_2,i_3\}} \right) \left(t_{\{i_2,i_3\}}-t_{\{i_2,i_3,i_4\}} \right) t_{\{i_3,i_4\}} \Big] t_{\{i_1+(r-e)\gamma,i_4+(r-e)\gamma\}}.
\end{align}
Case~9.3: The number of instances of $P_9$ with CNs $c_1$, $c_2$, $c_3$, and $c_4$ as defined in the previous paragraph, and all overlaps in two replicas s.t. $c_1-c_2$ and $c_2-c_3$ overlaps are in $\bold{R}_r$, and $c_3-c_4$ and $c_1-c_4$ overlaps are in $\bold{R}_e$, $r<e$, is:
\begin{align}\label{eq_p9_4}
\mathcal{C}_{P_9} &\big(t_{\{i_1,i_2\}}, t_{\{i_2,i_3\}}, t_{\{i_1,i_2,i_3\}}, t_{\{i_3+(r-e)\gamma,i_4+(r-e)\gamma\}}, t_{\{i_1+(r-e)\gamma,i_4+(r-e)\gamma\}}, t_{\{i_1+(r-e)\gamma,i_3+(r-e)\gamma,i_4+(r-e)\gamma\}} \big) 
\nonumber \\ 
&= \Big [ t_{\{i_1,i_2,i_3\}} \left( t_{\{i_2,i_3\}}-1 \right)^+ + \left ( t_{\{i_1,i_2\}} - t_{\{i_1,i_2,i_3\}} \right ) t_{\{i_2,i_3\}} \Big ] 
\nonumber \\ 
& \hspace{0.6em} \cdot \Big [  t_{\{i_1+(r-e)\gamma,i_3+(r-e)\gamma,i_4+(r-e)\gamma\}} \left( t_{\{i_1+(r-e)\gamma,i_4+(r-e)\gamma\}}-1 \right)^+ 
\nonumber \\ 
& \hspace{1.2em} + \left ( t_{\{i_3+(r-e)\gamma,i_4+(r-e)\gamma\}} - t_{\{i_1+(r-e)\gamma,i_3+(r-e)\gamma,i_4+(r-e)\gamma\}} \right ) t_{\{i_1+(r-e)\gamma,i_4+(r-e)\gamma\}} \Big ].
\end{align}
Case~9.4: The number of instances of $P_9$ with CNs $c_1$, $c_2$, $c_3$, and $c_4$ as defined in the previous paragraph, and all overlaps in two replicas s.t. $c_1-c_2$ and $c_3-c_4$ overlaps are in $\bold{R}_r$, and $c_2-c_3$ and $c_1-c_4$ overlaps are in $\bold{R}_e$, $r<e$, is:
\vspace{-0.1em}\begin{align}\label{eq_p9_5}
\mathcal{D}_{P_9} &\big(t_{\{i_1,i_2\}}, t_{\{i_3,i_4\}}, t_{\{i_1,i_2,i_3,i_4\}}, t_{\{i_2+(r-e)\gamma,i_3+(r-e)\gamma\}}, t_{\{i_1+(r-e)\gamma,i_4+(r-e)\gamma\}}, t_{\{i_1+(r-e)\gamma,i_2+(r-e)\gamma,i_3+(r-e)\gamma,i_4+(r-e)\gamma\}} \big) 
\nonumber \\ 
&= \Big [ t_{\{i_1,i_2,i_3,i_4\}} \left( t_{\{i_3,i_4\}}-1 \right)^+ + \left ( t_{\{i_1,i_2\}} - t_{\{i_1,i_2,i_3,i_4\}} \right ) t_{\{i_3,i_4\}} \Big ] 
\nonumber \\ 
& \hspace{0.6em} \cdot \Big [  t_{\{i_1+(r-e)\gamma,i_2+(r-e)\gamma,i_3+(r-e)\gamma,i_4+(r-e)\gamma\}} \left( t_{\{i_1+(r-e)\gamma,i_4+(r-e)\gamma\}}-1 \right)^+ 
\nonumber \\ 
& \hspace{1.2em} + \left ( t_{\{i_2+(r-e)\gamma,i_3+(r-e)\gamma\}} - t_{\{i_1+(r-e)\gamma,i_2+(r-e)\gamma,i_3+(r-e)\gamma,i_4+(r-e)\gamma\}} \right ) t_{\{i_1+(r-e)\gamma,i_4+(r-e)\gamma\}} \Big ].
\end{align}
Case~9.5: The number of instances of $P_9$ with CNs $c_1$, $c_2$, $c_3$, and $c_4$ as defined previously, and all overlaps in three replicas s.t. $c_1-c_2$ and $c_2-c_3$ overlaps are in $\bold{R}_r$, the $c_3-c_4$ overlap is in $\bold{R}_e$, and the $c_1-c_4$ overlap is in $\bold{R}_s$, $e < s$, is:
\begin{align}\label{eq_p9_6}
\mathcal{E}_{P_9} & \big(t_{\{i_1,i_2\}}, t_{\{i_2,i_3\}}, t_{\{i_1,i_2,i_3\}}, t_{\{i_3+(r-e)\gamma,i_4+(r-e)\gamma\}}, t_{\{i_1+(r-s)\gamma,i_4+(r-s)\gamma\}} \big) 
\nonumber \\ 
&= \Big [ t_{\{i_1,i_2,i_3\}} \left( t_{\{i_2,i_3\}}-1 \right)^+ + \left ( t_{\{i_1,i_2\}} - t_{\{i_1,i_2,i_3\}} \right ) t_{\{i_2,i_3\}} \Big ] t_{\{i_3+(r-e)\gamma,i_4+(r-e)\gamma\}} t_{\{i_1+(r-s)\gamma,i_4+(r-s)\gamma\}}.
\end{align}
Case~9.6: The number of instances of $P_9$ with CNs $c_1$, $c_2$, $c_3$, and $c_4$ as defined previously, and all overlaps in three replicas s.t. $c_1-c_2$ and $c_3-c_4$ overlaps are in $\bold{R}_r$, the $c_2-c_3$ overlap is in $\bold{R}_e$, and the $c_1-c_4$ overlap is in $\bold{R}_s$, $e < s$, is:
\begin{align}\label{eq_p9_7}
\mathcal{G}_{P_9} & \big(t_{\{i_1,i_2\}}, t_{\{i_3,i_4\}}, t_{\{i_1,i_2,i_3,i_4\}}, t_{\{i_2+(r-e)\gamma,i_3+(r-e)\gamma\}}, t_{\{i_1+(r-s)\gamma,i_4+(r-s)\gamma\}} \big) 
\nonumber \\ 
&= \Big [ t_{\{i_1,i_2,i_3,i_4\}} \left( t_{\{i_3,i_4\}}-1 \right)^+ + \left ( t_{\{i_1,i_2\}} - t_{\{i_1,i_2,i_3,i_4\}} \right ) t_{\{i_3,i_4\}} \Big ] t_{\{i_2+(r-e)\gamma,i_3+(r-e)\gamma\}} t_{\{i_1+(r-s)\gamma,i_4+(r-s)\gamma\}}.
\end{align}
Case~9.7: The number of instances of $P_9$ with CNs $c_1$, $c_2$, $c_3$, and $c_4$ as defined previously, and overlaps in four replicas s.t. the $c_1-c_2$ overlap is in $\bold{R}_r$, the $c_2-c_3$ overlap is in $\bold{R}_e$, the $c_3-c_4$ overlap is in $\bold{R}_s$, and the $c_1-c_4$ overlap is in $\bold{R}_u$, $r < e$, $e < u$, and $r < s$, is:
\begin{align}\label{eq_p9_8}
\mathcal{I}_{P_9} & \big(t_{\{i_1,i_2\}}, t_{\{i_2+(r-e)\gamma,i_3+(r-e)\gamma\}}, t_{\{i_3+(r-s)\gamma,i_4+(r-s)\gamma\}}, t_{\{i_1+(r-u)\gamma,i_4+(r-u)\gamma\}} \big) 
\nonumber \\ 
&= t_{\{i_1,i_2\}}t_{\{i_2+(r-e)\gamma,i_3+(r-e)\gamma\}}t_{\{i_3+(r-s)\gamma,i_4+(r-s)\gamma\}}t_{\{i_1+(r-u)\gamma,i_4+(r-u)\gamma\}}.
\end{align}
\end{lemma}

\begin{figure}[H]
\vspace{-1.5em}
\center
\includegraphics[width=7.0in]{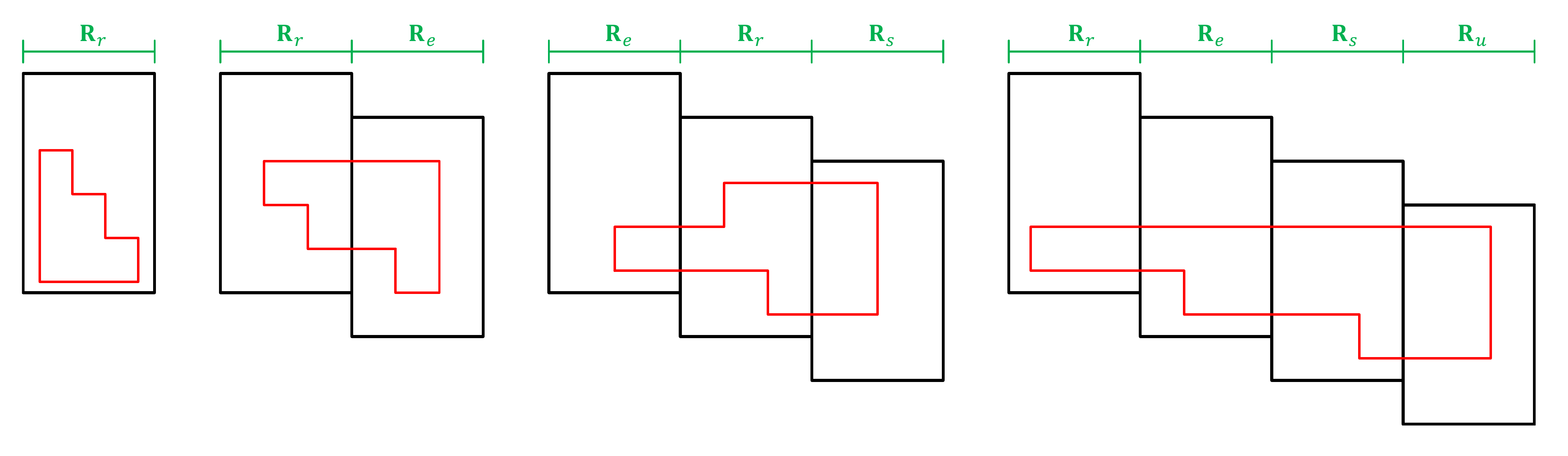}\vspace{-1.5em}
\caption{An instance of Pattern $P_9$ in Case~9.1, in Case~9.3, in Case~9.6, and in Case~9.7, from left to right. For simplicity, we have $e=r+y_1$, where $y_1 \in \{-1,1\}$, $s=e+y_2$, where $y_2 \in \{1,2\}$, and $u=s+1$.}
\label{Fig_pat9}
\vspace{-0.5em}
\end{figure}

Four of the seven cases are illustrated in Fig.~\ref{Fig_pat9}.

\begin{theorem}\label{thm_p9}
The total number of instances of Pattern $P_9$ in the binary protograph of an SC code that has parameters $\gamma \geq 3$, $\kappa$, $m$, $L \geq 2m+1$, and $\mathcal{O}$, is:
\begin{equation}\label{eq_p9_9}
F_{P_9} = \sum_{k=1}^{2m+1} (L-k+1) F^k_{P_9,1},
\end{equation}
where $F^k_{P_9,1}$, $k \in \{1, 2, \dots, 2m+1\}$, are given by:
\begin{align}\label{eq_p9_10}
F^1_{P_9,1} &= \hspace{0.3em} \frac{1}{2} \hspace{-16.2em} \sum_{\hspace{16.3em} \{i_1,i_3\} \subset \{0, \dots, (m+1)\gamma -1\}, \{i_2,i_4\} \subset \{0, \dots, (m+1)\gamma -1\}} \hspace{-16.1em} \mathcal{A}_{P_9}\big (t_{\{i_1,i_2\}}, t_{\{i_2,i_3\}}, t_{\{i_3,i_4\}}, t_{\{i_1,i_4\}}, t_{\{i_1,i_2,i_3\}}, t_{\{i_1,i_2,i_4\}}, t_{\{i_1,i_3,i_4\}}, t_{\{i_2,i_3,i_4\}}, t_{\{i_1,i_2,i_3,i_4\}} \big ), 
\nonumber \\
F^2_{P_9,1} &= \hspace{-22.0em} \sum_{\hspace{22.2em} \{i_1,i_4\} \subset \{\gamma, \dots, (m+1)\gamma -1\}, i_2 \in \{0, \dots, (m+1)\gamma -1\}, i_3 \in \{0, \dots, (m+1)\gamma -1\}} \hspace{-22.0em} \mathcal{B}_{P_9} \big (t_{\{i_1,i_2\}}, t_{\{i_2,i_3\}}, t_{\{i_3,i_4\}}, t_{\{i_1,i_2,i_3\}}, t_{\{i_2,i_3,i_4\}}, t_{\{i_1,i_2,i_3,i_4\}}, t_{\{i_1-\gamma,i_4-\gamma\}} \big ) 
\nonumber \\ 
&+  \hspace{-20.3em} \sum_{\hspace{20.8em} \{i_1,i_4\} \subset \{0, \dots, m\gamma -1\}, i_2 \in \{0, \dots, (m+1)\gamma -1\}, i_3 \in \{0, \dots, (m+1)\gamma -1\}} \hspace{-20.4em} \mathcal{B}_{P_9} \big (t_{\{i_1,i_2\}}, t_{\{i_2,i_3\}}, t_{\{i_3,i_4\}}, t_{\{i_1,i_2,i_3\}}, t_{\{i_2,i_3,i_4\}}, t_{\{i_1,i_2,i_3,i_4\}}, t_{\{i_1+\gamma,i_4+\gamma\}} \big ) 
\nonumber \\ 
&+ \hspace{-21.9em} \sum_{\hspace{22.3em} \{i_1,i_3\} \subset \{\gamma, \dots, (m+1)\gamma -1\}, i_2 \in \{0, \dots, (m+1)\gamma -1\}, i_4 \in \{\gamma, \dots, (m+2)\gamma -1\}} \hspace{-22.0em} \mathcal{C}_{P_9} \big (t_{\{i_1,i_2\}}, t_{\{i_2,i_3\}}, t_{\{i_1,i_2,i_3\}}, t_{\{i_3-\gamma,i_4-\gamma\}}, t_{\{i_1-\gamma,i_4-\gamma\}}, t_{\{i_1-\gamma,i_3-\gamma,i_4-\gamma\}} \big ) 
\nonumber \\ 
&+ \hspace{0.3em} \frac{1}{2} \hspace{-22.0em} \sum_{\hspace{22.2em} \{i_1,i_4\} \subset \{\gamma, \dots, (m+1)\gamma -1\}, i_2 \in \{\gamma, \dots, (m+1)\gamma -1\}, i_3 \in \{\gamma, \dots, (m+1)\gamma -1\}} \hspace{-22.1em} \mathcal{D}_{P_9} \big (t_{\{i_1,i_2\}}, t_{\{i_3,i_4\}}, t_{\{i_1,i_2,i_3,i_4\}}, t_{\{i_2-\gamma,i_3-\gamma\}}, t_{\{i_1-\gamma,i_4-\gamma\}}, t_{\{i_1-\gamma,i_2-\gamma,i_3-\gamma,i_4-\gamma\}} \big ), 
\nonumber \\
F^3_{P_9,1} &= \hspace{-22.6em} \sum_{\hspace{22.8em} \{i_1,i_4\} \subset \{2\gamma, \dots, (m+1)\gamma -1\}, i_2 \in \{0, \dots, (m+1)\gamma -1\}, i_3 \in \{0, \dots, (m+1)\gamma -1\}} \hspace{-22.3em} \mathcal{B}_{P_9} \big (t_{\{i_1,i_2\}}, t_{\{i_2,i_3\}}, t_{\{i_3,i_4\}}, t_{\{i_1,i_2,i_3\}}, t_{\{i_2,i_3,i_4\}}, t_{\{i_1,i_2,i_3,i_4\}}, t_{\{i_1-2\gamma,i_4-2\gamma\}} \big ) 
\nonumber \\ 
&+  \hspace{-22.0em} \sum_{\hspace{22.3em} \{i_1,i_4\} \subset \{0, \dots, (m-1)\gamma -1\}, i_2 \in \{0, \dots, (m+1)\gamma -1\}, i_3 \in \{0, \dots, (m+1)\gamma -1\}} \hspace{-21.9em} \mathcal{B}_{P_9} \big (t_{\{i_1,i_2\}}, t_{\{i_2,i_3\}}, t_{\{i_3,i_4\}}, t_{\{i_1,i_2,i_3\}}, t_{\{i_2,i_3,i_4\}}, t_{\{i_1,i_2,i_3,i_4\}}, t_{\{i_1+2\gamma,i_4+2\gamma\}} \big ) 
\nonumber \\ 
&+ \hspace{-22.9em} \sum_{\hspace{23.2em} \{i_1,i_3\} \subset \{2\gamma, \dots, (m+1)\gamma -1\}, i_2 \in \{0, \dots, (m+1)\gamma -1\}, i_4 \in \{2\gamma, \dots, (m+3)\gamma -1\}} \hspace{-22.8em} \mathcal{C}_{P_9} \big (t_{\{i_1,i_2\}}, t_{\{i_2,i_3\}}, t_{\{i_1,i_2,i_3\}}, t_{\{i_3-2\gamma,i_4-2\gamma\}}, t_{\{i_1-2\gamma,i_4-2\gamma\}}, t_{\{i_1-2\gamma,i_3-2\gamma,i_4-2\gamma\}} \big ) 
\nonumber \\ 
&+ \hspace{0.3em} \frac{1}{2} \hspace{-23.4em} \sum_{\hspace{23.7em} \{i_1,i_4\} \subset \{2\gamma, \dots, (m+1)\gamma -1\}, i_2 \in \{2\gamma, \dots, (m+1)\gamma -1\}, i_3 \in \{2\gamma, \dots, (m+1)\gamma -1\}} \hspace{-23.3em} \mathcal{D}_{P_9} \big (t_{\{i_1,i_2\}}, t_{\{i_3,i_4\}}, t_{\{i_1,i_2,i_3,i_4\}}, t_{\{i_2-2\gamma,i_3-2\gamma\}}, t_{\{i_1-2\gamma,i_4-2\gamma\}}, t_{\{i_1-2\gamma,i_2-2\gamma,i_3-2\gamma,i_4-2\gamma\}} \big )
\nonumber \\
&+ \hspace{-29.0em} \sum_{\hspace{29.3em} i_1 \in \{2\gamma, \dots, (m+1)\gamma -1\}, i_2 \in \{0, \dots, (m+1)\gamma -1\}, i_3 \in \{\gamma, \dots, (m+1)\gamma -1\}, i_4 \in \{2\gamma, \dots, (m+2)\gamma -1\}} \hspace{-28.8em} \mathcal{E}_{P_9} \big (t_{\{i_1,i_2\}}, t_{\{i_2,i_3\}}, t_{\{i_1,i_2,i_3\}}, t_{\{i_3-\gamma,i_4-\gamma\}}, t_{\{i_1-2\gamma,i_4-2\gamma\}} \big )
\nonumber \\
&+ \hspace{-24.9em} \sum_{\hspace{25.3em} i_1 \in \{\gamma, \dots, (m+1)\gamma -1\}, i_2 \in \{0, \dots, (m+1)\gamma -1\}, i_3 \in \{0, \dots, m\gamma -1\}, i_4 \in \{\gamma, \dots, m\gamma -1\}} \hspace{-24.8em} \mathcal{E}_{P_9} \big (t_{\{i_1,i_2\}}, t_{\{i_2,i_3\}}, t_{\{i_1,i_2,i_3\}}, t_{\{i_3+\gamma,i_4+\gamma\}},  t_{\{i_1-\gamma,i_4-\gamma\}} \big )
\nonumber \\
&+ \hspace{-27.0em} \sum_{\hspace{27.3em} i_1 \in \{0, \dots, m\gamma -1\}, i_2 \in \{0, \dots, (m+1)\gamma -1\}, i_3 \in \{0, \dots, (m-1)\gamma -1\}, i_4 \in \{-\gamma, \dots, (m-1)\gamma -1\}} \hspace{-26.9em} \mathcal{E}_{P_9} \big (t_{\{i_1,i_2\}}, t_{\{i_2,i_3\}}, t_{\{i_1,i_2,i_3\}}, t_{\{i_3+2\gamma,i_4+2\gamma\}}, t_{\{i_1+\gamma,i_4+\gamma\}} \big )
\nonumber \\
&+ \hspace{-22.7em} \sum_{\hspace{23.0em} \{i_1,i_4\} \subset \{2\gamma, \dots, (m+1)\gamma -1\}, i_2 \in \{\gamma, \dots, (m+1)\gamma -1\}, i_3 \in \{\gamma, \dots, (m+1)\gamma -1\}} \hspace{-22.5em} \mathcal{G}_{P_9} \big (t_{\{i_1,i_2\}}, t_{\{i_3,i_4\}}, t_{\{i_1,i_2,i_3,i_4\}}, t_{\{i_2-\gamma,i_3-\gamma\}}, t_{\{i_1-2\gamma,i_4-2\gamma\}} \big )
\nonumber \\
&+ \hspace{-18.8em} \sum_{\hspace{19.0em} \{i_1,i_4\} \subset \{\gamma, \dots, (m+1)\gamma -1\}, i_2 \in \{0, \dots, m\gamma -1\}, i_3 \in \{0, \dots, m\gamma -1\}} \hspace{-18.5em} \mathcal{G}_{P_9} \big (t_{\{i_1,i_2\}}, t_{\{i_3,i_4\}}, t_{\{i_1,i_2,i_3,i_4\}}, t_{\{i_2+\gamma,i_3+\gamma\}}, t_{\{i_1-\gamma,i_4-\gamma\}} \big )
\nonumber \\
&+ \hspace{-20.3em} \sum_{\hspace{20.5em} \{i_1,i_4\} \subset \{0, \dots, m\gamma -1\}, i_2 \in \{0, \dots, (m-1)\gamma -1\}, i_3 \in \{0, \dots, (m-1)\gamma -1\}} \hspace{-20.1em} \mathcal{G}_{P_9} \big (t_{\{i_1,i_2\}}, t_{\{i_3,i_4\}}, t_{\{i_1,i_2,i_3,i_4\}}, t_{\{i_2+2\gamma,i_3+2\gamma\}}, t_{\{i_1+\gamma,i_4+\gamma\}} \big ),
\nonumber \\
F^{k \geq 4}_{P_9,1} &= \hspace{-24.2em} \sum_{\hspace{24.2em} \{i_1,i_4\} \subset \{(k-1)\gamma, \dots, (m+1)\gamma -1\}, i_2 \in \{0, \dots, (m+1)\gamma -1\}, i_3 \in \{0, \dots, (m+1)\gamma -1\}} \hspace{-23.9em} \mathcal{B}_{P_9} \big (t_{\{i_1,i_2\}}, t_{\{i_2,i_3\}}, t_{\{i_3,i_4\}}, t_{\{i_1,i_2,i_3\}}, t_{\{i_2,i_3,i_4\}}, t_{\{i_1,i_2,i_3,i_4\}}, t_{\{i_1+(1-k)\gamma,i_4+(1-k)\gamma\}} \big ) 
\nonumber \\ 
&+  \hspace{-23.1em} \sum_{\hspace{23.2em} \{i_1,i_4\} \subset \{0, \dots, (m-k+2)\gamma -1\}, i_2 \in \{0, \dots, (m+1)\gamma -1\}, i_3 \in \{0, \dots, (m+1)\gamma -1\}} \hspace{-22.8em} \mathcal{B}_{P_9} \big (t_{\{i_1,i_2\}}, t_{\{i_2,i_3\}}, t_{\{i_3,i_4\}}, t_{\{i_1,i_2,i_3\}}, t_{\{i_2,i_3,i_4\}}, t_{\{i_1,i_2,i_3,i_4\}}, t_{\{i_1+(k-1)\gamma,i_4+(k-1)\gamma\}} \big ) 
\nonumber \\ 
&+ \hspace{-26.4em} \sum_{\hspace{26.5em} \{i_1,i_3\} \subset \{(k-1)\gamma, \dots, (m+1)\gamma -1\}, i_2 \in \{0, \dots, (m+1)\gamma -1\}, i_4 \in \{(k-1)\gamma, \dots, (m+k)\gamma -1\}} \hspace{-26.1em} \mathcal{C}_{P_9} \big (t_{\{i_1,i_2\}}, t_{\{i_2,i_3\}}, t_{\{i_1,i_2,i_3\}}, t_{\{i_3+(1-k)\gamma,i_4+(1-k)\gamma\}}, t_{\{i_1+(1-k)\gamma,i_4+(1-k)\gamma\}}, t_{\{i_1+(1-k)\gamma,i_3+(1-k)\gamma,i_4+(1-k)\gamma\}} \big ) 
\nonumber \\ 
&+ \hspace{0.3em} \frac{1}{2} \hspace{-28.2em} \sum_{\hspace{28.5em} \{i_1,i_4\} \subset \{(k-1)\gamma, \dots, (m+1)\gamma -1\}, i_2 \in \{(k-1)\gamma, \dots, (m+1)\gamma -1\}, i_3 \in \{(k-1)\gamma, \dots, (m+1)\gamma -1\}} \hspace{-28.3em} \mathcal{D}_{P_9} \big (t_{\{i_1,i_2\}}, t_{\{i_3,i_4\}}, t_{\{i_1,i_2,i_3,i_4\}}, t_{\{i_2+(1-k)\gamma,i_3+(1-k)\gamma\}}, t_{\{i_1+(1-k)\gamma,i_4+(1-k)\gamma\}}
\nonumber \\
&\hspace{7.0em}, t_{\{i_1+(1-k)\gamma,i_2+(1-k)\gamma,i_3+(1-k)\gamma,i_4+(1-k)\gamma\}} \big )
\nonumber \\
&+ \hspace{0.5em} \sum_{h=2}^{k-1} \hspace{-34.5em} \sum_{\hspace{35.0em} i_1 \in \{(k-1)\gamma, \dots, (m+1)\gamma -1\}, i_2 \in \{0, \dots, (m+1)\gamma -1\}, i_3 \in \{(h-1)\gamma, \dots, (m+1)\gamma -1\}, i_4 \in \{(k-1)\gamma, \dots, (m+h)\gamma -1\}} \hspace{-34.5em} \mathcal{E}_{P_9} \big (t_{\{i_1,i_2\}}, t_{\{i_2,i_3\}}, t_{\{i_1,i_2,i_3\}}, t_{\{i_3+(1-h)\gamma,i_4+(1-h)\gamma\}}, t_{\{i_1+(1-k)\gamma,i_4+(1-k)\gamma\}} \big )
\nonumber \\
&+ \hspace{0.5em} \sum_{h=2}^{k-1} \hspace{-34.5em} \sum_{\hspace{35.0em} i_1 \in \{(k-h)\gamma, \dots, (m+1)\gamma -1\}, i_2 \in \{0, \dots, (m+1)\gamma -1\}, i_3 \in \{0, \dots, (m-h+2)\gamma -1\}, i_4 \in \{(k-h)\gamma, \dots, (m-h+2)\gamma -1\}} \hspace{-34.5em} \mathcal{E}_{P_9} \big (t_{\{i_1,i_2\}}, t_{\{i_2,i_3\}}, t_{\{i_1,i_2,i_3\}}, t_{\{i_3+(h-1)\gamma,i_4+(h-1)\gamma\}}, t_{\{i_1+(h-k)\gamma,i_4+(h-k)\gamma\}} \big )
\nonumber \\
&+ \hspace{0.5em} \sum_{h=2}^{k-1} \hspace{-34.5em} \sum_{\hspace{35.0em} i_1 \in \{0, \dots, (m-k+h+1)\gamma -1\}, i_2 \in \{0, \dots, (m+1)\gamma -1\}, i_3 \in \{0, \dots, (m-k+2)\gamma -1\}, i_4 \in \{(h-k)\gamma, \dots, (m-k+2)\gamma -1\}} \hspace{-34.5em} \mathcal{E}_{P_9} \big (t_{\{i_1,i_2\}}, t_{\{i_2,i_3\}}, t_{\{i_1,i_2,i_3\}}, t_{\{i_3+(k-1)\gamma,i_4+(k-1)\gamma\}}, t_{\{i_1+(k-h)\gamma,i_4+(k-h)\gamma\}} \big )
\nonumber \\
&+ \hspace{0.5em} \sum_{h=2}^{k-1} \hspace{-28.4em} \sum_{\hspace{28.9em} \{i_1,i_4\} \subset \{(k-1)\gamma, \dots, (m+1)\gamma -1\}, i_2 \in \{(h-1)\gamma, \dots, (m+1)\gamma -1\}, i_3 \in \{(h-1)\gamma, \dots, (m+1)\gamma -1\}} \hspace{-28.4em} \mathcal{G}_{P_9} \big (t_{\{i_1,i_2\}}, t_{\{i_3,i_4\}}, t_{\{i_1,i_2,i_3,i_4\}}, t_{\{i_2+(1-h)\gamma,i_3+(1-h)\gamma\}}, t_{\{i_1+(1-k)\gamma,i_4+(1-k)\gamma\}} \big ) 
\nonumber \\
&+ \hspace{0.5em} \sum_{h=2}^{k-1} \hspace{-26.3em} \sum_{\hspace{26.8em} \{i_1,i_4\} \subset \{(k-h)\gamma, \dots, (m+1)\gamma -1\}, i_2 \in \{0, \dots, (m-h+2)\gamma -1\}, i_3 \in \{0, \dots, (m-h+2)\gamma -1\}} \hspace{-26.3em} \mathcal{G}_{P_9} \big (t_{\{i_1,i_2\}}, t_{\{i_3,i_4\}}, t_{\{i_1,i_2,i_3,i_4\}}, t_{\{i_2+(h-1)\gamma,i_3+(h-1)\gamma\}}, t_{\{i_1+(h-k)\gamma,i_4+(h-k)\gamma\}} \big )
\nonumber \\
&+ \hspace{0.5em} \sum_{h=2}^{k-1} \hspace{-26.3em} \sum_{\hspace{26.8em} \{i_1,i_4\} \subset \{0, \dots, (m-k+h+1)\gamma -1\}, i_2 \in \{0, \dots, (m-k+2)\gamma -1\}, i_3 \in \{0, \dots, (m-k+2)\gamma -1\}} \hspace{-26.2em} \mathcal{G}_{P_9} \big (t_{\{i_1,i_2\}}, t_{\{i_3,i_4\}}, t_{\{i_1,i_2,i_3,i_4\}}, t_{\{i_2+(k-1)\gamma,i_3+(k-1)\gamma\}}, t_{\{i_1+(k-h)\gamma,i_4+(k-h)\gamma\}} \big )
\nonumber \\ 
&+ \hspace{0.5em} \sum_{h=2}^{k-2} \hspace{0.3em} \sum_{w=h+1}^{k-1} \hspace{-37.0em} \sum_{\hspace{37.5em}  i_1 \in \{(k-1)\gamma, \dots, (m+1)\gamma -1\}, i_2 \in \{(h-1)\gamma, \dots, (m+1)\gamma -1\}, i_3 \in \{(w-1)\gamma, \dots, (m+h)\gamma -1\}, i_4 \in \{(k-1)\gamma, \dots, (m+w)\gamma -1\}} \hspace{-37.0em} \mathcal{I}_{P_9} \big( t_{\{i_1,i_2\}}, t_{\{i_2+(1-h)\gamma,i_3+(1-h)\gamma\}}, t_{\{i_3+(1-w)\gamma,i_4+(1-w)\gamma\}}, t_{\{i_1+(1-k)\gamma,i_4+(1-k)\gamma\}} \big) 
\nonumber \\ 
&+ \hspace{0.5em} \sum_{h=2}^{k-2} \hspace{0.3em} \sum_{w=h+1}^{k-1} \hspace{-37.0em} \sum_{\hspace{37.5em}  i_1 \in \{(k-1)\gamma, \dots, (m+1)\gamma -1\}, i_2 \in \{(w-1)\gamma, \dots, (m+1)\gamma -1\}, i_3 \in \{(w-1)\gamma, \dots, (m+h)\gamma -1\}, i_4 \in \{(k-1)\gamma, \dots, (m+h)\gamma -1\}} \hspace{-37.0em} \mathcal{I}_{P_9} \big( t_{\{i_1,i_2\}}, t_{\{i_2+(1-w)\gamma,i_3+(1-w)\gamma\}}, t_{\{i_3+(1-h)\gamma,i_4+(1-h)\gamma\}}, t_{\{i_1+(1-k)\gamma,i_4+(1-k)\gamma\}} \big) 
\nonumber \\ 
&+ \hspace{0.5em} \sum_{h=2}^{k-2} \hspace{0.3em} \sum_{w=h+1}^{k-1} \hspace{-37.0em} \sum_{\hspace{37.5em}  i_1 \in \{(w-1)\gamma, \dots, (m+1)\gamma -1\}, i_2 \in \{(h-1)\gamma, \dots, (m+1)\gamma -1\}, i_3 \in \{(k-1)\gamma, \dots, (m+h)\gamma -1\}, i_4 \in \{(k-1)\gamma, \dots, (m+w)\gamma -1\}} \hspace{-37.0em} \mathcal{I}_{P_9} \big( t_{\{i_1,i_2\}}, t_{\{i_2+(1-h)\gamma,i_3+(1-h)\gamma\}}, t_{\{i_3+(1-k)\gamma,i_4+(1-k)\gamma\}}, t_{\{i_1+(1-w)\gamma,i_4+(1-w)\gamma\}} \big),
\end{align}
with $\overline{i_1} \neq \overline{i_2}$, $i_1 \neq i_3$, $\overline{i_1} \neq \overline{i_4}$, $\overline{i_2} \neq \overline{i_3}$, $i_2 \neq i_4$, and $\overline{i_3} \neq \overline{i_4}$.
\end{theorem}

After deriving the expressions of $F_{P_\ell}$, $\forall \ell$, as functions of the overlap parameters in $\mathcal{O}$, we use (\ref{eq_ftot}), (\ref{eq_oind}), and \cite[Lemma~3]{homa_boo} to express $F_{\textup{sum}}$ as a function of the parameters in $\mathcal{O}_{\textup{ind}}$ (which is the set of all independent non-zero overlap parameters). Thus, our \textit{\textbf{discrete optimization problem}} is:
\begin{equation}\label{eq_opt_pb}
F^*_{\textup{sum}} = \min_{\mathcal{O}_{\textup{ind}}} F_{\textup{sum}}.
\end{equation}

The constraints of the optimization problem in (\ref{eq_opt_pb}) are linear constraints capturing the interval constraints under which the resultant partitioning is valid. We also add the balanced partitioning constraint, which guarantees a balanced distribution of the non-zero circulants among the $(m+1)$ component matrices. (see also \cite{homa_boo} and \cite{ahh_nboo}). A balanced partitioning is preferred in order to prevent the situation where a group of non-zero elements in a particular component matrix are involved in significantly more cycles than the remaining non-zero elements. This constraint, although it might result in a sub-optimal solution in the protograph (in a few cases), is observed to be very beneficial when we apply the CPO to construct the final code.

As with the set $\mathcal{O}_{\textup{ind}}$, the optimization constraints depend only on code parameters, and not on the common substructure of interest (which depends on the channel). For example, in the case of $\gamma=3$, $m=1$, and any $\kappa$, $\mathcal{O}_{\textup{ind}}=\{t_0, t_1, t_2, t_{\{0,1\}}, t_{\{0,2\}}, \allowbreak t_{\{1,2\}}, t_{\{0,1,2\}}\}$, and the optimization constraints are (see also \cite{homa_boo} and \cite{ahh_nboo}):
\begin{align}\label{eq_const}
&0 \leq t_0 \leq \kappa, \hspace{2.0em} 0 \leq t_{\{0,1\}} \leq t_0, \hspace{2.0em} t_{\{0,1\}} \leq t_1 \leq \kappa-t_0+t_{\{0,1\}}, 
\nonumber \\
&0 \leq t_{\{0,1,2\}} \leq t_{\{0,1\}}, \hspace{6.9em} t_{\{0,1,2\}} \leq t_{\{0,2\}} \leq t_0-t_{\{0,1\}}+t_{\{0,1,2\}}, 
\nonumber \\
&t_{\{0,1,2\}} \leq t_{\{1,2\}} \leq t_1-t_{\{0,1\}}+t_{\{0,1,2\}}, 
\nonumber \\
&t_{\{0,2\}}+t_{\{1,2\}}-t_{\{0,1,2\}} \leq t_2 \leq \kappa-t_0-t_1+t_{\{0,1\}}+t_{\{0,2\}}+t_{\{1,2\}}-t_{\{0,1,2\}}, 
\nonumber \\ 
&\textit{and }\left \lfloor {3\kappa}/{2} \right \rfloor \leq t_0+t_1+t_2 \leq \left \lceil {3\kappa}/{2} \right \rceil.
\end{align}

The solution of this optimization problem is not unique. However, since all the solutions have the same performance (e.g., they all achieve $F^*_{\textup{sum}}$, see also \cite{ahh_nboo}), we work with one of these solutions, and call it an optimal vector, $\bold{t}^*$.

\section{CPO: Customization for PR Systems}\label{sec_cpo}

Using an optimal vector $\bold{t}^*$, computed as described in the previous section, $\bold{H}^{\textup{p}}$ is partitioned and the protograph matrix of the SC code, $\bold{H}^{\textup{p}}_{\textup{SC}}$, is constructed. The next step is preventing as many objects in the protograph as possible from being reflected in the unlabeled graph of the SC code, via optimizing the circulant powers using the CPO. Here, the CPO is customized for the $(4, 4(\gamma-2))$ object, which is the common substructure for detrimental configurations in the case of PR systems (see also Fig.~\ref{Fig_denom}).

From the previous analysis, a Pattern $P_\ell$ spans at most either $m+1$ or $2m+1$ consecutive replicas, depending on the value of $\ell$. Let $\xi = 2m+1$. Thus, in the CPO, it suffices to operate on the PM $\bold{\Pi}^{\xi,\textup{p}}_1$, which is the non-zero part of the first $\xi$ replicas in $\bold{H}^{\textup{p}}_{\textup{SC}}$, and has the size $(\xi+m) \gamma \times \xi \kappa$. The circulant powers associated with the $1$'s in $\bold{H}^{\textup{p}}$ are defined as $f_{i,j}$, where $0 \leq i \leq \gamma-1$ and $0 \leq j \leq \kappa-1$. Let the circulant powers associated with the $1$'s in $\bold{\Pi}^{\xi,\textup{p}}_1$ be $f'_{i',j'}$, where $0 \leq i' \leq (\xi+m)\gamma-1$ and $0 \leq j' \leq \xi\kappa-1$. From the repetitive nature of the PM $\bold{\Pi}^{\xi,\textup{p}}_1$, $f'_{i',j'}=f_{\overline{i'},\widetilde{j'}}$, where $\overline{i'}=(i' \textup{ mod } \gamma)$ and $\widetilde{j'}=(j' \textup{ mod } \kappa)$. Define our cycle-$8$ candidate in the graph of $\bold{\Pi}^{\xi,\textup{p}}_1$ as $c_1-v_1-c_2-v_2-c_3-v_3-c_4-v_4$, which is a particular way of traversing a pattern and not necessarily a protograph cycle (see also Figures~\ref{Fig_denom} and \ref{Fig_pat}). This candidate results in $z$ (or $z/2$ in the case of $P_1$ only) cycles of length $8$ after lifting if and only if \cite{fos_cyc}:
\begin{align}\label{eq_power8}
f'_{c_1,v_1}+f'_{c_2,v_2}+f'_{c_3,v_3}+f'_{c_4,v_4} \equiv f'_{c_1,v_2}+f'_{c_2,v_3}+f'_{c_3,v_4}+f'_{c_4,v_1} \textup{ (mod $z$)}.
\end{align}

The goal is to prevent as many cycle-$8$ candidates in the graph of $\bold{H}^{\textup{p}}_{\textup{SC}}$ as possible from being converted into $z$ (or $z/2$ in the case of $P_1$) $(4, 4(\gamma-2))$ UASs/UTSs in the graph of $\bold{H}_{\textup{SC}}$, which is the unlabeled graph of the SC code. In other words, a cycle-$8$ candidate in the graph of $\bold{H}^{\textup{p}}_{\textup{SC}}$ is allowed to be converted into multiple $(4, 4(\gamma-2)-2\delta)$ UASs/UTSs, with $\delta \in \{1,2\}$, as long as they are not $(4, 0)$ UASs, in the unlabeled graph since these are not instances of the common substructure of interest. These $(4, 4(\gamma-2)-2\delta)$ UASs/UTSs, $\delta \in \{1,2\}$, are cycles of length $8$ with \textit{\textbf{internal connections}}, which means $v_1$ and $v_3$ are adjacent or/and $v_2$ and $v_4$ are adjacent (see Fig.~\ref{Fig_denom}). For the cycle-$8$ candidate in the graph of $\bold{\Pi}^{\xi,\textup{p}}_1$ that is described in the previous paragraph and has a CN, say $c_5$, connecting $v_1$ and $v_3$, in order to have this internal connection in the lifted cycles, the following condition for a cycle of length $6$ must be satisfied in addition to (\ref{eq_power8}):
\begin{equation}\label{eq_power61}
f'_{c_1,v_1}+f'_{c_2,v_2}+f'_{c_5,v_3} \equiv f'_{c_1,v_2}+f'_{c_2,v_3}+f'_{c_5,v_1} \textup{ (mod $z$)}.
\end{equation}
Similarly, for that cycle-$8$ candidate in the graph of $\bold{\Pi}^{\xi,\textup{p}}_1$ that has a CN, say $c_6$, connecting $v_2$ and $v_4$, in order to have this internal connection in the lifted cycles, the following condition for a cycle of length $6$ must be satisfied in addition to (\ref{eq_power8}):
\begin{equation}\label{eq_power62}
f'_{c_1,v_1}+f'_{c_6,v_2}+f'_{c_4,v_4} \equiv f'_{c_1,v_2}+f'_{c_6,v_4}+f'_{c_4,v_1} \textup{ (mod $z$)}.
\end{equation}
Note that the two CNs, $c_5$ and $c_6$, have to be different from the CNs the pattern encompasses in order that we consider them in the CPO algorithm as possible internal connections. The reason is that the final unlabeled graphs of our codes must have no cycles of length $4$ (which is also why (\ref{eq_power8}) is applied for $P_1$ since $f'_{c_1,v_1}+f'_{c_2,v_2} \equiv f'_{c_1,v_2}+f'_{c_2,v_1} \textup{ (mod $z$)}$ is not allowed for any protograph cycle of length $4$, $c_1 - v_1 - c_2 - v_2$).

The following lemma discusses the internal connections for different patterns in the protograph. 

\begin{lemma}\label{lem_int_conn}
Let $\eta_{P_\ell}$ be the maximum number of internal connections Pattern $P_\ell$ can have (multiple internal connections between the same two VNs are only counted once). Then,
\begin{align}
\eta_{P_\ell}=\left\{\begin{matrix}0, \textup{ } &\ell \in \{1, 3, 5\},
\\ 1, \textup{ } &\ell \in \{2, 6, 8\},
\\ 2, \textup{ } &\ell \in \{4, 7, 9\}.
\end{matrix}\right.
\end{align}
\end{lemma}

\begin{IEEEproof}
A protograph pattern, $P_\ell$, with only two VNs ($\ell \in \{1, 3, 5\}$) cannot have any internal connections. A protograph pattern with three VNs ($\ell \in \{2, 6, 8\}$) can have at most one internal connection. A protograph pattern with four VNs ($\ell \in \{4, 7, 9\}$) can have up to two internal connection, which completes the proof.
\end{IEEEproof}

The case of multiple internal connections between the same two VNs is addressed in the CPO algorithm.

The steps of the customized CPO algorithm for SC codes that have parameters $\gamma \geq 3$, $\kappa$, $m$, and $L \geq 2m+1$, are:
\begin{enumerate}
\item Assign initial circulant powers to all the $\gamma \kappa$ $1$'s in $\bold{H}^{\textup{p}}$. In this work, our initial powers are as in SCB codes. For example, $f_{i,j} = (i^2)(2j)$, $0 \leq i \leq \gamma-1$ and $0 \leq j \leq \kappa-1$  (initially, no cycles of length $4$ are in $\bold{H}_{\textup{SC}}$).
\item Construct $\bold{\Pi}^{\xi,\textup{p}}_1$ via $\bold{H}^{\textup{p}}$ and $\bold{t}^*$. Circulant powers of the $1$'s in $\bold{\Pi}^{\xi,\textup{p}}_1$, $f'_{i',j'}$, are obtained from the $1$'s in $\bold{H}^{\textup{p}}$.
\item Define a counting variable $\psi_{i,j}$, $0 \leq i \leq \gamma-1$ and $0 \leq j \leq \kappa-1$, for each of the $1$'s in $\bold{H}^{\textup{p}}$. Define another counting variable $\psi'_{i',j'}$, $0 \leq i' \leq (\xi+m)\gamma-1$ and $0 \leq j' \leq \xi\kappa-1$, for each of the elements in $\bold{\Pi}^{\xi,\textup{p}}_1$. Initialize all the variables in this step with zeros. Only $\xi \gamma \kappa$ counting variables of the form $\psi'_{i',j'}$ are associated with $1$'s in $\bold{\Pi}^{\xi,\textup{p}}_1$. The other variables remain zeros. 
\item Locate all instances of the nine patterns in $\bold{\Pi}^{\xi,\textup{p}}_1$. Note that locating $P_1$ means also locating all cycles of length $4$ in $\bold{\Pi}^{\xi,\textup{p}}_1$, which is needed.
\item Determine the $\zeta_{P_\ell}$ ways to traverse each instance of $P_\ell$, $\forall \ell$, to reach $(4, 4(\gamma-2))$ UASs/UTSs in the unlabeled graph, which are the $\zeta_{P_\ell}$ cycle-$8$ candidates.
\item Specify all internal connections (CNs) in each candidate determined in Step~5 if they can exist.
\item For each cycle-$8$ candidate in $\bold{\Pi}^{\xi,\textup{p}}_1$, check whether (\ref{eq_power8}) is satisfied for its circulant powers or not.
\item If (\ref{eq_power8}) is satisfied, and the candidate has no internal connections, or (\ref{eq_power8}) is satisfied and the candidate has internal connection(s) but neither (\ref{eq_power61}) nor (\ref{eq_power62}) is satisfied for any internal connection, mark this cycle-$8$ candidate as an \textit{\textbf{active candidate}}.  
\item Let $F^{k,\textup{a}}_{P_\ell,1}$, where $k \in \{1, 2, \dots, \xi\}$, be the number of active candidates of $P_\ell$ starting at the first replica and spanning $k$ consecutive replicas in $\bold{\Pi}^{\xi,\textup{p}}_1$. Thus, the number of active candidates of $P_\ell$ spanning $k$ consecutive replicas in $\bold{\Pi}^{\xi,\textup{p}}_1$ is $(\xi-k+1)F^{k,\textup{a}}_{P_\ell,1}$. \textit{(For example, for $k=1$, $\xi F^{1,\textup{a}}_{P_\ell,1}$ is the number of active candidates of $P_\ell$, for any value of $\ell$, spanning one replica in $\bold{\Pi}^{\xi,\textup{p}}_1$.)}
\item Compute the number of $(4, 4(\gamma-2))$ UASs/UTSs in $\bold{H}_{\textup{SC}}$ using the following formula (see also \cite{homa_boo}):
\begin{equation}
F_{\textup{SC}} = \sum_{\ell=1}^{9} \hspace{0.3em} \sum_{k=1}^{\xi} \left ( (L-k+1)F^{k,\textup{a}}_{P_\ell,1} \right ) z_{P_\ell},
\end{equation}
where $z_{P_\ell}=z/2$ if $\ell=1$, and $z_{P_\ell}=z$ otherwise. Recall that $\xi = 2m+1$.
\item Count the number of active candidates each $1$ in $\bold{\Pi}^{\xi,\textup{p}}_1$ is involved in. Assign weight $w_k = (L-k+1)/(\xi-k+1)$ to the number of active candidates spanning $k$ consecutive replicas in $\bold{\Pi}^{\xi,\textup{p}}_1$ (see also \cite{homa_boo}). Multiply $w_k$ by $1/2$ if the candidate is associated to $P_1$. \textit{(For example, for $k=\xi$, the weight of the number of active candidates spanning $\xi$ consecutive replicas is $(L-\xi+1)$.)}
\item Store the weighted count associated with each $1$ in $\bold{\Pi}^{\xi,\textup{p}}_1$, which is indexed by $(i', j')$, in $\psi'_{i',j'}$.
\item Calculate the counting variables $\psi_{i,j}$, $\forall i,j$, associated with the $1$'s in $\bold{H}^{\textup{p}}$ from the counting variables $\psi'_{i',j'}$ associated with the $1$'s in $\bold{\Pi}^{\xi,\textup{p}}_1$ (computed in Steps~11 and 12) using the following formula:
\begin{equation}
\psi_{i,j} = \sum_{i': \overline{i'}=i} \sum_{\substack{j': \widetilde{j'}=j \\ \bold{\Pi}^{\xi,\textup{p}}_1 [i'][j'] \neq 0}} \psi'_{i',j'},
\end{equation}
\item Sort these $\gamma \kappa$ $1$'s of $\bold{H}^{\textup{p}}$ in a list descendingly according to the counts in $\psi_{i,j}$, $\forall i,j$.
\item Pick a subset of $1$'s from the top of this list, and change the circulant powers associated with them.
\item Using these interim powers, do Steps~7, 8, 9, and 10.
\item If $F_{\textup{SC}}$ is reduced while maintaining no cycles of length $4$ and no $(4, 0)$ objects (in the case of $\gamma=3$) in $\bold{H}_{\textup{SC}}$, update $F_{\textup{SC}}$ and the circulant powers, then go to Step~11.
\item Otherwise, return to Step~15 to pick a different set of circulant powers or/and a different subset of $1$'s (from the $1$'s in $\bold{H}^{\textup{p}}$).
\item Iterate until the target $F_{\textup{SC}}$ (set by the code designer) is achieved, or the reduction in $F_{\textup{SC}}$ approaches zero.
\end{enumerate}
Step~15 in the CPO algorithm is performed heuristically. The number of $1$'s to work with depend on the circulant size, the values of the counts, and how these values are distributed. Moreover, tracking the counts of active candidates and the distribution of their values over different $1$'s in $\bold{H}^{\textup{p}}$ is the main factor to decide which $1$'s to select in each iteration.

\begin{figure}[H]
\vspace{-1.7em}
\center
\includegraphics[width=5.0in]{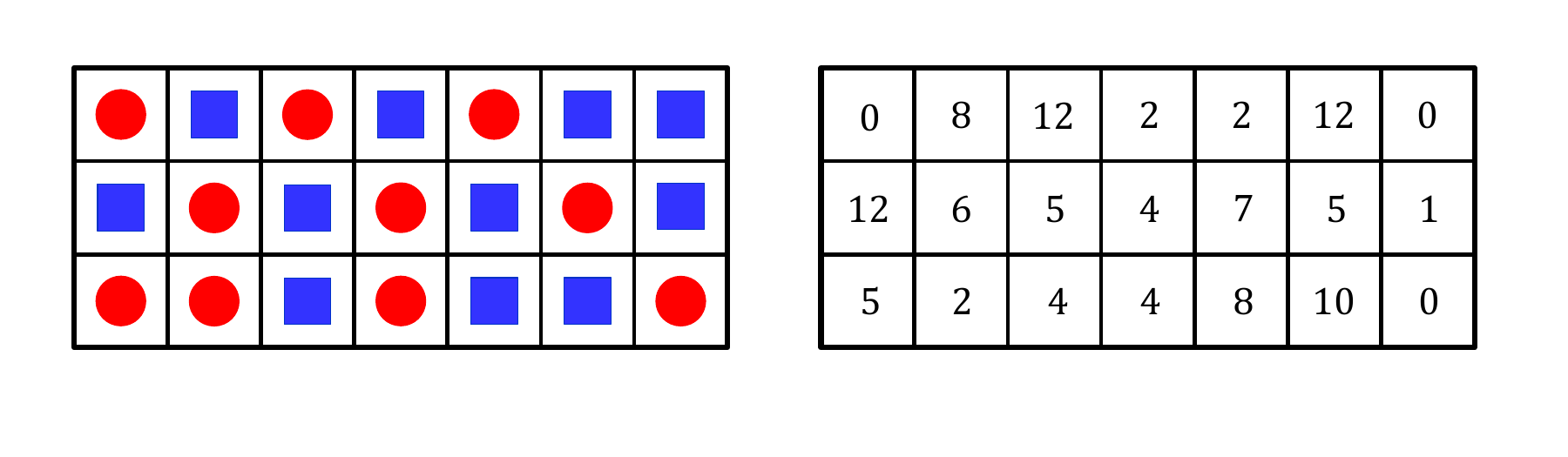}\vspace{-1.8em}
\text{\hspace{-0em}\footnotesize{(a) \hspace{19.5em} (b)}}
\caption{(a) The OO partitioning of $\bold{H}^{\textup{p}}$ (or $\bold{H}$) of the SC code in Example~\ref{ex_oo_cpo}. Entries with circles (resp., squares) are assigned to $\bold{H}^{\textup{p}}_0$ (resp., $\bold{H}^{\textup{p}}_1$). (b) The circulant power arrangement for the circulants in $\bold{H}$.}
\label{Fig_OOp7}
\vspace{-0.5em}
\end{figure}

\begin{example}\label{ex_oo_cpo}
Suppose we are designing an SC code with $\gamma = 3$, $\kappa=7$, $z=13$, $m=1$, and $L=10$ using the OO-CPO approach for PR systems. Solving the optimization problem in (\ref{eq_opt_pb}) gives an optimal vector $\bold{t}^*=[t^*_0 \text{ } t^*_1 \text{ } t^*_2 \text{ } t^*_{\{0,1\}} \text{ } t^*_{\{0,2\}} \text{ } t^*_{\{1,2\}} \allowbreak \text{ } t^*_{\{0,1,2\}}]^{\textup{T}}=[3 \text{ } 3 \text{ } 4 \text{ } 0 \text{ } 1 \text{ } 2 \text{ } 0]^{\textup{T}}$, with $F^*_{\textup{sum}} = 5{,}170$ patterns (rounded weighted sum) in the graph of $\bold{H}^{\textup{p}}_{\textup{SC}}$. Fig.~\ref{Fig_OOp7}(a) shows how the partitioning is applied on $\bold{H}^{\textup{p}}$ (or $\bold{H}$). Next, applying the CPO results in $2{,}613$ $(4, 4)$ UASs in the graph of $\bold{H}_{\textup{SC}}$. Fig.~\ref{Fig_OOp7}(b) shows the final circulant power arrangement for all circulants in $\bold{H}$.
\end{example}

\begin{remark}\label{rmk_extend}
After introducing the concept of patterns in this work, the OO-CPO approach can be easily extended to target other common substructures if needed.
\end{remark}

\section{Experimental Results}\label{sec_exp}

In this section, we propose experimental results demonstrating the effectiveness of the OO-CPO approach compared with other code design techniques in PR (1-D MR) systems.

\begin{remark}
In this section, all the codes used have no cycles of length $4$. Moreover, we opted to work with circulant sizes $z > \kappa$ in order to give more freedom to the CPO, which results in less detrimental objects. 
\end{remark}

First, we compare the total number of instances of the common substructure of interest in the unlabeled graphs of SC codes designed using various techniques. We present results for two groups of codes.

All the codes in the first group have $\gamma = 3$ (i.e., the common substructure of interest is the $(4, 4)$ UAS in Fig.~\ref{Fig_denom}) and $m \in \{1, 2\}$. We also choose $L=10$ for this group. In addition to the uncoupled setting ($\bold{H}_0=\bold{H}$ and $\bold{H}_1=\bold{0}$), we show results for the following five SC code design techniques:
\begin{enumerate}
\item The CV technique (see \cite{homa_sc}) with $m=1$.
\item The OO technique with no CPO applied and with $m=1$.
\item The OO technique with circulant powers optimized via the CPO (the OO-CPO approach) and with $m=1$.
\item The OO technique with no CPO applied and with $m=2$.
\item The OO technique with circulant powers optimized via the CPO (the OO-CPO approach) and with $m=2$.
\end{enumerate}
In the uncoupled setting in addition to the first, second, and fourth techniques, circulant powers as in SCB codes, $f_{i,j} = f(i)f(j) = (i^2)(2j)$, are used. This choice of circulant powers guarantees no cycles of length $4$.

The results of the first group of codes for different choices of $\kappa$ and $z$ are listed in Table~\ref{table_1}. For a particular choice of $\kappa$, $z$, $m$, and $L$, SC codes designed using these different techniques all have block length $=\kappa z L \log_2(q)$ bits and rate $\approx \left[1- \frac{3 (L+m)}{\kappa L} \right]$. Table~\ref{table_1} demonstrates the significant gains achieved by the OO-CPO approach compared with other techniques. In particular, for $m=1$, the proposed OO-CPO approach achieves a reduction in the number of $(4, 4)$ UASs that ranges between $85\%$ and $92\%$ compared with the uncoupled setting, and between $61\%$ and $72\%$ compared with the CV technique. The table also illustrates the positive effect of increasing the memory of the SC code. In particular, the OO-CPO approach with $m=2$ achieves a reduction in the number of $(4, 4)$ UASs that ranges between $54\%$ and $69\%$ compared with the OO-CPO approach with $m=1$. Moreover, the importance of the two stages (the OO and the CPO) is highlighted by the numbers in Table~\ref{table_1}.

\vspace{-0.2em}
\begin{table}[H]
\caption{Number of $(4, 4)$ UASs in SC codes with $\gamma = 3$, $m \in \{1, 2\}$, and $L=10$ designed using different techniques.
}
\vspace{-0.5em}
\centering
\scalebox{1.00}
{
\begin{tabular}{|c|c|c|c|c|}
\hline

\multirow{3}{*}{Design technique} & \multicolumn{4}{|c|}{\makecell{Number of $(4, 4)$ UASs}} \\
\cline{2-5}
{} & \makecell{$\kappa=7$, \\ $z=13$} & \makecell{$\kappa=11$, \\ $z=23$} & \makecell{$\kappa=13$, \\ $z=29$} & \makecell{$\kappa=17$, \\ $z=37$} \\
\hline
Uncoupled with SCB & $32{,}370$ & $254{,}610$ & $540{,}850 $ & $1{,}700{,}890$ \\
\hline
SC CV with SCB and $m=1$ & $9{,}464$ & $91{,}333$ & $197{,}084$ & $652{,}347$ \\
\hline
SC OO with SCB and $m=1$ & $6{,}500$ & $53{,}130$ & $123{,}395$  & $440{,}818$ \\
\hline
SC OO-CPO and $m=1$ & $2{,}613$ & $32{,}361$ & $70{,}151$ & $254{,}005$ \\
\hline
SC OO with SCB and $m=2$ & $3{,}172$ & $27{,}508$ & $60{,}233$  & $194{,}176$ \\
\hline
SC OO-CPO and $m=2$ & $819$ & $13{,}110$ & $32{,}074$ & $117{,}697$ \\
\hline
\end{tabular}}
\label{table_1}
\end{table}
\vspace{-0.2em}

As for the second group, all the codes have $\gamma = 4$ (i.e., the common substructure of interest is the $(4, 8)$ UTS in Fig.~\ref{Fig_denom}) and $m=1$. We also choose $L=10$ for this group. In addition to the uncoupled setting ($\bold{H}_0=\bold{H}$ and $\bold{H}_1=\bold{0}$), we show results for the following three SC code design techniques:
\begin{enumerate}
\item The CV technique (see \cite{homa_sc}).
\item The OO technique with no CPO applied.
\item The OO technique with circulant powers optimized via the CPO (the OO-CPO approach).
\end{enumerate}
In the uncoupled setting in addition to the first and second techniques, circulant powers as in SCB codes, $f_{i,j} = f(i)f(j) = (i^2)(2j)$, are used. This choice of circulant powers guarantees no cycles of length $4$.

The results of the second group of codes for different choices of $\kappa$ and $z$ are listed in Table~\ref{table_2}. For a particular choice of $\kappa$, $z$, and $L$, SC codes designed using these different techniques all have block length $=\kappa z L \log_2(q)$ bits and rate $\approx \left[1- \frac{4 (L+1)}{\kappa L} \right]$. Table~\ref{table_2} again demonstrates the significant gains achieved by the OO-CPO approach compared with other techniques. In particular, the proposed OO-CPO approach achieves a reduction in the number of $(4, 8)$ UTSs that ranges between $82\%$ and $87\%$ compared with the uncoupled setting, and between $55\%$ and $64\%$ compared with the CV technique. Moreover, the importance of the two stages (the OO and the CPO) is again highlighted by the numbers in Table~\ref{table_2}.

\vspace{-0.4em}
\begin{table}[H]
\caption{Number of $(4, 8)$ UTSs in SC codes with $\gamma = 4$, $m = 1$, and $L = 10$ designed using different techniques.
}
\vspace{-0.5em}
\centering
\scalebox{1.00}
{
\begin{tabular}{|c|c|c|c|c|}
\hline

\multirow{3}{*}{Design technique} & \multicolumn{4}{|c|}{\makecell{Number of $(4, 8)$ UTSs}} \\
\cline{2-5}
{} & \makecell{$\kappa=7$, \\ $z=13$} & \makecell{$\kappa=11$, \\ $z=23$} & \makecell{$\kappa=13$, \\ $z=29$} & \makecell{$\kappa=17$, \\ $z=37$} \\
\hline
Uncoupled with SCB & $131{,}820$ & $1{,}034{,}310$ & $2{,}193{,}850$ & $7{,}081{,}430$ \\
\hline
SC CV with SCB & $48{,}074$ & $396{,}474$ & $843{,}233$ & $2{,}782{,}844$ \\
\hline
SC OO with SCB & $27{,}729$ & $230{,}230$ & $508{,}544$  & $1{,}667{,}886$ \\
\hline
SC OO-CPO & $17{,}095$ & $165{,}071$ & $366{,}212$ & $1{,}253{,}745$ \\
\hline
\end{tabular}}
\label{table_2}
\end{table}
\vspace{-0.2em}

Second, we present simulation results of binary and non-binary SC codes designed using various techniques over the PR channel. We present results for three groups of codes. We use the PR channel described in \cite{ahh_bas}. This channel incorporates inter-symbol interference (intrinsic memory), jitter, and electronic noise. The normalized channel density \cite{shafa_2d, tomv_pr} we use is $1.4$, and the PR equalization target is $[8$~$14$~$2]$. The receiver consists of filtering units followed by a Bahl Cocke Jelinek Raviv (BCJR) detector \cite{bcjr}, which is based on pattern-dependent noise prediction (PDNP) \cite{pdnp}, in addition to a fast Fourier transform based $q$-ary sum-product algorithm (FFT-QSPA) LDPC decoder \cite{dec_fft}, with $q$ being set to $2$ in the case of binary codes. The number of global (detector-decoder) iterations is $10$, and the number of local (decoder only) iterations is $20$. Unless a codeword is reached, the decoder performs its prescribed number of local iterations for each global iteration. More details about this PR system can be found in \cite{ahh_bas}.

\begin{figure}[H]
\vspace{-1.2em}
\center
\includegraphics[trim={0.3in 1.0in 0.5in 0.4in},clip,width=5.0in]{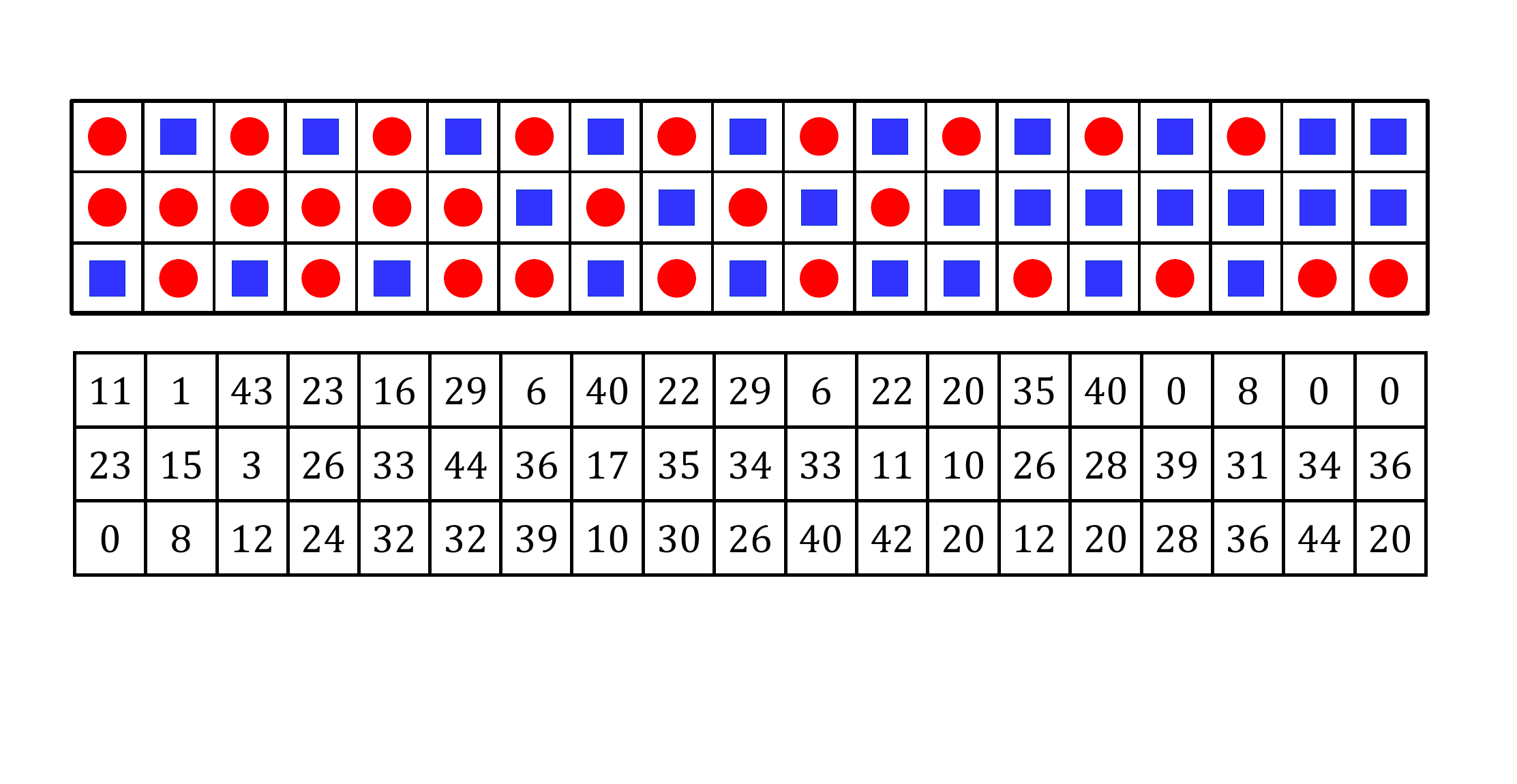}
\vspace{-0.7em}
\caption{Upper panel: the OO partitioning of $\bold{H}^{\textup{p}}$ (or $\bold{H}$) of Code~3. Entries with circles (resp., squares) are assigned to $\bold{H}^{\textup{p}}_0$ (resp., $\bold{H}^{\textup{p}}_1$). Lower panel: the circulant power arrangement for the circulants in $\bold{H}$ of Code~3.
}
\label{Fig_OOs1}
\vspace{-0.5em}
\end{figure}

The first group of simulated codes contains five different codes. All the five codes are defined over GF($4$). Codes~1, 2, 3, and 4 have $\gamma = 3$, $\kappa=19$, $z=46$, $m=1$, and $L=5$. Thus, these codes have block length $=8{,}740$ bits, and the SC codes have rate $\approx 0.81$. Code~1 is uncoupled. Code~2 is an SC code designed using the CV technique for PR channels as described in \cite{homa_sc}. The optimal cutting vector used for Code~2 is $[4 \text{ } 9 \text{ } 15]$. Codes~1 and 2 have SCB circulant powers of the form $f_{i,j}=(i^2)(2j)$. Code~3 is an SC code designed using the OO-CPO approach. The partitioning and the circulant power arrangement of Code~3 are given in Fig.~\ref{Fig_OOs1}. Codes~1, 2, and 3 have unoptimized edge weights. Code~4 is the result of applying the WCM framework to Code~3 in order to optimize its edge weights. The numbers of $(4, 4)$ UASs in the unlabeled graphs of Codes~1, 2, and 3 are $2{,}425{,}120$, $845{,}434$, and $184{,}667$, respectively. Code~5 is a block (BL) code, which is also protograph-based (PB), designed as in \cite{ahh_jsac} and \cite{ahh_tit}. Code~5 has column weight $=3$, circulant size $=46$, block length $= 8{,}832$ bits, rate $\approx 0.81$ (same as all SC codes), and unoptimized weights (similar to all codes except Code~4).

\begin{figure}[H]
\vspace{-0.8em}
\center
\includegraphics[trim={0.4in 0.0in 0.5in 0.2in},clip,width=4.2in]{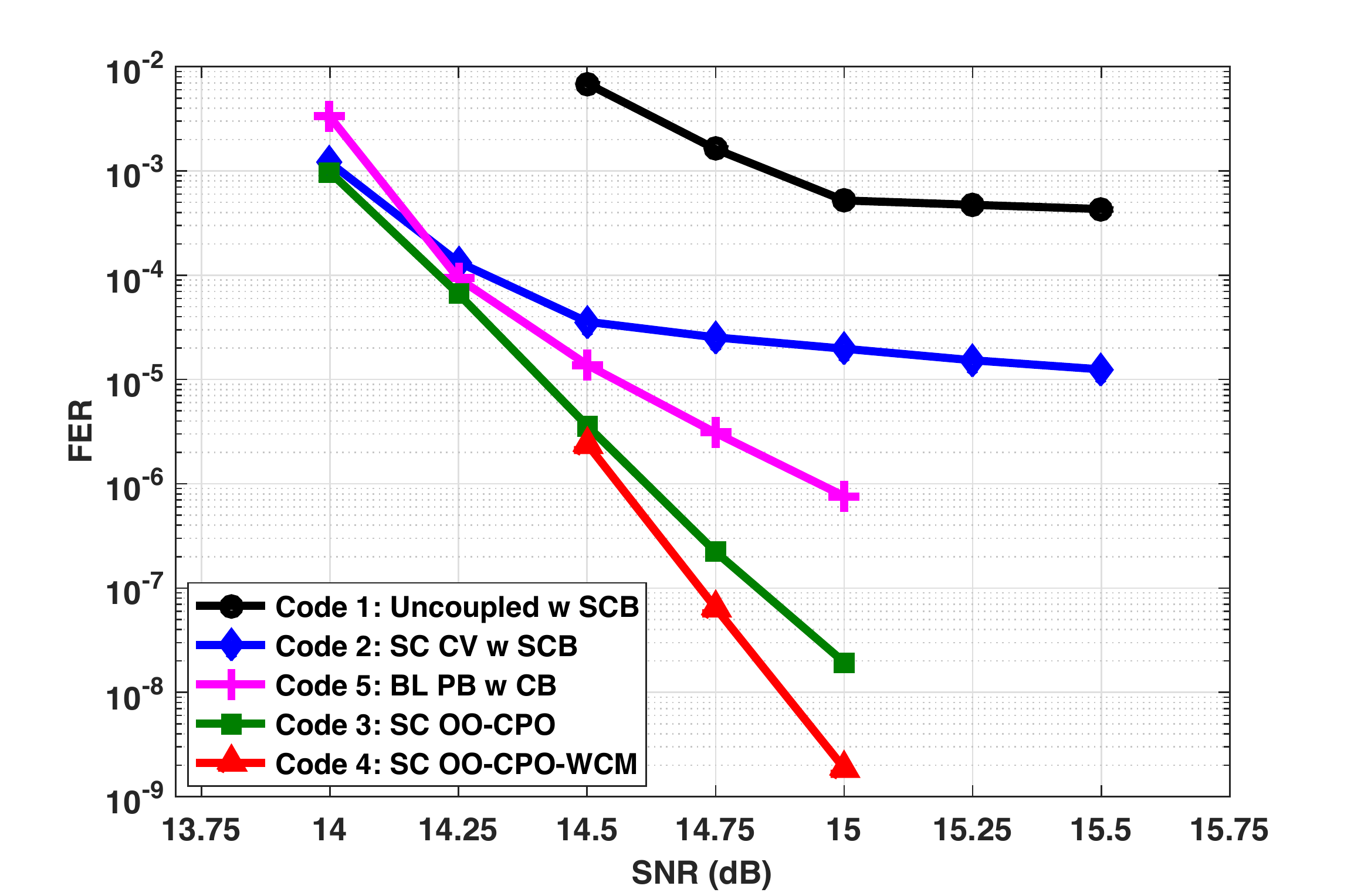}
\vspace{-0.5em}
\caption{Simulation results over the PR channel for SC codes having $\gamma=3$ and $m=1$ designed using different techniques and a BL code of the same length and rate.}
\label{Fig_fer1}
\vspace{-0.7em}
\end{figure}

Fig.~\ref{Fig_fer1} demonstrates the effectiveness of the proposed OO-CPO approach in designing high performance SC codes for PR channels. In particular, Code~3 (designed using the OO-CPO approach) outperforms Code~2 (designed using the CV technique) by about $3$ orders of magnitude at signal-to-noise ratio (SNR) $=15$ dB, and by about $1.1$ dB at FER $= 10^{-5}$. More intriguingly, Code~3 outperforms Code~5 (the block code) by about $1.6$ orders of magnitude at SNR $=15$ dB, and by almost $0.4$ dB at FER $= 10^{-6}$. The performance of Code~3 is better than the performance of Code~5 not only in the error floor region, but also in the waterfall region. An interesting observation is that, in the error profile of Code~3, we found no codewords of weights $\in\{6, 8\}$ (which are $(6,0,0,9,0)$ and $(8,0,0,12,0)$ BASTs) despite the dominant presence of such low weight codewords in the error profiles of Codes~1, 2, and 5 (see also \cite{ahh_bas}, \cite{ahh_tit}, and \cite{homa_sc}). From Fig.~\ref{Fig_fer1}, the WCM framework achieves $1$ order of magnitude additional gain.

An important reason behind the improved waterfall performance of Code~3 is the significant reduction in the multiplicity of low weight codewords achieved by the OO-CPO approach. This reduction is a result of the fact that such low weight codewords also have the $(4, 4)$ UAS as a common substructure in their configurations (see Fig.~\ref{Fig_denom}).

\begin{figure}[H]
\vspace{-1.4em}
\center
\includegraphics[trim={0.3in 1.0in 1.2in 0.4in},clip,width=4.7in]{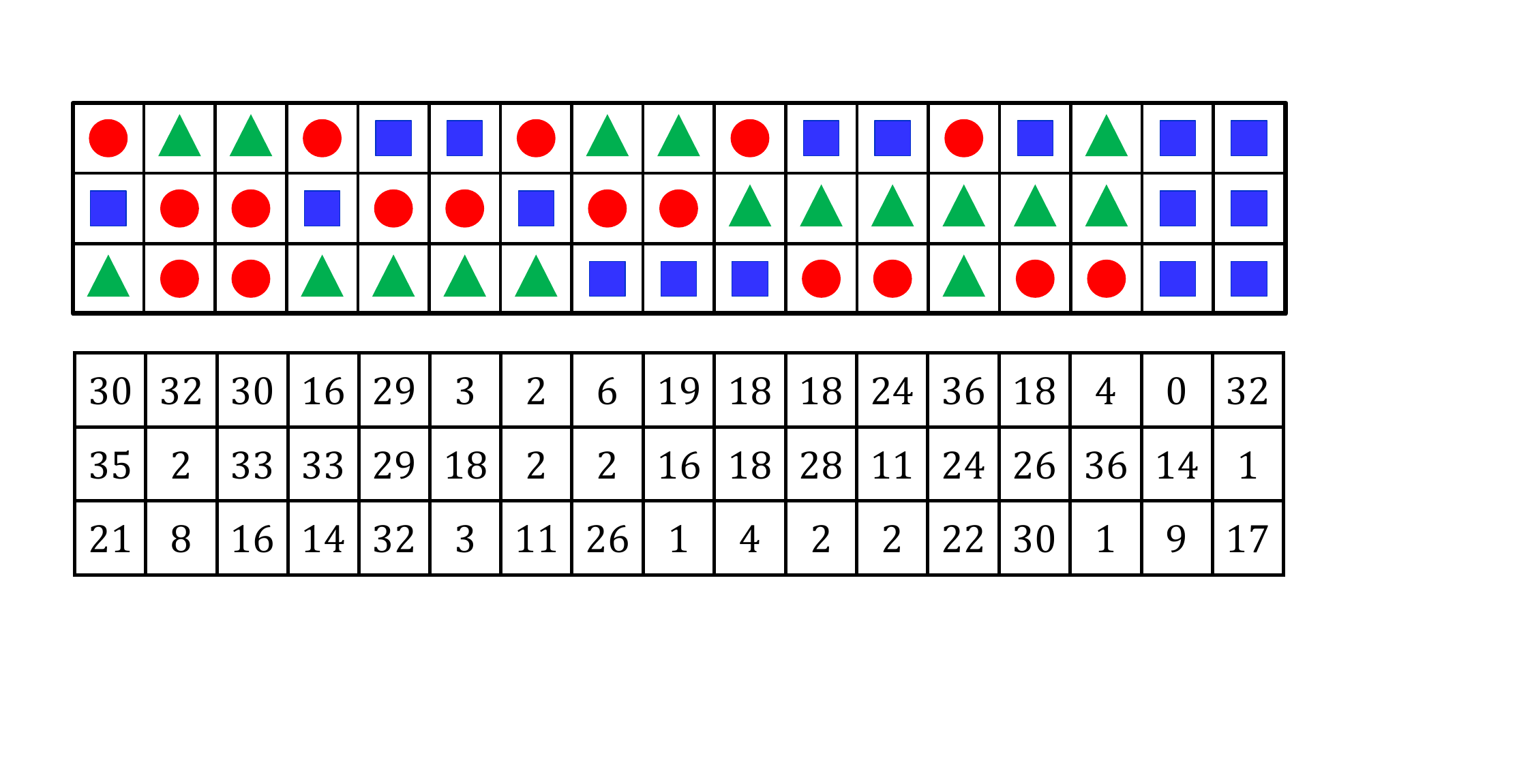}
\vspace{-0.9em}
\caption{Upper panel: the OO partitioning of $\bold{H}^{\textup{p}}$ (or $\bold{H}$) of Code~6. Entries with circles (resp., squares and triangles) are assigned to $\bold{H}^{\textup{p}}_0$ (resp., $\bold{H}^{\textup{p}}_1$ and $\bold{H}^{\textup{p}}_2$). Lower panel: the circulant power arrangement for the circulants in $\bold{H}$ of Code~6.
}
\label{Fig_OOs2}
\vspace{-0.4em}
\end{figure}

\begin{figure}[H]
\vspace{-0.8em}
\center
\includegraphics[trim={0.4in 1.8in 3.3in 0.2in},clip,width=4.2in]{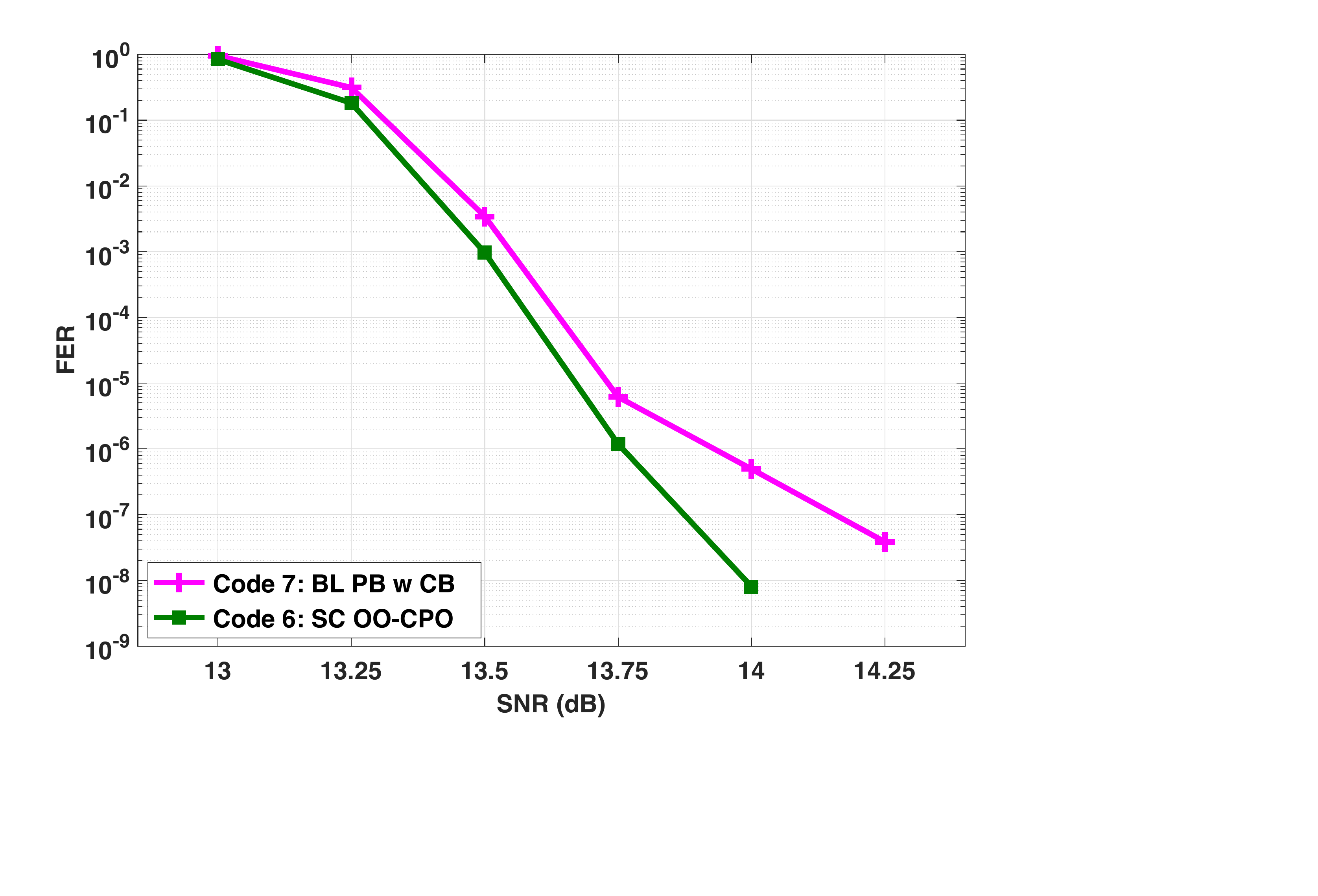}
\vspace{-0.5em}
\caption{Simulation results over the PR channel for an SC code having $\gamma=3$ and $m=2$ designed using the OO-CPO approach and a BL code of the same length and rate.}
\label{Fig_fer2}
\vspace{-0.5em}
\end{figure}

The second group of simulated codes contains two different codes. The two codes are defined over GF($4$). Code~6 has $\gamma = 3$, $\kappa=17$, $z=37$, $m=2$, and $L=7$. Thus, this code has block length $=8{,}806$ bits and rate $\approx 0.77$. Code~6 is an SC code designed using the OO-CPO approach. The partitioning and the circulant power arrangement of Code~6 are given in Fig.~\ref{Fig_OOs2}. Code~6 has unoptimized edge weights. The number of $(4, 4)$ UASs in the unlabeled graph of Code~6 is reduced to $75{,}850$ via the OO-CPO approach. Code~7 is a BL PB code designed as in \cite{ahh_jsac} and \cite{ahh_tit}. Code~7 has column weight $=3$, circulant size $=43$, block length $= 8{,}944$ bits, rate $\approx 0.77$ (same as the SC code), and unoptimized weights (similar to the SC code).

The purpose of Fig.~\ref{Fig_fer2} is to stress on the intriguing conclusion that SC codes designed using the OO-CPO approach outperform block codes having the same parameters. In particular, Code~6 (designed using the OO-CPO approach) outperforms Code~7 (the block code) by about $1.8$ orders of magnitude at SNR $=14$ dB, and by about $0.3$ dB at FER $= 10^{-7}$. These gains are projected to be significantly bigger as we go deeper in FER noting that we could not collect a single error after simulating around $10^9$ frames of Code~6 at SNR $=14.25$ dB. Moreover, the performance of Code~6 is better than the performance of Code~7 not only in the error floor region, but also in the waterfall region.

\begin{figure}[H]
\vspace{-0.8em}
\center
\includegraphics[trim={0.1in 1.0in 5.2in 0.0in},clip,width=4.7in]{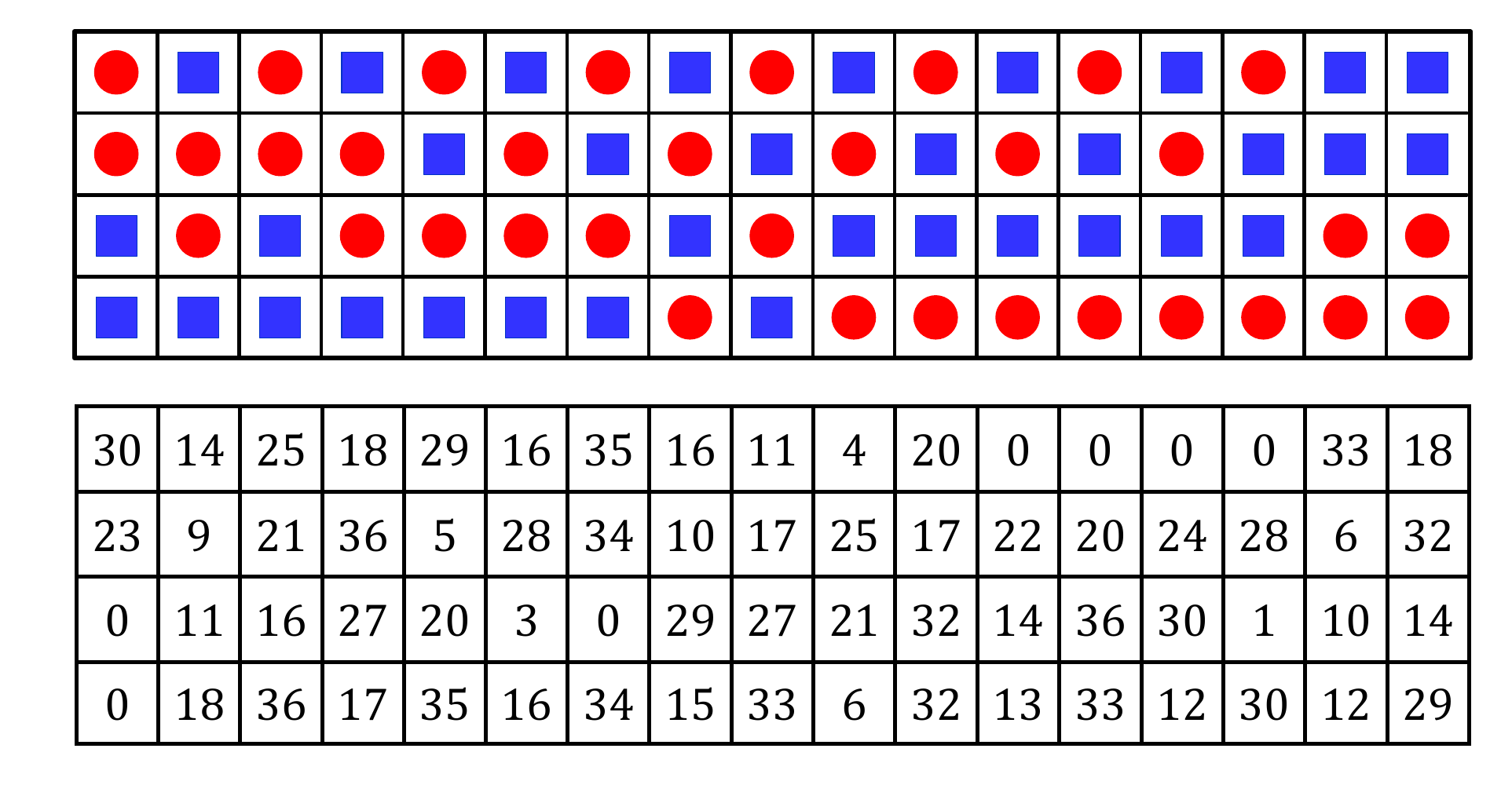}
\vspace{-0.8em}
\caption{Upper panel: the OO partitioning of $\bold{H}^{\textup{p}}$ (or $\bold{H}$) of Code~10. Entries with circles (resp., squares) are assigned to $\bold{H}^{\textup{p}}_0$ (resp., $\bold{H}^{\textup{p}}_1$). Lower panel: the circulant power arrangement for the circulants in $\bold{H}$ of Code~10.
}
\label{Fig_OOs3}
\vspace{-0.5em}
\end{figure}

\begin{figure}[H]
\vspace{-0.9em}
\center
\includegraphics[trim={0.4in 1.8in 3.3in 0.2in},clip,width=4.2in]{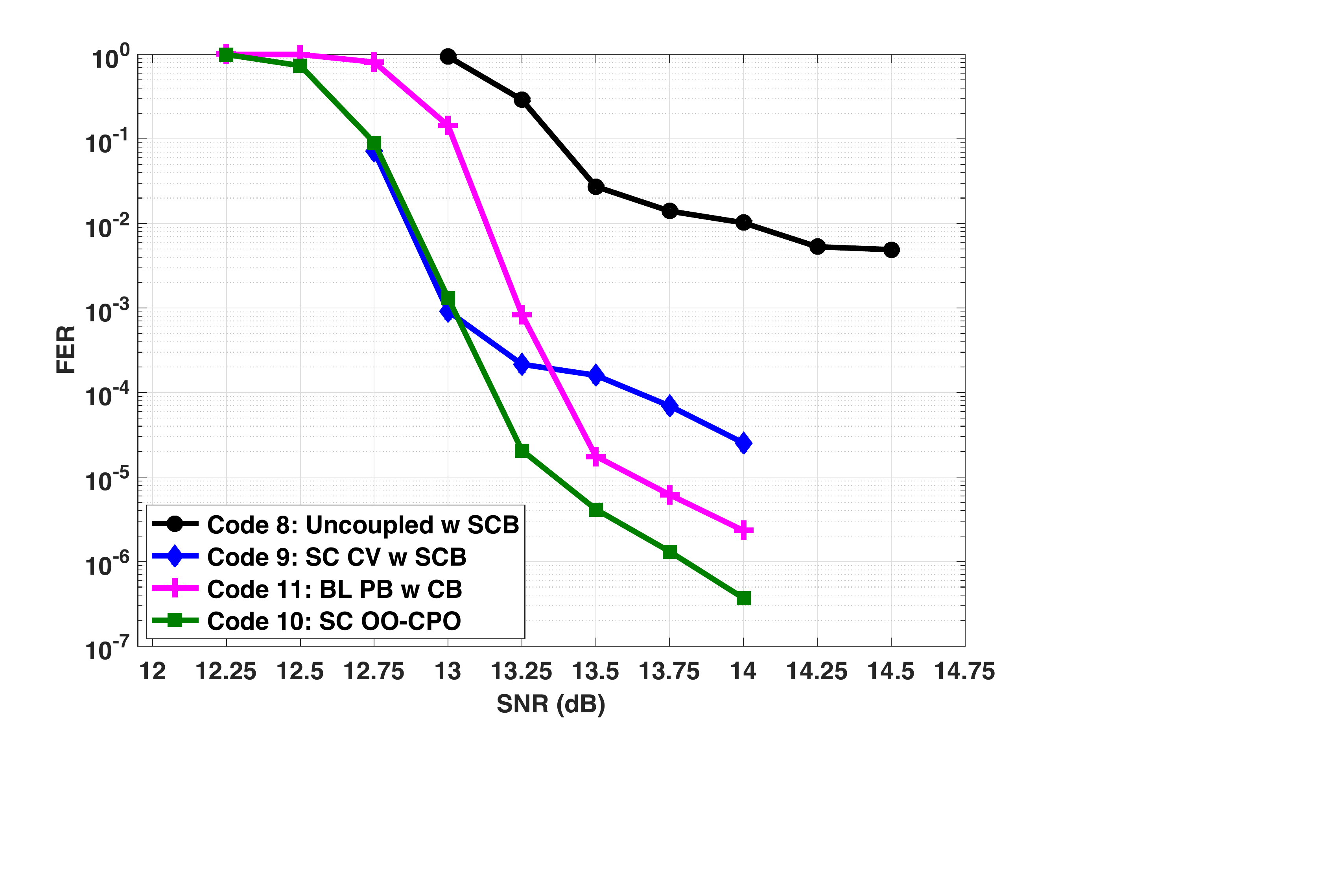}
\vspace{-0.5em}
\caption{Simulation results over the PR channel for SC codes having $\gamma=4$ and $m=1$ designed using different techniques and a BL code of the same length and rate.}
\label{Fig_fer3}
\vspace{-0.5em}
\end{figure}

The third group of simulated codes contains four different codes. All the four codes are defined over GF($2$), i.e., binary codes. Codes~8, 9, and 10 have $\gamma = 4$, $\kappa=17$, $z=37$, $m=1$, and $L=6$. Thus, these codes have block length $=3{,}774$ bits, and the SC codes have rate $\approx 0.73$. Code~8 is uncoupled. Code~9 is an SC code designed using the CV technique for PR channels as described in \cite{homa_sc}. The optimal cutting vector used for Code~9 is $[3 \text{ } 7 \text{ } 11 \text{ } 14]$. Codes~8 and 9 have SCB circulant powers of the form $f_{i,j}=(i^2)(2j)$. Code~10 is an SC code designed using the OO-CPO approach. The partitioning and the circulant power arrangement of Code~10 are given in Fig.~\ref{Fig_OOs3} The numbers of $(4, 8)$ UTSs in the unlabeled graphs of Codes~8, 9, and 10 are $4{,}248{,}858$, $1{,}589{,}816$, and $705{,}849$, respectively. Code~11 is a BL PB code designed as in \cite{ahh_jsac} and \cite{ahh_tit}. Code~11 has column weight $=4$, circulant size $=41$, block length $= 3{,}690$ bits and rate $\approx 0.73$ (same as all SC codes).

Fig.~\ref{Fig_fer3} again demonstrates the effectiveness of the OO-CPO approach in designing high performance SC codes with various parameters for PR channels. In particular, Code~10 (designed using the OO-CPO approach) outperforms Code~9 (designed using the CV technique) by more than $1.8$ orders of magnitude at SNR $=14$ dB, and by nearly $0.75$ dB at FER $= 3 \times 10^{-5}$. More intriguingly, Code~10 outperforms Code~11 (the block code) by about $0.8$ of an order of magnitude at SNR $=14$ dB, and by about $0.35$ dB at FER $= 3 \times 10^{-6}$. A very interesting observation here is that Code~10 achieves an early waterfall gain of about $0.25$ dB compared with Code~11 (see, for example, the performance of the two codes at FER $=10^{-1}$). In other words, Code~10 has a $0.25$ dB threshold improvement compared with Code~11. Note that the codes here have a relatively low rate, which demonstrates the gains achieved by the OO-CPO approach for a diverse range of rates.

There are two key takeaways from these experimental results. First, SC codes constructed using the proposed OO-CPO approach significantly outperform SC codes constructed using the techniques currently available in the literature. Second, and most importantly, properly exploiting the additional degree of freedom provided by partitioning in the construction of SC codes enables a design of SC codes that outperform block codes of the same total length, which conclusively answers an open question about whether SC codes can outperform block codes under equal total length. This proper exploitation is performed exclusively through taking into account the characteristics of the channel of interest, which is what we do in this work.

\begin{remark}
Unlike a lot of literature works that compare an SC code to a block code having a length equal to the constraint length of the SC code, which is $\kappa z (m+1) \log_2(q)$ bits, we compare an SC code to a block code having the same length of the SC code in total, which is $\kappa z L \log_2(q)$ bits, approximately. Moreover, while our high performance SC codes designed using the OO-CPO approach do outperform block codes of the same parameters, other SC codes available in the literature do not. An example demonstrating this statement is presented in Fig.~\ref{Fig_fer1} and Fig.~\ref{Fig_fer3}, where block codes outperform SC codes of the same parameters and designed using the CV technique.
\end{remark}

\begin{remark}
Because our main focus in this work is the performance, a relatively small to average values of $L$ ($5 \leq L \leq 7$) along with block decoding are used for all SC Codes.
\end{remark}

\section{Conclusion}\label{sec_conc}

We proposed the OO-CPO approach to optimally design binary and non-binary SC codes for PR channels, via minimizing the number of detrimental objects in the graph of the code. A common substructure was first identified in the graphs of the detrimental configurations in the case of PR systems. We graphically determined the protograph patterns that are capable of generating instances of this common substructure in the final graph of the code. Next, through combinatorial techniques, we built a discrete optimization problem in which the weighted sum of the total number of instances of these patterns is expressed in terms of the partitioning parameters. The partitioning that achieves the minimum weighted sum was obtained. Then, the lifting parameters were optimized in order to achieve more reduction in the number of detrimental objects of interest. SC codes designed using the proposed OO-CPO approach were shown to significantly outperform SC codes designed using techniques from the literature over PR channels. More importantly, our channel-aware combinatorial approach demonstrated that appropriate exploitation of the available degrees of freedom in the SC code design can give SC codes significant performance advantages over structured block codes having the same parameters. We believe this research will open the door for engineers to deploy high performance SC codes in a wide variety of applications in addition to data storage.

\begin{appendices}

\section{Proofs of Pattern $P_1$}\label{sec_appa}

\subsection{Proof of Lemma \ref{lem_p1}}

\begin{IEEEproof}
In Case~1.1, the number of instances is the number of ways to choose $2$ overlaps out of $t_{\{i_1,i_2\}}$ overlaps (the pattern has two $c_1 - c_2$ overlaps), which is given by (\ref{eq_p1_1}). In Case~1.2, the number of instances is the number of ways to choose $1$ overlap out of $t_{\{i_1,i_2\}}$ and $1$ overlap out of $t_{\{i_1+(r-e)\gamma, i_2+(r-e)\gamma \}}$, which is given by (\ref{eq_p1_2}).
\end{IEEEproof}

\subsection{Proof of Theorem \ref{thm_p1}}

\begin{IEEEproof}
To compute $F_{P_1}$, we use the formula in \cite[Theorem~1]{homa_boo}, with $\chi$, which is the maximum number of replicas the pattern can span, equals $m+1$. Since the overlaps of $P_1$ can exist in up to $2$ replicas, we need to find expressions only for $F^1_{P_1,1}$ (overlaps are in $1$ replica) and $F^{k \geq 2}_{P_1,1}$ (overlaps are in $2$ replicas).

Then, $F^1_{P_1,1}$ is the sum of function $\mathcal{A}_{P_1}$ in (\ref{eq_p1_1}), with $r=1$, over all possible values of $\{i_1,i_2\}$. Here, $\{i_1,i_2\}$ can take any distinct two values in the range from the start to the end of $\bold{R}_1$, i.e., from $0$ to $(m+1)\gamma-1$ (see Fig.~\ref{Fig_pat1}).

Moreover, $F^{k \geq 2}_{P_1,1}$ is the sum of function $\mathcal{B}_{P_1}$ in (\ref{eq_p1_2}), with $r=1$ and $e=k$, over all possible values of $\{i_1,i_2\}$. Here, $\{i_1,i_2\}$ can take any distinct two values in the range from the start of $\bold{R}_k$ to the end of $\bold{R}_1$, i.e., from $(k-1)\gamma$ to $(m+1)\gamma-1$ (see also Fig.~\ref{Fig_pat1}).
\end{IEEEproof}

\section{Proofs of Pattern $P_2$}\label{sec_appb}

\subsection{Proof of Lemma \ref{lem_p2}}

\begin{IEEEproof}
In Case~2.1, the number of instances is the number of ways to choose $3$ overlaps out of $t_{\{i_1,i_2\}}$ overlaps (the pattern has three $c_1 - c_2$ overlaps), which is given by (\ref{eq_p2_1}). In Case~2.2, the number of instances is the number of ways to choose $2$ overlap out of $t_{\{i_1,i_2\}}$ and $1$ overlap out of $t_{\{i_1+(r-e)\gamma, i_2+(r-e)\gamma \}}$, which is given by (\ref{eq_p2_2}). In Case~2.3, the number of instances is the number of ways to choose $1$ overlap out of $t_{\{i_1,i_2\}}$, $1$ overlap out of $t_{\{i_1+(r-e)\gamma, i_2+(r-e)\gamma \}}$, and $1$ overlap out of $t_{\{i_1+(r-s)\gamma, i_2+(r-s)\gamma \}}$, which is given by (\ref{eq_p2_3}).
\end{IEEEproof}

\subsection{Proof of Theorem \ref{thm_p2}}

\begin{IEEEproof}
To compute $F_{P_2}$, we use the formula in \cite[Theorem~1]{homa_boo}, with $\chi=m+1$. Since the overlaps of $P_2$ can exist in up to $3$ replicas, we need to find expressions only for $F^1_{P_2,1}$, $F^2_{P_2,1}$, and $F^{k \geq 3}_{P_2,1}$.

Then, $F^1_{P_2,1}$ is the sum of function $\mathcal{A}_{P_2}$ in (\ref{eq_p2_1}), with $r=1$, over all possible values of $\{i_1,i_2\}$. Here, $\{i_1,i_2\}$ can take any distinct two values in the range from the start to the end of $\bold{R}_1$, i.e., from $0$ to $(m+1)\gamma-1$ (see Fig.~\ref{Fig_pat2}).

Regarding $F^2_{P_2,1}$, we need to distinguish between two situations; when $r < e$ (i.e., replica $\bold{R}_r$, which has two overlaps, comes before replica $\bold{R}_e$), and when $r > e$ (i.e., replica $\bold{R}_r$ comes after replica $\bold{R}_e$). This distinction gives the two summations of function $\mathcal{B}_{P_2}$ in $F^2_{P_2,1}$. For the first summation, $\mathcal{B}_{P_2}$ in (\ref{eq_p2_2}) has $r=1$ and $e=2$. Thus, $\{i_1,i_2\}$ can take any distinct two values in the range from the start of $\bold{R}_2$ to the end of $\bold{R}_1$, i.e., from $\gamma$ to $(m+1)\gamma-1$ (see Fig.~\ref{Fig_pat2} for more illustration). For the second summation, $\mathcal{B}_{P_2}$ in (\ref{eq_p2_2}) has $r=2$ and $e=1$. Thus, $\{i_1,i_2\}$ can take any distinct two values in the range from the start of $\bold{R}_2$ (which is now $\bold{R}_r$) to the end of $\bold{R}_1$, i.e., from $0$ to $m\gamma-1$.

As for $F^{k \geq 3}_{P_2,1}$, the overlaps can be in $2$ replicas (the first two summations in $F^{k \geq 3}_{P_2,1}$) or $3$ replicas (the third summation in $F^{k \geq 3}_{P_2,1}$). The first two summations are derived in a way similar to what we did for $F^2_{P_2,1}$, with a change in the summation indices; $\bold{R}_2$ is replaced by $\bold{R}_k$ here. For the third (double) summation, $\mathcal{C}_{P_2}$ in (\ref{eq_p2_3}) has $r=1$, $e=h$, and $s=k$. Thus, $\{i_1,i_2\}$ can take any distinct two values in the range from the start of $\bold{R}_k$ to the end of $\bold{R}_1$, i.e., from $(k-1)\gamma$ to $(m+1)\gamma-1$ (see Fig.~\ref{Fig_pat2}). The outer summation is over all possible values of $h$, and we have $1 < h < k$.
\end{IEEEproof}

\section{Proofs of Pattern $P_3$}\label{sec_appc}

\subsection{Proof of Lemma \ref{lem_p3}}

\begin{IEEEproof}
In Case~3.1, the number of instances is the number of ways to choose $2$ overlaps out of $t_{\{i_1,i_2,i_3\}}$ (the pattern has two $c_1 - c_2 - c_3$ overlaps), which is given by (\ref{eq_p3_1}). In Case~3.2, the number of instances is the number of ways to choose $1$ overlap out of $t_{\{i_1,i_2,i_3\}}$ and $1$ overlap out of $t_{\{i_1+(r-e)\gamma, i_2+(r-e)\gamma, i_3+(r-e)\gamma \}}$, which is given by (\ref{eq_p3_2}).
\end{IEEEproof}

\subsection{Proof of Theorem \ref{thm_p3}}

\begin{IEEEproof}
To compute $F_{P_3}$, we use the formula in \cite[Theorem~1]{homa_boo}, with $\chi=m+1$. Since the overlaps of $P_3$ can exist in up to $2$ replicas, we need to find expressions only for $F^1_{P_3,1}$ and $F^{k \geq 2}_{P_3,1}$.

Then, $F^1_{P_3,1}$ is the sum of function $\mathcal{A}_{P_3}$ in (\ref{eq_p3_1}), with $r=1$, over all possible values of $\{i_1,i_2,i_3\}$. Here, $\{i_1,i_2,i_3\}$ can take any distinct three values in the range from the start to the end of $\bold{R}_1$, i.e., from $0$ to $(m+1)\gamma-1$ (see Fig.~\ref{Fig_pat3}).

Moreover, $F^{k \geq 2}_{P_3,1}$ is the sum of function $\mathcal{B}_{P_3}$ in (\ref{eq_p3_2}), with $r=1$ and $e=k$, over all possible values of $\{i_1,i_2,i_3\}$. Here, $\{i_1,i_2,i_3\}$ can take any distinct three values in the range from the start of $\bold{R}_k$ to the end of $\bold{R}_1$, i.e., from $(k-1)\gamma$ to $(m+1)\gamma-1$ (see also Fig.~\ref{Fig_pat3}).
\end{IEEEproof}

\section{Proofs of Pattern $P_4$}\label{sec_appd}

\subsection{Proof of Lemma \ref{lem_p4}}

\begin{IEEEproof}
In Case~4.1, the number of instances is the number of ways to choose $4$ overlaps out of $t_{\{i_1,i_2\}}$ (the pattern has four $c_1 - c_2$ overlaps), which is given by (\ref{eq_p4_1}). In Case~4.2, the number of instances is the number of ways to choose $3$ overlaps out of $t_{\{i_1,i_2\}}$ and $1$ overlap out of $t_{\{i_1+(r-e)\gamma, i_2+(r-e)\gamma \}}$, which is given by (\ref{eq_p4_2}). In Case~4.3, the number of instances is the number of ways to choose $2$ overlaps out of $t_{\{i_1,i_2\}}$ and $2$ overlaps out of $t_{\{i_1+(r-e)\gamma, i_2+(r-e)\gamma \}}$, which is given by (\ref{eq_p4_3}). In Case~4.4, the number of instances is the number of ways to choose $2$ overlaps out of $t_{\{i_1,i_2\}}$, $1$ overlap out of $t_{\{i_1+(r-e)\gamma, i_2+(r-e)\gamma \}}$, and $1$ overlap out of $t_{\{i_1+(r-s)\gamma, i_2+(r-s)\gamma \}}$, which is given by (\ref{eq_p4_4}). In Case~4.5, the number of instances is the number of ways to choose $1$ overlap out of $t_{\{i_1,i_2\}}$, $1$ overlap out of $t_{\{i_1+(r-e)\gamma, i_2+(r-e)\gamma \}}$, $1$ overlap out of $t_{\{i_1+(r-s)\gamma, i_2+(r-s)\gamma \}}$, and $1$ overlap out of $t_{\{i_1+(r-u)\gamma, i_2+(r-u)\gamma \}}$, which is given by (\ref{eq_p4_5}).
\end{IEEEproof}

\subsection{Proof of Theorem \ref{thm_p4}}

\begin{IEEEproof}
To compute $F_{P_4}$, we use the formula in \cite[Theorem~1]{homa_boo}, with $\chi=m+1$. Since the overlaps of $P_4$ can exist in up to $4$ replicas, we need to find expressions only for $F^1_{P_4,1}$, $F^2_{P_4,1}$, $F^3_{P_4,1}$, and $F^{k \geq 4}_{P_4,1}$.

Then, $F^1_{P_4,1}$ is the sum of function $\mathcal{A}_{P_4}$ in (\ref{eq_p4_1}), with $r=1$, over all possible values of $\{i_1,i_2\}$. Here, $\{i_1,i_2\}$ can take any distinct two values in the range from $0$ to $(m+1)\gamma-1$ (see Fig.~\ref{Fig_pat4}).

Regarding $F^2_{P_4,1}$, we need to account for Case~4.2 and Case~4.3. For Case~4.2, we need to distinguish between two situations; when $r < e$ (i.e., replica $\bold{R}_r$, which has three overlaps, comes before replica $\bold{R}_e$), and when $r > e$ (i.e., replica $\bold{R}_r$ comes after replica $\bold{R}_e$). This distinction gives the two summations of function $\mathcal{B}_{P_4}$ in $F^2_{P_4,1}$. For the first summation, $\mathcal{B}_{P_4}$ in (\ref{eq_p4_2}) has $r=1$ and $e=2$. Thus, $\{i_1,i_2\}$ can take any distinct two values in the range from $\gamma$ to $(m+1)\gamma-1$. For the second summation, $\mathcal{B}_{P_4}$ in (\ref{eq_p4_2}) has $r=2$ and $e=1$. Thus, $\{i_1,i_2\}$ can take any distinct two values in the range from $0$ to $m\gamma-1$. The above distinction is not needed for Case~4.3 since the two replicas have the same number of degree-$2$ overlaps. For the third summation, $\mathcal{C}_{P_4}$ in (\ref{eq_p4_3}) has $r=1$ and $e=2$. Thus, $\{i_1,i_2\}$ can take any distinct two values in the range from $\gamma$ to $(m+1)\gamma-1$ (see Fig.~\ref{Fig_pat4} for more illustration).

As for $F^3_{P_4,1}$, the overlaps can be in $2$ replicas (the first three summations in $F^3_{P_4,1}$) or $3$ replicas (the following three summations in $F^3_{P_4,1}$). The first three summations are derived in a way similar to what we did for $F^2_{P_4,1}$, with a change in the summation indices; $\bold{R}_2$ is replaced by $\bold{R}_3$ here. The following three summations are related to Case~4.4. For Case~4.4, we need to distinguish between three situations; when $r < e < s$ (i.e., replica $\bold{R}_r$, which has two overlaps, comes before replicas $\bold{R}_e$ and $\bold{R}_s$ as in Fig.~\ref{Fig_pat4}), when $e < r < s$ (i.e., replica $\bold{R}_r$ comes between replicas $\bold{R}_e$ and $\bold{R}_s$), and when $e < s < r$ (i.e., replica $\bold{R}_r$ comes after replicas $\bold{R}_e$ and $\bold{R}_s$). This distinction gives the three summations of function $\mathcal{D}_{P_4}$ in $F^3_{P_4,1}$. For the fourth summation in $F^3_{P_4,1}$, $\mathcal{D}_{P_4}$ in (\ref{eq_p4_4}) has $r=1$, $e=2$, and $s=3$. Thus, $\{i_1,i_2\}$ can take any distinct two values in the range from the start of $\bold{R}_3$ to the end of $\bold{R}_1$, i.e., from $2\gamma$ to $(m+1)\gamma-1$. For the fifth summation, $\mathcal{D}_{P_4}$ in (\ref{eq_p4_4}) has $r=2$, $e=1$, and $s=3$. Thus, $\{i_1,i_2\}$ can take any distinct two values in the range from the start of $\bold{R}_3$ to the end of $\bold{R}_1$ ($\bold{R}_r$ now is $\bold{R}_2$), i.e., from $\gamma$ to $m\gamma-1$. For the sixth summation, $\mathcal{D}_{P_4}$ in (\ref{eq_p4_4}) has $r=3$, $e=1$, and $s=2$. Thus, $\{i_1,i_2\}$ can take any distinct two values in the range from the start of $\bold{R}_3$ (which is $\bold{R}_r$ now) to the end of $\bold{R}_1$, i.e., from $0$ to $(m-1)\gamma-1$.

Regarding $F^{k \geq 4}_{P_4,1}$, the overlaps can be in $2$ replicas (the first three summations in $F^{k \geq 4}_{P_4,1}$), $3$ replicas (the following three summations in $F^{k \geq 4}_{P_4,1}$), or $4$ replicas (the seventh summation in $F^{k \geq 4}_{P_4,1}$). The first three summations are derived in a way similar to what we did for $F^2_{P_4,1}$, with a change in the summation indices; $\bold{R}_2$ is replaced by $\bold{R}_k$. The following three summations are derived in a way similar to what we did for $\mathcal{D}_{P_4}$ in $F^3_{P_4,1}$, with a change in the summation indices; $\bold{R}_2$ and $\bold{R}_3$ are replaced by $\bold{R}_h$ and $\bold{R}_k$, respectively, which also requires changing these three summations of $\mathcal{D}_{P_4}$ to be double summations. For the seventh (triple) summation, $\mathcal{E}_{P_4}$ in (\ref{eq_p4_5}) has $r=1$, $e=h$, $s=w$, and $u=k$. Thus, $\{i_1,i_2\}$ can take any distinct two values in the range from the start of $\bold{R}_k$ to the end of $\bold{R}_1$, i.e., from $(k-1)\gamma$ to $(m+1)\gamma-1$ (see Fig.~\ref{Fig_pat4}). The outer two summations are over all possible values of $h$ and $w$, and we have $1 < h < k-1$ and $h < w < k$.
\end{IEEEproof}

\section{Proofs of Pattern $P_5$}\label{sec_appe}

\subsection{Proof of Lemma \ref{lem_p5}}

\begin{IEEEproof}
In Case~5.1, the number of instances is the number of ways to choose $2$ overlaps out of $t_{\{i_1,i_2,i_3,i_4\}}$ (the pattern has two $c_1 - c_2 - c_3 - c_4$ overlaps), which is given by (\ref{eq_p5_1}). In Case~5.2, the number of instances is the number of ways to choose $1$ overlap out of $t_{\{i_1,i_2,i_3,i_4\}}$ and $1$ overlap out of $t_{\{i_1+(r-e)\gamma, i_2+(r-e)\gamma, i_3+(r-e)\gamma, i_4+(r-e)\gamma \}}$, which is given in (\ref{eq_p5_2}).
\end{IEEEproof}

\subsection{Proof of Theorem \ref{thm_p5}}

\begin{IEEEproof}
To compute $F_{P_5}$, we use the formula in \cite[Theorem~1]{homa_boo}, with $\chi=m+1$. Since the overlaps of $P_5$ can exist in up to $2$ replicas, we need to find expressions only for $F^1_{P_5,1}$ and $F^{k \geq 2}_{P_5,1}$.

Then, $F^1_{P_5,1}$ is the sum of function $\mathcal{A}_{P_5}$ in (\ref{eq_p5_1}), with $r=1$, over all possible values of $\{i_1,i_2,i_3,i_4\}$. Here, $\{i_1,i_2,i_3,i_4\}$ can take any distinct four values in the range from the start to the end of $\bold{R}_1$, i.e., from $0$ to $(m+1)\gamma-1$ (see Fig.~\ref{Fig_pat5}).

Moreover, $F^{k \geq 2}_{P_5,1}$ is the sum of function $\mathcal{B}_{P_5}$ in (\ref{eq_p5_2}), with $r=1$ and $e=k$, over all possible values of $\{i_1,i_2,i_3,i_4\}$. Here, $\{i_1,i_2,i_3,i_4\}$ can take any distinct four values in the range from the start of $\bold{R}_k$ to the end of $\bold{R}_1$, i.e., from $(k-1)\gamma$ to $(m+1)\gamma-1$ (see also Fig.~\ref{Fig_pat5}).
\end{IEEEproof}

\section{Proofs of Pattern $P_6$}\label{sec_appf}

\subsection{Proof of Lemma \ref{lem_p6}}

\begin{IEEEproof}
In Case~6.1, the number of instances is the number of ways to choose $1$ overlap from each family in $\bold{R}_r$ (there exist three different families for $P_6$; $c_1 - c_2 - c_3$, $c_1 - c_2$, and $c_1 - c_3$). We choose the $c_1 - c_2 - c_3$ degree-$3$ overlap first. Then, in order to avoid over-counting, it is required to distinguish between the two situations when the $c_1 - c_2$ degree-$2$ overlap is part of a $c_1 - c_2 - c_3$ degree-$3$ overlap, and when this is not the case. Taking this requirement into account yields the two added terms in (\ref{eq_p6_1}). The same applies for Case~6.2, with the exception that here the degree-$3$ overlap is chosen from $t_{\{i_1+(r-e)\gamma, i_2+(r-e)\gamma, i_3+(r-e)\gamma\}}$ overlaps, resulting in (\ref{eq_p6_2}). Following the same logic of Case~6.1 for Case~6.3, with the exception that the $c_1 - c_3$ overlap is chosen from $t_{\{i_1+(r-e)\gamma, i_3+(r-e)\gamma\}}$ overlaps, gives (\ref{eq_p6_3}). In Case~6.4, the number of instances is the number of ways to choose $1$ overlap out of $t_{\{i_1,i_2\}}$, $1$ overlap out of $t_{\{i_1+(r-e)\gamma, i_3+(r-e)\gamma \}}$, and $1$ overlap out of $t_{\{i_1+(r-s)\gamma, i_2+(r-s)\gamma, i_3+(r-s)\gamma \}}$, which is given by (\ref{eq_p6_4}).
\end{IEEEproof}

\subsection{Proof of Theorem \ref{thm_p6}}

\begin{IEEEproof}
To compute $F_{P_6}$, we use the formula in \cite[Theorem~1]{homa_boo}, with $\chi=m+1$. Since the overlaps of $P_6$ can exist in up to $3$ replicas, we need to find expressions only for $F^1_{P_6,1}$, $F^2_{P_6,1}$, and $F^{k \geq 3}_{P_6,1}$.

Then, $F^1_{P_6,1}$ is the sum of function $\mathcal{A}_{P_6}$ in (\ref{eq_p6_1}), with $r=1$, over all possible values of $i_1$ and $\{i_2,i_3\}$. In Pattern $P_6$, CN $c_1$, which connects all three VNs, is different from the other two CNs. Moreover, in a group of three CNs that can form $P_6$, $c_1$ can be any one of these three CNs, which means we have three possible ways to form $P_6$ from these three CNs. These facts combined are the reason why $i_1$ of $c_1$ has to be separated from $\{i_2,i_3\}$, despite having the same range, in the expression of $F^1_{P_6,1}$ (this applies for other expressions too). Here, $i_1$ (resp., $\{i_2,i_3\}$) can take any value (resp., distinct two values) in the range from the start to the end of $\bold{R}_1$, i.e., from $0$ to $(m+1)\gamma-1$ (see also Fig.~\ref{Fig_pat6}).

Regarding $F^2_{P_6,1}$, we need to account for Case~6.2 and Case~6.3. For each of the two cases, we need to distinguish between two situations; when $r < e$ and when $r > e$. This distinction gives the two summations of $\mathcal{B}_{P_6}$ and the two summations of $\mathcal{C}_{P_6}$ in $F^2_{P_6,1}$. In Case~6.2, each of the three CNs of $P_6$ connects overlaps in $\bold{R}_r$ and $\bold{R}_e$ (because the degree-$3$ overlap is moved to $\bold{R}_e$). For the first summation in $F^2_{P_6,1}$, $\mathcal{B}_{P_6}$ in (\ref{eq_p6_2}) has $r=1$ and $e=2$. Thus, $i_1$ (resp., $\{i_2,i_3\}$) can take any value (resp., distinct two values) in the range from the start of $\bold{R}_2$ to the end of $\bold{R}_1$, i.e., from $\gamma$ to $(m+1)\gamma-1$. For the second summation, $\mathcal{B}_{P_6}$ in (\ref{eq_p6_2}) has $r=2$ and $e=1$. Thus, $i_1$ (resp., $\{i_2,i_3\}$) can take any value (resp., distinct two values) in the range from $0$ to $m\gamma-1$. In Case~6.3, and as shown in Fig.~\ref{Fig_pat6}, $c_1$ and $c_3$ each connects overlaps in $\bold{R}_r$ and $\bold{R}_e$, while $c_2$ connects overlaps in $\bold{R}_r$ only (because the $c_1 - c_3$ overlap is moved to $\bold{R}_e$ here). For the third summation in $F^2_{P_6,1}$, $\mathcal{C}_{P_6}$ in (\ref{eq_p6_3}) has $r=1$ and $e=2$. Thus, $i_1$ (resp., $i_2$ and $i_3$) can take any value in the range from the start of $\bold{R}_2$ (resp., $\bold{R}_1$ and $\bold{R}_2$) to the end of $\bold{R}_1$, i.e., from $\gamma$ (resp., $0$ and $\gamma$) to $(m+1)\gamma-1$. For the fourth summation, $\mathcal{C}_{P_6}$ in (\ref{eq_p6_3}) has $r=2$ and $e=1$ (see Fig.~\ref{Fig_pat6}). Thus, $i_1$ (resp., $i_2$ and $i_3$) can take any value in the range from the start of $\bold{R}_2$ to the end of $\bold{R}_1$ (resp., $\bold{R}_2$ and $\bold{R}_1$), i.e., from $0$ to $m\gamma-1$ (resp., $(m+1)\gamma-1$ and $m\gamma-1$). Note that the ranges of $i_2$ and $i_3$ are different in Case~6.3, unlike Case~6.2, which is the reason why $i_2$ and $i_3$ are not in a set in the summations of $\mathcal{C}_{P_6}$.

As for $F^{k \geq 3}_{P_6,1}$, the overlaps can be in $2$ replicas (the first four summations in $F^{k \geq 3}_{P_6,1}$) or $3$ replicas (the following three summations in $F^{k \geq 3}_{P_6,1}$). The first four summations are derived in a way similar to what we did for $F^2_{P_6,1}$, with a change in the summation indices; $\bold{R}_2$ is replaced by $\bold{R}_k$ here. The following three summations are associated with Case~6.4. In Case~6.4, $c_1$ connects overlaps in $\bold{R}_r$, $\bold{R}_e$, and $\bold{R}_s$. On the other hand, $c_2$ (resp., $c_3$) connects overlaps in $\bold{R}_r$ (resp., $\bold{R}_e$) and $\bold{R}_s$. For the fifth (double) summation, $\mathcal{D}_{P_6}$ in (\ref{eq_p6_4}) has $r=1$, $e=h$, and $s=k$ (see Fig.~\ref{Fig_pat6}). Thus, $i_1$ (resp., $i_2$ and $i_3$) can take any value in the range from the start of $\bold{R}_k$ to the end of $\bold{R}_1$ (resp., $\bold{R}_1$ and $\bold{R}_h$), i.e., from $(k-1)\gamma$ to $(m+1)\gamma-1$ (resp., $(m+1)\gamma-1$ and $(m+h)\gamma-1$). For the sixth (double) summation, $\mathcal{D}_{P_6}$ in (\ref{eq_p6_4}) has $r=1$, $e=k$, and $s=h$. Thus, $i_1$ (resp., $i_2$ and $i_3$) can take any value in the range from the start of $\bold{R}_k$ (resp., $\bold{R}_h$ and $\bold{R}_k$) to the end of $\bold{R}_1$ (resp., $\bold{R}_1$ and $\bold{R}_h$), i.e., from $(k-1)\gamma$ (resp., $(h-1)\gamma$ and $(k-1)\gamma$) to $(m+1)\gamma-1$ (resp., $(m+1)\gamma-1$ and $(m+h)\gamma-1$). For the seventh (double) summation, $\mathcal{D}_{P_6}$ in (\ref{eq_p6_4}) has $r=h$, $e=k$, and $s=1$. Thus, $i_1$ (resp., $i_2$ and $i_3$) can take any value in the range from the start of $\bold{R}_k$ (resp., $\bold{R}_h$ and $\bold{R}_k$) to the end of $\bold{R}_1$, i.e., from $(k-h)\gamma$ (resp., $0$ and $(k-h)\gamma$) to $(m-h+2)\gamma-1$. The outer summation is over all possible values of $h$, and we have $1 < h < k$.
\end{IEEEproof}

\section{Proofs of Pattern $P_7$}\label{sec_appg}

\subsection{Proof of Lemma \ref{lem_p7}}

\begin{IEEEproof}
In Case~7.1, the number of instances is the number of ways to choose $2$ overlaps from each family in $\bold{R}_r$ (the pattern has two $c_1 - c_2$ overlaps and two $c_1 - c_3$ overlaps). In order to avoid over-counting, it is required to distinguish between the three situations when the two $c_1 - c_2$ overlaps are each part of a $c_1 - c_2 - c_3$ overlap, when only one $c_1 - c_2$ overlap is part of a $c_1 - c_2 - c_3$ overlap, and when neither of them is. Taking this requirement into account yields the three added terms in (\ref{eq_p7_1}). The same applies for Case~7.2, with the exception that here, one $c_1 - c_3$ overlap is chosen from $t_{\{i_1+(r-e)\gamma, i_3+(r-e)\gamma\}}$ overlaps. In Case~7.3, there is no need to make this distinction since both $c_1 - c_3$ overlaps are chosen from $t_{\{i_1+(r-e)\gamma, i_3+(r-e)\gamma\}}$ overlaps (they are in $\bold{R}_e$), and the result is in (\ref{eq_p7_3}). In Case~7.4, the distinction is applied separately on the $c_1 - c_2$ overlap in $\bold{R}_r$ and the $c_1 - c_2$ overlap in $\bold{R}_e$ to give (\ref{eq_p7_4}). Case~7.5 is similar to Case~7.3, with the exception that one of the two $c_1 - c_3$ overlaps is chosen from $t_{\{i_1+(r-s)\gamma, i_3+(r-s)\gamma\}}$ overlaps since it is now in $\bold{R}_s$. Case~7.6 is similar to Case~7.4, with the exception that one $c_1 - c_3$ overlap is chosen from $t_{\{i_1+(r-s)\gamma, i_3+(r-s)\gamma\}}$ overlaps since it is now in $\bold{R}_s$ (was the $c_1 - c_3$ overlap in $\bold{R}_e$ in Case~7.4). Consequently, the above distinction is only applied to the $c_1 - c_2$ overlap in $\bold{R}_r$, which results in (\ref{eq_p7_6}). In Case~7.7, the number of instances is the number of ways to choose $1$ overlap out of $t_{\{i_1,i_2\}}$, $1$ overlap out of $t_{\{i_1+(r-e)\gamma, i_2+(r-e)\gamma \}}$, $1$ overlap out of $t_{\{i_1+(r-s)\gamma, i_3+(r-s)\gamma \}}$, and $1$ overlap out of $t_{\{i_1+(r-u)\gamma, i_3+(r-u)\gamma\}}$, which is given by (\ref{eq_p7_7}).
\end{IEEEproof}

\subsection{Proof of Theorem \ref{thm_p7}}

\begin{IEEEproof}
To compute $F_{P_7}$, we use the formula in \cite[Theorem~1]{homa_boo}, with $\chi=m+1$. Since the overlaps of $P_7$ can exist in up to $4$ replicas, we need to find expressions only for $F^1_{P_7,1}$, $F^2_{P_7,1}$, $F^3_{P_7,1}$, and $F^{k \geq 4}_{P_7,1}$.

Then, $F^1_{P_7,1}$ is the sum of function $\mathcal{A}_{P_7}$ in (\ref{eq_p7_1}), with $r=1$, over all possible values of $i_1$ and $\{i_1,i_2\}$. In Pattern $P_7$, CN $c_1$, which connects all four VNs, is different from the other two CNs. Moreover, in a group of three CNs that can form $P_7$, $c_1$ can be any one of these three CNs, which means we have three possible ways to form $P_7$ from these three CNs. These facts combined are the reason why $i_1$ of $c_1$ has to be separated from $\{i_2, i_3\}$, despite having the same range, in the expression of $F^1_{P_7,1}$ (this applies for other expressions too). Here, $i_1$ (resp., $\{i_2,i_3\}$) can take any value (resp., distinct two values) in the range from the start to the end of $\bold{R}_1$, i.e., from $0$ to $(m+1)\gamma-1$ (see also Fig.~\ref{Fig_pat7}).

Regarding $F^2_{P_7,1}$, we need to account for Case~7.2, Case~7.3, and Case~7.4. For Case~7.2, we need to distinguish between two situations; when $r < e$ and when $r > e$, which gives the two summations of $\mathcal{B}_{P_7}$ in $F^2_{P_7,1}$. In Case~7.2, and as shown in Fig.~\ref{Fig_pat7}, $c_1$ and $c_3$ each connects overlaps in $\bold{R}_r$ and $\bold{R}_e$, while $c_2$ connects overlaps in $\bold{R}_r$ only. For the first summation in $F^2_{P_7,1}$, $\mathcal{B}_{P_7}$ in (\ref{eq_p7_2}) has $r=1$ and $e=2$. Thus, $i_1$ (resp., $i_2$ and $i_3$) can take any value in the range from the start of $\bold{R}_2$ (resp., $\bold{R}_1$ and $\bold{R}_2$) to the end of $\bold{R}_1$, i.e., from $\gamma$ (resp., $0$ and $\gamma$) to $(m+1)\gamma-1$. For the second summation, $\mathcal{B}_{P_7}$ in (\ref{eq_p7_2}) has $r=2$ and $e=1$. Thus, $i_1$ (resp., $i_2$ and $i_3$) can take any value in the range from the start of $\bold{R}_2$ to the end of $\bold{R}_1$ (resp., $\bold{R}_2$ and $\bold{R}_1$), i.e., from $0$ to $m\gamma-1$ (resp., $(m+1)\gamma-1$ and $m\gamma-1$). Note that the ranges of $i_2$ and $i_3$ are different in Case~7.2. The above distinction is not needed for neither Case~7.3 nor Case~7.4 since the two replicas have the same number of degree-$2$ overlaps with similar connectivity. In Case~7.3, $c_1$ connects overlaps in $\bold{R}_r$ and $\bold{R}_e$, while $c_2$ (resp., $c_3$) connects overlaps in $\bold{R}_r$ (resp., $\bold{R}_e$) only. For the third summation in $F^2_{P_7,1}$, $\mathcal{C}_{P_7}$ in (\ref{eq_p7_3}) has $r=1$ and $e=2$. Thus, $i_1$ (resp., $i_2$ and $i_3$) can take any value in the range from the start of $\bold{R}_2$ (resp., $\bold{R}_1$ and $\bold{R}_2$) to the end of $\bold{R}_1$ (resp., $\bold{R}_1$ and $\bold{R}_2$), i.e., from $\gamma$ (resp., $0$ and $\gamma$) to $(m+1)\gamma-1$ (resp., $(m+1)\gamma-1$ and $(m+2)\gamma-1$). Note that the ranges of $i_2$ and $i_3$ are also different in Case~7.3. In Case~7.4, all the CNs connect overlaps in $\bold{R}_r$ and $\bold{R}_e$. For the fourth summation in $F^2_{P_7,1}$, $\mathcal{D}_{P_7}$ in (\ref{eq_p7_4}) has $r=1$ and $e=2$. Thus, $i_1$ (resp., $\{i_2,i_3\}$) can take any value (resp., distinct two values) in the range from the start of $\bold{R}_2$ to the end of $\bold{R}_1$, i.e., from $\gamma$ to $(m+1)\gamma-1$. Note that the ranges of $i_2$ and $i_3$ are the same in Case~7.4 (similar to Case~7.1).

As for $F^3_{P_7,1}$, the overlaps can be in $2$ replicas (the first four summations in $F^3_{P_7,1}$) or $3$ replicas (the following six summations in $F^3_{P_7,1}$). The first four summations are derived in a way similar to what we did for $F^2_{P_7,1}$, with a change in the summation indices; $\bold{R}_2$ is replaced by $\bold{R}_3$ here. Then, we need to account for Case~7.5 (fifth to seventh summations) and Case~7.6 (eighth to tenth summations). In Case~7.5, $c_1$ connects overlaps in $\bold{R}_r$, $\bold{R}_e$, and $\bold{R}_s$. On the other hand, $c_2$ (resp., $c_3$) connects overlaps in $\bold{R}_r$ only (resp., $\bold{R}_e$ and $\bold{R}_s$). For the fifth summation, $\mathcal{E}_{P_7}$ in (\ref{eq_p7_5}) has $r=1$, $e=2$, and $s=3$ (see Fig.~\ref{Fig_pat7}). Thus, $i_1$ (resp., $i_2$ and $i_3$) can take any value in the range from the start of $\bold{R}_3$ (resp., $\bold{R}_1$ and $\bold{R}_3$) to the end of $\bold{R}_1$ (resp., $\bold{R}_1$ and $\bold{R}_2$), i.e., from $2\gamma$ (resp., $0$ and $2\gamma$) to $(m+1)\gamma-1$ (resp., $(m+1)\gamma-1$ and $(m+2)\gamma-1$). For the sixth summation, $\mathcal{E}_{P_7}$ in (\ref{eq_p7_5}) has $r=2$, $e=1$, and $s=3$. Thus, $i_1$ (resp., $i_2$ and $i_3$) can take any value in the range from the start of $\bold{R}_3$ (resp., $\bold{R}_2$ and $\bold{R}_3$) to the end of $\bold{R}_1$ (resp., $\bold{R}_2$ and $\bold{R}_1$), i.e., from $\gamma$ (resp., $0$ and $\gamma$) to $m\gamma-1$ (resp., $(m+1)\gamma-1$ and $m\gamma-1$). For the seventh summation, $\mathcal{E}_{P_7}$ in (\ref{eq_p7_5}) has $r=3$, $e=1$, and $s=2$. Thus, $i_1$ (resp., $i_2$ and $i_3$) can take any value in the range from the start of $\bold{R}_3$ (resp., $\bold{R}_3$ and $\bold{R}_2$) to the end of $\bold{R}_1$ (resp., $\bold{R}_3$ and $\bold{R}_1$), i.e., from $0$ (resp., $0$ and $-\gamma$) to $(m-1)\gamma-1$ (resp., $(m+1)\gamma-1$ and $(m-1)\gamma-1$). In Case~7.6, $c_1$ connects overlaps in $\bold{R}_r$, $\bold{R}_e$, and $\bold{R}_s$. On the other hand, $c_2$ (resp., $c_3$) connects overlaps in $\bold{R}_r$ and $\bold{R}_e$ (resp., $\bold{R}_s$). For the eighth summation, $\mathcal{G}_{P_7}$ in (\ref{eq_p7_6}) has $r=1$, $e=2$, and $s=3$. Thus, $i_1$ (resp., $i_2$ and $i_3$) can take any value in the range from the start of $\bold{R}_3$ (resp., $\bold{R}_2$ and $\bold{R}_3$) to the end of $\bold{R}_1$, i.e., from $2\gamma$ (resp., $\gamma$ and $2\gamma$) to $(m+1)\gamma-1$. For the ninth summation, $\mathcal{G}_{P_7}$ in (\ref{eq_p7_6}) has $r=2$, $e=1$, and $s=3$. Thus, $i_1$ (resp., $i_2$ and $i_3$) can take any value in the range from the start of $\bold{R}_3$ (resp., $\bold{R}_2$ and $\bold{R}_3$) to the end of $\bold{R}_1$ (resp., $\bold{R}_1$ and $\bold{R}_2$), i.e., from $\gamma$ (resp., $0$ and $\gamma$) to $m\gamma-1$ (resp., $m\gamma-1$ and $(m+1)\gamma-1$). For the tenth summation, $\mathcal{G}_{P_7}$ in (\ref{eq_p7_6}) has $r=3$, $e=1$, and $s=2$. Thus, $i_1$ (resp., $i_2$ and $i_3$) can take any value in the range from the start of $\bold{R}_3$ to the end of $\bold{R}_1$ (resp., $\bold{R}_1$ and $\bold{R}_2$), i.e., from $0$ to $(m-1)\gamma-1$ (resp., $(m-1)\gamma-1$ and $m\gamma-1$).

Regarding $F^{k \geq 4}_{P_7,1}$, the overlaps can be in $2$ replicas (the first four summations in $F^{k \geq 4}_{P_7,1}$), $3$ replicas (the following six summations in $F^{k \geq 4}_{P_7,1}$), or $4$ replicas (the last three summations in $F^{k \geq 4}_{P_7,1}$). The first four summations are derived in a way similar to what we did for $F^2_{P_7,1}$, with a change in the summation indices; $\bold{R}_2$ is replaced by $\bold{R}_k$. The following six summations are derived in a way similar to what we did for $F^3_{P_7,1}$, with a change in the summation indices; $\bold{R}_2$ and $\bold{R}_3$ are replaced by $\bold{R}_h$ and $\bold{R}_k$, respectively, which also requires changing these six summations of $\mathcal{E}_{P_7}$ and $\mathcal{G}_{P_7}$ to be double summations. The following three summations are associated with Case~7.7. In Case~7.7, $c_1$ connects overlaps in $\bold{R}_r$, $\bold{R}_e$, $\bold{R}_s$, and $\bold{R}_u$. On the other hand, $c_2$ (resp., $c_3$) connects overlaps in $\bold{R}_r$ and $\bold{R}_e$ (resp., $\bold{R}_s$ and $\bold{R}_u$). See Fig.~\ref{Fig_pat7} for more illustration. There are three situations to distinguish between; the two $c_1 - c_2$ overlaps are in the first and second replicas, in the first and third replicas, and in the first and last replicas. The ordering of replicas here is with respect to the four replicas in which the overlaps of $P_7$ exist. For the eleventh (triple) summation, $\mathcal{I}_{P_7}$ in (\ref{eq_p7_7}) has $r=1$, $e=h$, $s=w$, and $u=k$. Thus, $i_1$ (resp., $i_2$ and $i_3$) can take any value in the range from the start of $\bold{R}_k$ (resp., $\bold{R}_h$ and $\bold{R}_k$) to the end of $\bold{R}_1$ (resp., $\bold{R}_1$ and $\bold{R}_w$), i.e., from $(k-1)\gamma$ (resp., $(h-1)\gamma$ and $(k-1)\gamma$) to $(m+1)\gamma-1$ (resp., $(m+1)\gamma-1$ and $(m+w)\gamma-1$). For the twelfth (triple) summation, $\mathcal{I}_{P_7}$ in (\ref{eq_p7_7}) has $r=1$, $e=w$, $s=h$, and $u=k$. Thus, $i_1$ (resp., $i_2$ and $i_3$) can take any value in the range from the start of $\bold{R}_k$ (resp., $\bold{R}_w$ and $\bold{R}_k$) to the end of $\bold{R}_1$ (resp., $\bold{R}_1$ and $\bold{R}_h$), i.e., from $(k-1)\gamma$ (resp., $(w-1)\gamma$ and $(k-1)\gamma$) to $(m+1)\gamma-1$ (resp., $(m+1)\gamma-1$ and $(m+h)\gamma-1$). For the thirteenth (triple) summation, $\mathcal{I}_{P_7}$ in (\ref{eq_p7_7}) has $r=1$, $e=k$, $s=h$, and $u=w$. Thus, $i_1$ (resp., $i_2$ and $i_3$) can take any value in the range from the start of $\bold{R}_k$ (resp., $\bold{R}_k$ and $\bold{R}_w$) to the end of $\bold{R}_1$ (resp., $\bold{R}_1$ and $\bold{R}_h$), i.e., from $(k-1)\gamma$ (resp., $(k-1)\gamma$ and $(w-1)\gamma$) to $(m+1)\gamma-1$ (resp., $(m+1)\gamma-1$ and $(m+h)\gamma-1$). The outer two summations are over all possible values of $h$ and $w$, and we have $1 < h < k-1$ and $h < w < k$ (similar to Pattern $P_4$).

Note that $c_2$ and $c_3$ are not adjacent (no path of only one VN connects them) in $P_7$, which means it is possible to have $\overline{i_2} = \overline{i_3}$, but not $i_2 = i_3$, for that pattern.
\end{IEEEproof}

\section{Proofs of Pattern $P_8$}\label{sec_apph}

\subsection{Proof of Lemma \ref{lem_p8}}

\begin{IEEEproof}
In Case~8.1, the number of instances is the number of ways to choose $1$ overlap from each family in $\bold{R}_r$ (there exist three different families for $P_8$; $c_1 - c_2 - c_3 - c_4$, $c_1 - c_2$, and $c_3 - c_4$). We choose the $c_1 - c_2 - c_3 - c_4$ degree-$4$ overlap first. Then, in order to avoid over-counting, it is required to distinguish between the two situations when the $c_1 - c_2$ degree-$2$ overlap is part of a $c_1 - c_2 - c_3 - c_4$ degree-$4$ overlap, and when this is not the case. Taking this requirement into account yields the two added terms in (\ref{eq_p8_1}). The same applies for Case~8.2, with the exception that here the degree-$4$ overlap is chosen from $t_{\{i_1+(r-e)\gamma, i_2+(r-e)\gamma, i_3+(r-e)\gamma\, i_4+(r-e)\gamma\}}$ overlaps, resulting in (\ref{eq_p8_2}). Following the same logic of Case~8.1 for Case~8.3, with the exception that the $c_3 - c_4$ overlap is chosen from $t_{\{i_3+(r-e)\gamma, i_4+(r-e)\gamma\}}$ overlaps, gives (\ref{eq_p8_3}). In Case~8.4, the number of instances is the number of ways to choose $1$ overlap out of $t_{\{i_1,i_2\}}$, $1$ overlap out of $t_{\{i_3+(r-e)\gamma, i_4+(r-e)\gamma \}}$, and $1$ overlap out of $t_{\{i_1+(r-s)\gamma, i_2+(r-s)\gamma, i_3+(r-s)\gamma, i_4+(r-s)\gamma \}}$, which is given by (\ref{eq_p8_4}).
\end{IEEEproof}

\subsection{Proof of Theorem \ref{thm_p8}}

\begin{IEEEproof}
To compute $F_{P_8}$, we use the formula in \cite[Theorem~1]{homa_boo}, with $\chi=2m+1$. Since the overlaps of $P_8$ can exist in up to $3$ replicas, we need to find expressions only for $F^1_{P_8,1}$, $F^2_{P_8,1}$, and $F^{k \geq 3}_{P_8,1}$.

Then, $F^1_{P_8,1}$ is the sum of function $\mathcal{A}_{P_8}$ in (\ref{eq_p8_1}), with $r=1$, over all possible values of $\{i_1,i_2\}$ and $\{i_3,i_4\}$. In Pattern $P_8$, CNs $c_1$ and $c_2$ are directly connected twice, and CNs $c_3$ and $c_4$ are directly connected twice, which creates two separate groups of CNs. Moreover, in a group of four CNs that can form $P_8$, $c_1$ and $c_2$ can be any two of these four CNs. These facts combined are the reason why the set $\{i_1,i_2\}$ has to be separated from the set $\{i_3,i_4\}$, despite having the same range, in the expression of $F^1_{P_8,1}$ (this applies for other expressions too). We have $\binom{4}{2}=6$ possible ways to choose $\{i_1, i_2\}$, i.e., to choose $c_1$ and $c_2$ out of the four CNs. However, it does not matter for the count of $\mathcal{A}_{P_8}$ whether the set $\{i_1,i_2\}$ or the set $\{i_3,i_4\}$ is chosen first. Thus, we only have three possible ways to form $P_8$ from these four CNs, and the remaining three ways are repetitive. This fact is the reason why we multiply by $\frac{1}{2}$ in the expression $F^1_{P_8,1}$. Here, $\{i_1,i_2\}$ (resp., $\{i_3,i_4\}$) can take any distinct two values in the range from the start to the end of $\bold{R}_1$, i.e., from $0$ to $(m+1)\gamma-1$ (see also Fig.~\ref{Fig_pat8}).

Regarding $F^2_{P_8,1}$, we need to account for Case~8.2 and Case~8.3. For each of the two cases, we need to distinguish between two situations; when $r < e$ and when $r > e$. This distinction gives the two summations of $\mathcal{B}_{P_8}$ and the two summations of $\mathcal{C}_{P_8}$ in $F^2_{P_8,1}$. In Case~8.2, the multiplication by $\frac{1}{2}$ for the counts of $\mathcal{B}_{P_8}$ is also to account for repetitions (as with $\mathcal{A}_{P_8}$). Moreover, in Case~8.2, each of the four CNs of $P_8$ connects overlaps in $\bold{R}_r$ and $\bold{R}_e$ (because the degree-$4$ overlap is moved to $\bold{R}_e$), as shown in Fig.~\ref{Fig_pat8}. For the first summation in $F^2_{P_8,1}$, $\mathcal{B}_{P_8}$ in (\ref{eq_p8_2}) has $r=1$ and $e=2$ (see also Fig.~\ref{Fig_pat8}). Thus, $\{i_1,i_2\}$ (resp., $\{i_3,i_4\}$) can take any distinct two values in the range from the start of $\bold{R}_2$ to the end of $\bold{R}_1$, i.e., from $\gamma$ to $(m+1)\gamma-1$. For the second summation, $\mathcal{B}_{P_8}$ in (\ref{eq_p8_2}) has $r=2$ and $e=1$. Thus, $\{i_1,i_2\}$ (resp., $\{i_3,i_4\}$) can take any distinct two values in the range from $0$ to $m\gamma-1$. In Case~8.3, and contrarily to Case~8.2, it does matter for the count of $\mathcal{C}_{P_8}$ whether the set $\{i_1,i_2\}$ or the set $\{i_3,i_4\}$ is chosen first because the degree-$2$ overlaps, $c_1 - c_2$ and $c_3 - c_4$, are in two different replicas. Consequently, the multiplication by $\frac{1}{2}$ is not needed in this case. Moreover, in Case~8.3, $c_1$ and $c_2$ each connects overlaps in $\bold{R}_r$ only, while $c_3$ and $c_4$ each connects overlaps in $\bold{R}_r$ and $\bold{R}_e$ (because the $c_3 - c_4$ overlap is moved to $\bold{R}_e$ here). For the third summation in $F^2_{P_8,1}$, $\mathcal{C}_{P_8}$ in (\ref{eq_p8_3}) has $r=1$ and $e=2$. Thus, $\{i_1,i_2\}$ (resp., $\{i_3,i_4\}$) can take any distinct two values in the range from the start of $\bold{R}_1$ (resp., $\bold{R}_2$) to the end of $\bold{R}_1$, i.e., from $0$ (resp., $\gamma$) to $(m+1)\gamma-1$. For the fourth summation, $\mathcal{C}_{P_8}$ in (\ref{eq_p8_3}) has $r=2$ and $e=1$. Thus, $\{i_1,i_2\}$ (resp., $\{i_3,i_4\}$) can take any distinct two values in the range from the start of $\bold{R}_2$ to the end of $\bold{R}_2$ (resp., $\bold{R}_1$), i.e., from $0$ to $(m+1)\gamma-1$ (resp., $m\gamma-1$).

As for $F^{k \geq 3}_{P_8,1}$, the overlaps can be in $2$ replicas (the first four summations in $F^{k \geq 3}_{P_8,1}$) or $3$ replicas (the following three summations in $F^{k \geq 3}_{P_8,1}$). The first four summations are derived in a way similar to what we did for $F^2_{P_8,1}$, with a change in the summation indices; $\bold{R}_2$ is replaced by $\bold{R}_k$ here. The following three summations are associated with Case~8.4. In Case~8.4, $c_1$ and $c_2$ each connects overlaps in $\bold{R}_r$ and $\bold{R}_s$. On the other hand, $c_3$ and $c_4$ each connects overlaps in $\bold{R}_e$ and $\bold{R}_s$. For the fifth (double) summation, $\mathcal{D}_{P_8}$ in (\ref{eq_p8_4}) has $r=1$, $e=h$, and $s=k$. Thus, $\{i_1,i_2\}$ (resp., $\{i_3,i_4\}$) can take any distinct two values in the range from the start of $\bold{R}_k$ to the end of $\bold{R}_1$ (resp., $\bold{R}_h$), i.e., from $(k-1)\gamma$ to $(m+1)\gamma-1$ (resp., $(m+h)\gamma-1$). For the sixth (double) summation, $\mathcal{D}_{P_6}$ in (\ref{eq_p8_4}) has $r=1$, $e=k$, and $s=h$ (see Fig.~\ref{Fig_pat8}). Thus, $\{i_1,i_2\}$ (resp., $\{i_3,i_4\}$) can take any distinct two values in the range from the start of $\bold{R}_h$ (resp., $\bold{R}_k$) to the end of $\bold{R}_1$ (resp., $\bold{R}_h$), i.e., from $(h-1)\gamma$ (resp., $(k-1)\gamma$) to $(m+1)\gamma-1$ (resp., $(m+h)\gamma-1$). For the seventh (double) summation, $\mathcal{D}_{P_8}$ in (\ref{eq_p8_4}) has $r=h$, $e=k$, and $s=1$. Thus, $\{i_1,i_2\}$ (resp., $\{i_3,i_4\}$) can take any distinct two values in the range from the start of $\bold{R}_h$ (resp., $\bold{R}_k$) to the end of $\bold{R}_1$, i.e., from $0$ (resp., $(k-h)\gamma$) to $(m-h+2)\gamma-1$. The outer summation is over all possible values of $h$, and we have $1 < h < k$.
\end{IEEEproof}

\section{Proofs of Pattern $P_9$}\label{sec_appi}

\subsection{Proof of Lemma \ref{lem_p9}}

\begin{IEEEproof}
In Case~9.1, the number of instances is the number of ways to choose $1$ overlap from each family in $\bold{R}_r$ (there exist four different families for $P_9$; $c_1 - c_2$, $c_2 - c_3$, $c_3 - c_4$, and $c_1 - c_4$). In order to avoid over-counting, multiple distinctions need to be performed. For the degree-$2$ overlap $c_1 - c_2$, it is required to distinguish between the four situations when that overlap is part of a $c_1 - c_2 - c_3 - c_4$ degree-$4$ overlap, when that overlap is part of a $c_1 - c_2 - c_3$ degree-$3$ overlap that is not itself part of a $c_1 - c_2 - c_3 - c_4$ degree-$4$ overlap, when that overlap is part of a $c_1 - c_2 - c_4$ degree-$3$ overlap that is not itself part of a $c_1 - c_2 - c_3 - c_4$ degree-$4$ overlap, and when neither of these previous three situations holds. This particular distinction results in having four functions, $\mathcal{A}_{P_9,1}$, $\mathcal{A}_{P_9,2}$, $\mathcal{A}_{P_9,3}$, and $\mathcal{A}_{P_9,4}$. Next, only the degree-$2$ overlaps $c_2 - c_3$, $c_3 - c_4$, and $c_1 - c_4$ need to be chosen. Consequently, for the degree-$2$ overlap $c_2 - c_3$, it is required to distinguish between only three situations; when that overlap is part of a $c_1 - c_2 - c_3 - c_4$ degree-$4$ overlap, when that overlap is part of a $c_2 - c_3 - c_4$ degree-$3$ overlap that is not itself part of a $c_1 - c_2 - c_3 - c_4$ degree-$4$ overlap, and when neither of these previous two situations holds. As for the degree-$2$ overlap $c_3 - c_4$, it is required to distinguish between only two situations; when that overlap is part of a $c_1 - c_3 - c_4$ degree-$3$ overlap, and when this is not the case. Addressing all these distinctions results in (\ref{eq_p9_1}) and (\ref{eq_p9_2}), with six added terms for each of the four functions constituting $\mathcal{A}_{P_9}$.

The same applies for Case~9.2, with the exception that here the degree-$2$ overlap $c_1 - c_4$ is chosen from $t_{\{i_1+(r-e)\gamma, i_4+(r-e)\gamma\}}$ overlaps, which divides the number of added terms in (\ref{eq_p9_2}) by four to reach (\ref{eq_p9_3}). In Case~9.3, the distinction is applied separately on the $c_1 - c_2$ overlap in $\bold{R}_r$ and the $c_3 - c_4$ overlap in $\bold{R}_e$ to give (\ref{eq_p9_4}). The distinction here is between two situations; when the degree-$2$ overlap is part of a degree-$3$ overlap, and when this is not the case. In Case~9.4, the distinction is applied separately on the $c_1 - c_2$ overlap in $\bold{R}_r$ and the $c_2 - c_3$ overlap in $\bold{R}_e$ to give (\ref{eq_p9_5}). The distinction here is between two situations; when the degree-$2$ overlap is part of a degree-$4$ overlap, and when this is not the case. Case~9.5 is similar to Case~6.2, with the exception that here there are two degree-$2$ overlaps outside $\bold{R}_r$, and they are distributed over $\bold{R}_e$ (for the $c_3 - c_4$ overlap) and $\bold{R}_s$ (for the $c_1 - c_4$ overlap). Case~9.6 is similar to Case~8.2, with the exception that here there are two degree-$2$ overlaps outside $\bold{R}_r$, and they are distributed over $\bold{R}_e$ (for the $c_2 - c_3$ overlap) and $\bold{R}_s$ (for the $c_1 - c_4$ overlap). In Case~9.7, the number of instances is the number of ways to choose $1$ overlap out of $t_{\{i_1,i_2\}}$, $1$ overlap out of $t_{\{i_2+(r-e)\gamma, i_3+(r-e)\gamma \}}$, $1$ overlap out of $t_{\{i_3+(r-s)\gamma, i_4+(r-s)\gamma \}}$, and $1$ overlap out of $t_{\{i_1+(r-u)\gamma, i_4+(r-u)\gamma\}}$, which is given by (\ref{eq_p9_8}).
\end{IEEEproof}

\subsection{Proof of Theorem \ref{thm_p9}}

\begin{IEEEproof}
To compute $F_{P_9}$, we use the formula in \cite[Theorem~1]{homa_boo}, with $\chi=2m+1$. Since the overlaps of $P_9$ can exist in up to $4$ replicas, we need to find expressions only for $F^1_{P_9,1}$, $F^2_{P_9,1}$, $F^3_{P_9,1}$, and $F^{k \geq 4}_{P_9,1}$.

Then, $F^1_{P_9,1}$ is the sum of function $\mathcal{A}_{P_9}$ in (\ref{eq_p9_1}), with $r=1$, over all possible values of $\{i_1,i_3\}$ and $\{i_2,i_4\}$. In a group of four CNs, say $c_{x_1}$, $c_{x_2}$, $c_{x_3}$, and $c_{x_4}$, there exist $3$ unique ways to form $P_9$, which is a cycle of length $8$, among them. These $3$ ways are: $c_{x_1} - c_{x_2} - c_{x_3} - c_{x_4}$, $c_{x_1} - c_{x_2} - c_{x_4} - c_{x_3}$, and $c_{x_1} - c_{x_3} - c_{x_2} - c_{x_4}$. In these ways, VNs are omitted for convenience, and the last CN in each way is connected to the first CN through a VN. These facts combined are the reason why we separate $\{i_1,i_3\}$ from $\{i_2,i_4\}$, despite having the same range, in the expression of $F^1_{P_9,1}$ (this applies for other expressions too). Since this separation gives $\binom{4}{2}=6$ options, we multiply by $\frac{1}{2}$ in the expression of $F^1_{P_9,1}$ to account for repetitions. Here, $\{i_1,i_3\}$ (resp., $\{i_2,i_4\}$) can take any distinct two values in the range from the start to the end of $\bold{R}_1$, i.e., from $0$ to $(m+1)\gamma-1$ (see also Fig.~\ref{Fig_pat9}).

Regarding $F^2_{P_9,1}$, we need to account for Case~9.2, Case~9.3, and Case~9.4. For Case~9.2, we need to distinguish between two situations; when $r < e$ and when $r > e$, which gives the two summations of $\mathcal{B}_{P_9}$ in $F^2_{P_9,1}$. In Case~9.2, $c_1$ and $c_4$ each connects overlaps in $\bold{R}_r$ and $\bold{R}_e$, while $c_2$ and $c_3$ each connects overlaps in $\bold{R}_r$ only. For the first summation in $F^2_{P_9,1}$, $\mathcal{B}_{P_9}$ in (\ref{eq_p9_3}) has $r=1$ and $e=2$. Thus, $\{i_1,i_4\}$ can take any distinct two values in the range from the start of $\bold{R}_2$ to the end of $\bold{R}_1$, i.e., from $\gamma$ to $(m+1)\gamma-1$. Moreover, $i_2$ (resp., $i_3$) can take any value in the range from the start to the end of $\bold{R}_1$, i.e., from $0$ to $(m+1)\gamma-1$. For the second summation, $\mathcal{B}_{P_9}$ in (\ref{eq_p9_3}) has $r=2$ and $e=1$. Thus, $\{i_1,i_4\}$ can take any distinct two values in the range from the start of $\bold{R}_2$ to the end of $\bold{R}_1$, i.e., from $0$ to $m\gamma-1$. Moreover, $i_2$ (resp., $i_3$) can take any value in the range from the start to the end of $\bold{R}_2$, i.e., from $0$ to $(m+1)\gamma-1$. Note that the ranges of $i_2$ and $i_3$ are the same in Case~9.2. However, $i_2$ and $i_3$ still need to be separated in order to count all the ways of forming $P_9$. The above distinction is not needed for neither Case~9.3 nor Case~9.4 since the two replicas have the same number of degree-$2$ overlaps with similar connectivity. In Case~9.3, and as shown in Fig.~\ref{Fig_pat9}, $c_1$ and $c_3$ each connects overlaps in $\bold{R}_r$ and $\bold{R}_e$, while $c_2$ (resp., $c_4$) connects overlaps in $\bold{R}_r$ (resp., $\bold{R}_e$) only. For the third summation in $F^2_{P_9,1}$, $\mathcal{C}_{P_9}$ in (\ref{eq_p9_4}) has $r=1$ and $e=2$. Thus, $\{i_1,i_3\}$ can take any distinct two values in the range from the start of $\bold{R}_2$ to the end of $\bold{R}_1$, i.e., from $\gamma$ to $(m+1)\gamma-1$. Moreover, $i_2$ (resp., $i_4$) can take any value in the range from the start to the end of $\bold{R}_1$ (resp., $\bold{R}_2$), i.e., from $0$ (resp., $\gamma$) to $(m+1)\gamma-1$ (resp., $(m+2)\gamma-1$). Note that the ranges of $i_2$ and $i_4$ are different in Case~9.3. In Case~9.4, all the CNs connect overlaps in $\bold{R}_r$ and $\bold{R}_e$. For the fourth summation in $F^2_{P_9,1}$, $\mathcal{D}_{P_9}$ in (\ref{eq_p9_5}) has $r=1$ and $e=2$. Thus, $\{i_1,i_4\}$ can take any distinct two values in the range from the start of $\bold{R}_2$ to the end of $\bold{R}_1$, i.e., from $\gamma$ to $(m+1)\gamma-1$. Moreover, $i_2$ (resp., $i_3$) can take any value in the range from the start of $\bold{R}_2$ to the end of $\bold{R}_1$, i.e., from $\gamma$ to $(m+1)\gamma-1$. Note that the ranges of $i_2$ and $i_3$ are the same in Case~9.4, but they are still separated in order to count all the ways of forming $P_9$. Moreover, the multiplication by $\frac{1}{2}$ for the counts of $\mathcal{D}_{P_9}$ is also to account for repetitions (as with $\mathcal{A}_{P_9}$).

As for $F^3_{P_9,1}$, the overlaps can be in $2$ replicas (the first four summations in $F^3_{P_9,1}$) or $3$ replicas (the following six summations in $F^3_{P_9,1}$). The first four summations are derived in a way similar to what we did for $F^2_{P_9,1}$, with a change in the summation indices; $\bold{R}_2$ is replaced by $\bold{R}_3$ here. Then, we need to account for Case~9.5 (fifth to seventh summations) and Case~9.6 (eighth to tenth summations). In Case~9.5, $c_1$ (resp., $c_2$) connects overlaps in $\bold{R}_r$ and $\bold{R}_s$ (resp., $\bold{R}_r$ only). On the other hand, $c_3$ (resp., $c_4$) connects overlaps in $\bold{R}_r$ and $\bold{R}_e$ (resp., $\bold{R}_e$ and $\bold{R}_s$). For the fifth summation, $\mathcal{E}_{P_9}$ in (\ref{eq_p9_6}) has $r=1$, $e=2$, and $s=3$. Thus, $i_1$ (resp., $i_2$, $i_3$, and $i_4$) can take any value in the range from the start of $\bold{R}_3$ (resp., $\bold{R}_1$, $\bold{R}_2$, and $\bold{R}_3$) to the end of $\bold{R}_1$ (resp., $\bold{R}_1$, $\bold{R}_1$, and $\bold{R}_2$), i.e., from $2\gamma$ (resp., $0$, $\gamma$, and $2\gamma$) to $(m+1)\gamma-1$ (resp., $(m+1)\gamma-1$, $(m+1)\gamma-1$, and $(m+2)\gamma-1$). For the sixth summation, $\mathcal{E}_{P_9}$ in (\ref{eq_p9_6}) has $r=2$, $e=1$, and $s=3$. Thus, $i_1$ (resp., $i_2$, $i_3$, and $i_4$) can take any value in the range from the start of $\bold{R}_3$ (resp., $\bold{R}_2$, $\bold{R}_2$, $\bold{R}_3$) to the end of $\bold{R}_2$ (resp., $\bold{R}_2$, $\bold{R}_1$, and $\bold{R}_1$), i.e., from $\gamma$ (resp., $0$, $0$, and $\gamma$) to $(m+1)\gamma-1$ (resp., $(m+1)\gamma-1$, $m\gamma-1$, and $m\gamma-1$). For the seventh summation, $\mathcal{E}_{P_9}$ in (\ref{eq_p9_6}) has $r=3$, $e=1$, and $s=2$. Thus, $i_1$ (resp., $i_2$, $i_3$, and $i_4$) can take any value in the range from the start of $\bold{R}_3$ (resp., $\bold{R}_3$, $\bold{R}_3$, and $\bold{R}_2$) to the end of $\bold{R}_2$ (resp., $\bold{R}_3$, $\bold{R}_1$, and $\bold{R}_1$), i.e., from $0$ (resp., $0$, $0$, and $-\gamma$) to $m\gamma-1$ (resp., $(m+1)\gamma-1$, $(m-1)\gamma-1$, and $(m-1)\gamma-1$). In Case~9.6, and as shown in  Fig.~\ref{Fig_pat9}, $c_1$ and $c_4$ each connects overlaps in $\bold{R}_r$ and $\bold{R}_s$. On the other hand, $c_2$ and $c_3$ each connects overlaps in $\bold{R}_r$ and $\bold{R}_e$. For the eighth summation, $\mathcal{G}_{P_9}$ in (\ref{eq_p9_7}) has $r=1$, $e=2$, and $s=3$. Thus, $\{i_1,i_4\}$ can take any distinct two values in the range from the start of $\bold{R}_3$ to the end of $\bold{R}_1$, i.e., from $2\gamma$ to $(m+1)\gamma-1$. Moreover, $i_2$ (resp., $i_3$) can take any value in the range from the start of $\bold{R}_2$ to the end of $\bold{R}_1$, i.e., from $\gamma$ to $(m+1)\gamma-1$. For the ninth summation, $\mathcal{G}_{P_9}$ in (\ref{eq_p9_7}) has $r=2$, $e=1$, and $s=3$ (see Fig.~\ref{Fig_pat9}). Thus, $\{i_1,i_4\}$ can take any distinct two values in the range from the start of $\bold{R}_3$ to the end of $\bold{R}_2$, i.e., from $\gamma$ to $(m+1)\gamma-1$. Moreover, $i_2$ (resp., $i_3$) can take any value in the range from the start of $\bold{R}_2$ to the end of $\bold{R}_1$, i.e., from $0$ to $m\gamma-1$. For the tenth summation, $\mathcal{G}_{P_9}$ in (\ref{eq_p9_7}) has $r=3$, $e=1$, and $s=2$. Thus, $\{i_1,i_4\}$ can take any distinct two values in the range from the start of $\bold{R}_3$ to the end of $\bold{R}_2$, i.e., from $0$ to $m\gamma-1$. Moreover, $i_2$ (resp., $i_3$) can take any value in the range from the start of $\bold{R}_3$ to the end of $\bold{R}_1$, i.e., from $0$ to $(m-1)\gamma-1$.

Regarding $F^{k \geq 4}_{P_9,1}$, the overlaps can be in $2$ replicas (the first four summations in $F^{k \geq 4}_{P_9,1}$), $3$ replicas (the following six summations in $F^{k \geq 4}_{P_9,1}$), or $4$ replicas (the last three summations in $F^{k \geq 4}_{P_9,1}$). The first four summations are derived in a way similar to what we did for $F^2_{P_9,1}$, with a change in the summation indices; $\bold{R}_2$ is replaced by $\bold{R}_k$. The following six summations are derived in a way similar to what we did for $F^3_{P_9,1}$, with a change in the summation indices; $\bold{R}_2$ and $\bold{R}_3$ are replaced by $\bold{R}_h$ and $\bold{R}_k$, respectively, which also requires changing these six summations of $\mathcal{E}_{P_9}$ and $\mathcal{G}_{P_9}$ to be double summations. The following three summations are associated with Case~9.7. In Case~9.7, $c_1$ (resp., $c_2$) connects overlaps in $\bold{R}_r$ and $\bold{R}_u$ (resp., $\bold{R}_r$ and $\bold{R}_e$). On the other hand, $c_3$ (resp., $c_4$) connects overlaps in $\bold{R}_e$ and $\bold{R}_s$ (resp., $\bold{R}_s$ and $\bold{R}_u$). See Fig.~\ref{Fig_pat9} for more illustration. The two overlaps connected to the $c_1 - c_2$ overlap through CNs are the $c_2 - c_3$ and the $c_1 - c_4$ overlaps. There are three situations to distinguish between; these two overlaps are in the second and last replicas, in the third and last replicas, and in the second and third replicas. The ordering of replicas here is with respect to the four replicas in which the overlaps of $P_9$ exist. For the eleventh (triple) summation, $\mathcal{I}_{P_9}$ in (\ref{eq_p9_8}) has $r=1$, $e=h$, $s=w$, and $u=k$ (see Fig.~\ref{Fig_pat9}). Thus, $i_1$ (resp., $i_2$, $i_3$, and $i_4$) can take any value in the range from the start of $\bold{R}_k$ (resp., $\bold{R}_h$, $\bold{R}_w$, and $\bold{R}_k$) to the end of $\bold{R}_1$ (resp., $\bold{R}_1$, $\bold{R}_h$, and $\bold{R}_w$), i.e., from $(k-1)\gamma$ (resp., $(h-1)\gamma$, $(w-1)\gamma$, and $(k-1)\gamma$) to $(m+1)\gamma-1$ (resp., $(m+1)\gamma-1$, $(m+h)\gamma-1$, and $(m+w)\gamma-1$). For the twelfth (triple) summation, $\mathcal{I}_{P_9}$ in (\ref{eq_p9_8}) has $r=1$, $e=w$, $s=h$, and $u=k$. Thus, $i_1$ (resp., $i_2$, $i_3$, and $i_4$) can take any value in the range from the start of $\bold{R}_k$ (resp., $\bold{R}_w$, $\bold{R}_w$, and $\bold{R}_k$) to the end of $\bold{R}_1$ (resp., $\bold{R}_1$, $\bold{R}_h$, and $\bold{R}_h$), i.e., from $(k-1)\gamma$ (resp., $(w-1)\gamma$, $(w-1)\gamma$, and $(k-1)\gamma$) to $(m+1)\gamma-1$ (resp., $(m+1)\gamma-1$, $(m+h)\gamma-1$, and $(m+h)\gamma-1$). For the thirteenth (triple) summation, $\mathcal{I}_{P_9}$ in (\ref{eq_p9_8}) has $r=1$, $e=h$, $s=k$, and $u=w$. Thus, $i_1$ (resp., $i_2$, $i_3$, and $i_4$) can take any value in the range from the start of $\bold{R}_w$ (resp., $\bold{R}_h$, $\bold{R}_k$, and $\bold{R}_k$) to the end of $\bold{R}_1$ (resp., $\bold{R}_1$, $\bold{R}_h$, and $\bold{R}_w$), i.e., from $(w-1)\gamma$ (resp., $(h-1)\gamma$, $(k-1)\gamma$, and $(k-1)\gamma$) to $(m+1)\gamma-1$ (resp., $(m+1)\gamma-1$, $(m+h)\gamma-1$, and $(m+w)\gamma-1$). The outer two summations are over all possible values of $h$ and $w$, and we have $1 < h < k-1$ and $h < w < k$ (similar to Patterns $P_4$ and $P_7$).

Note that $c_1$ and $c_3$ are not adjacent in $P_9$, and the same applies for $c_2$ and $c_4$. Thus, it is possible to have $\overline{i_1} = \overline{i_3}$ and $\overline{i_2} = \overline{i_4}$, but not $i_1 = i_3$ nor $i_2 = i_4$, for that pattern.
\end{IEEEproof}

\end{appendices}

\section*{Acknowledgement}

The authors would like to thank Homa Esfahanizadeh and Andrew Tan for their assistance in carrying out this research. The work is supported in part by an ASTC-IDEMA grant.


\end{document}